\documentclass{aa}
\bibpunct{(}{)}{;}{a}{}{,} 

\let\jnlstyle=\rm
\def\refjnl#1{{\jnlstyle#1}}

\def\apjl{\refjnl{ApJ}}                
\def\apjs{\refjnl{ApJS}}               
\def\ao{\refjnl{Appl.~Opt.}}           
\def\aap{\refjnl{A\&A}}                

\input{psfig.sty}

\usepackage{graphicx}
\usepackage{epsfig}
\usepackage{amssymb}
\usepackage{amsmath}
\usepackage{amsfonts}
\usepackage{txfonts}
\usepackage{multirow}
\usepackage{afterpage}
\usepackage{color}

\definecolor{mypink1}{rgb}{0.858, 0.188, 0.478}
\definecolor{mygreen1}{rgb}{0.258, 0.788, 0.878}
\definecolor{myorange1}{rgb}{0.5, 0.2, 0.2}

\title{Cosmic Shear Covariance Matrix in $w$CDM: Cosmology Matters}
\author{J. Harnois-D\'{e}raps \inst{\ref{inst1}}\thanks{jharno@roe.ac.uk} \and  B. Giblin \inst{\ref{inst1}} \and B. Joachimi \inst{\ref{inst2}}}
\institute{
Scottish Universities Physics Alliance, Institute for Astronomy, University of Edinburgh, Blackford Hill, Scotland, UK \label{inst1} \and Department of Physics and Astronomy, University College London, Gower Street, London, WC1E 6BT, UK \label{inst2}}

\begin{document}

\date{Received 10$^{\rm th}$ May 2019 / Accepted 27$^{\rm th}$ September 2019}


\label{firstpage}

\abstract{
We present here the {\it cosmo}-SLICS, a new suite of simulations specially designed for the analysis of current and upcoming weak lensing data beyond the standard two-point cosmic shear. We sampled the $[\Omega_{\rm m}, \sigma_8, h, w_0]$ parameter space at 25 points organised in a Latin hyper-cube, spanning a range that contains most of the $2\sigma$ posterior distribution from ongoing lensing surveys. At each of these nodes we evolved a pair of $N$-body simulations in which the sampling variance is highly suppressed, and ray-traced the volumes 800 times to further increase the effective sky coverage. We extracted a lensing covariance matrix from these {\it pseudo}-independent light-cones and show that it closely matches a brute-force construction based on an ensemble of 800 truly independent $N$-body runs. More precisely, a Fisher analysis reveals that both methods yield marginalized two-dimensional constraints that vary by less than 6\% in area, a result that holds under different survey specifications and that matches to within 15\% the area obtained from an analytical covariance calculation. Extending this comparison with our 25 $w$CDM models, we probed  the cosmology dependence of the lensing covariance directly from numerical simulations, reproducing remarkably well the Fisher results from the analytical models at most cosmologies.
We demonstrate that varying the cosmology at which the covariance matrix is evaluated in the first place might have an order of magnitude greater impact  on the parameter constraints than varying the choice of covariance estimation technique.  We present a test case in which we generate fast predictions for both the lensing signal and its associated variance  with a flexible Gaussian process regression emulator, achieving an accuracy of a few percent on the former and 10\% on the latter.
}


\maketitle
 

\section{Introduction}

Weak lensing has recently emerged as an accurate probe of cosmology, exploiting the high-quality photometric data recorded by dedicated surveys such as the Canada-France-Hawaii Telescope Lensing Survey\footnote{http://www.cfhtlens.org} (CFHTLenS hereafter), the Kilo Degree Survey\footnote{http://kids.strw.leidenuniv.nl} (KiDS), the Dark Energy Survey\footnote{http://darkenergysurvey.org} (DES) and the Hyper Suprime-Cam Survey\footnote{https://hsc.mtk.nao.ac.jp/ssp/} (HSC).
These collaborations have  developed a number of tools to model, extract and analyse the cosmic shear signal -- the weak lensing distortions imprinted on the image of background galaxies by the foreground large scale structures \citep[see][for reviews]{2001PhR...340..291B,2015RPPh...78h6901K}. 

Given a catalogue of galaxies with shear and redshift estimates, there exist many ways to extract the lensing information that is required to constrain the underlying cosmological parameters that describe our Universe at its largest scales. The central approach adopted by the above-mentioned surveys starts with the measurement of a two-point summary statistics, either the configuration-space correlation function \citep[as in][]{Kilbinger2013,KiDS450,DES1_Troxel, KV450} or the Fourier-space power spectra \citep[as in][]{2015PhRvD..91f3507L, KiDS450_QE,2018arXiv180909148H}.

The motivations for choosing these statistics are multiple and compelling:  
the accuracy of the signal predictions is better than a percent over many scales \citep[see e.g.][]{MeadFit}, while the effect of most known systematic effects can be either modelled, measured, mitigated, self-calibrated, or suppressed with simple cuts applied on the data vector. Examples of such effects include the secondary signal caused by the intrinsic alignment of galaxies \citep{2015SSRv..193....1J,2015SSRv..193...67K, 2015SSRv..193..139K}, the strong  baryon feedback processes that modify the lensing signal at small and intermediate scales \citep{Semboloni11} or the relatively large uncertainty on the source redshift distribution and on the shape measurement. For a recent review of the many systematics that affect weak lensing measurements, see \citet[][]{2017arXiv171003235M}. 

In the case of  two-point functions, it has been possible to model or parameterise most of these effects in a way that allows for an efficient  marginalisation, and therefore leads to a potentially unbiased estimation of the cosmological parameters \citep{2018arXiv180309795M}.
These statistics benefit from another key advantage, which is that there exist analytical calculations that describe the covariance of the signal \citep[see, e.g.,][]{HEPT,Takada2009a, cosmolike}. In addition to its reduced computational cost compared to the  simulation-based ensemble approach, this estimate is noise-free, providing a significant gain in stability during the inversion process that occurs within the cosmological inference segment of the analysis. For these reasons, the analytical approach  stands out as a prime method for evaluating the statistical uncertainties in cosmic shear analyses \citep{KiDS450,KV450, 2018arXiv180909148H, DES1_Troxel}. 
The caveat is that its accuracy is not well established, and comparisons with the ensemble approach yield discrepancies. \citet{KiDS450}, for example, show that swapping the covariance matrix from a simulation-based to the analytic method shifts the cosmological results by more than 0.5$\sigma$. This clearly calls for further investigations in both methods, which have yet to come.

Although two-point functions are powerful and clean summary statistics, they do not capture  all the cosmological information contained within the lensing data, and hence they are sub-optimal in that sense. The situation would be different if the matter distribution resembled a Gaussian random field, however 
gravity introduces a variety of non-Gaussian features  that can only be captured by  higher-order statistics. 
%
%
%
Accessing this additional information generally results in an improved constraining power on the cosmological parameters  with the same data, as demonstrated in lensing data analyses based on alternative estimators such as the  bispectrum \citep{Fu2014}, the peak count statistics \citep{ 2015PhRvD..91f3507L, 2015MNRAS.450.2888L, 2016MNRAS.463.3653K, 2017arXiv170907678M, 2018MNRAS.474.1116S}, the Minkowski functionals \citep{2015PhRvD..91j3511P},  clipped lensing \citep{Giblin18}, the  density-split lensing  statistics \citep{Brouwer, Gruen2017}  or convolutional neural networks \citep{Fluri19}. Recent studies further suggest that some of these new methods on their own could outperform the two-point cosmic shear at constraining the sum of neutrino masses, and further help in constraining many other parameters (notably $\Omega_{\rm m}$ and $\sigma_8$) when analysed jointly with the two-point functions \citep{MassiveNu1, MassiveNu2, MassiveNu3, MassiveNu4}.  Moreover, there is growing evidence that some of these methods could be particularly helpful for probing modifications to the theory of General Relativity  \citep[see][for modified gravity analyses with peak counts and machine learning methods]{2016PhRvL.117e1101L,2018arXiv181011030P,2018A&A...619A..38P}.
These are all compelling reasons to further refine such promising tools, but at the moment they are often regarded as immature alternatives to the standard two-point functions
for a number of reasons. 

%

Indeed, developing a new analysis strategy relies heavily on weak lensing numerical simulations for modelling the primary and secondary signals, for covariance estimation and for understanding the impact of residual systematics in the data. Furthermore, these simulations must meet a number of requirements: the redshift distribution of the mock source galaxies has to match that of the data; 
the noise properties must be closely reproduced; the cosmology coverage of the simulations must be wide enough for the likelihood analysis\footnote{This precise requirement has been a severe limitation for cosmic emulators based on the {\it Coyote Universe} \citep{Heitmann2013} or the {\it Mira Titan} simulations \citep{MiraTitan}, which span a parameter space that is too restricted for current lensing data.};
the overall accuracy in the non-linear growth of structure has to be sufficiently high to correctly  model the physical scales involved in the measurement. For instance, the \citet[][DH10 hereafter]{DH10} simulations were used a number of times \citep{2016MNRAS.463.3653K,2017arXiv170907678M, Giblin18} and  have been shown by the latest of these analyses to be only 5-10\% accurate on the cosmic shear correlation functions, a level that is problematic given the increasing statistical power of lensing surveys. Other limitations such as the box size and the mass resolution must further be taken into account in the calibration, carefully understanding what parts of a given lensing estimator are affected by these. To illustrate this point, consider the {\it DarkMatter} simulation suite\footnote{http://columbialensing.org/\#dm}  described in \citet{Matilla2017},  where $512^3$ particles were evolved in volumes of $240h^{-1}$Mpc on the side (see Table \ref{table:prior} for more details on existing lensing simulation suites). Such a small box size significantly affects the measurement of shear correlation functions at the degree scale, but has negligible impact on the lensing power spectrum, peak counts or PDF count analyses. Understanding these properties is therefore an integral part of the development of new lensing estimators.

In this paper we introduce a new suite of simulations, the {\it cosmo}-SLICS, which are primarily designed to calibrate novel weak lensing measurement statistics and enable competitive cosmological analyses with  current weak lensing data. We followed the global numerical setup of the SLICS simulations\footnote{SLICS: https://slics.roe.ac.uk} \citep[][HD18 hereafter]{SLICS} in terms of volume and particle number, which accurately model the cosmic shear signal and covariance over a wide range of scales and are central to many CFHTLenS and KiDS data analyses \citep[e.g.][]{2017MNRAS.465.2033J, KiDS450, 2017arXiv170706627J, 2017arXiv170605004V, Amon2017, Giblin18}. We varied four cosmological parameters over a range informed by current constraints from weak lensing experiments: the matter density $\Omega_{\rm m}$, a combination of the matter density and clumpiness $S_8\equiv \sigma_8 \sqrt{\Omega_{\rm m}/0.3}$, the dark energy equation of state $w_{\rm 0}$ and the reduced Hubble parameter $h$. We sampled this four-dimensional volume at 25 points organised in a Latin hyper-cube, and developed a general cosmic shear emulator based on Gaussian process regression, similar to the tool discussed in e.g. \citet{2008PhRvD..78f3529S, 2010ApJ...713.1322L} and \citet{MassiveNuS}, but in principle applicable to any statistics.
We show in the Appendix that with as few as 25 nodes, the interpolation accuracy is at the percent level over the scales relevant to lensing analyses with two-point statistics, for most of the four-dimensional parameter volume. 
Our emulator is fast, flexible and easily interfaces with an MCMC sampler. 
 
When calibrating an estimator with a small number of $N$-body simulations, one needs to consider the impact of sampling variance. This becomes an important issue especially when the measurement is sensitive to large angular scales that fluctuate the most. We suppressed this effect with a mode-cancellation technique that preserves Gaussianity in the initial density fields, unlike the method presented in  \citet{AnguloPontzen} that sacrifice this statistical property, but achieve a higher level of cancellation. Our approach has a significant advantage that becomes clear in the following use. 


As a first application, we investigate the accuracy of a weak lensing covariance matrix estimated from the {\it cosmo}-SLICS, when compared to the results from 800 truly independent simulations. 
We revisit and reinforce the  findings from \citet{Petri16}, according to which the lensing covariance matrix can be estimated from a reduced number of independent realisations. 
We discuss  the reasons why this works so well with the {\it cosmo}-SLICS, and how this can be put to use. In particular, the smaller computational cost allows us to  explore the cosmological dependence of the  covariance matrices in a four-dimensional parameter space, eventually for any lensing estimator.  The variations with cosmology are known to matter to some level, and its impact on the inferred cosmological parameters  could lead to important biases if neglected \citep{2009A&A...502..721E,2017arXiv170605004V}. A recent forecast by \citet{2018arXiv181111584K} suggests that the impact on a LSST-like survey would be negligible provided that the fixed covariance is evaluated at the true cosmology, which is {\it a priori} unknown. 
Indeed, under assumption of Gaussian field, a Gaussian likelihood approximation with fixed covariance recovers the mode and second moments of the true likelihood, as shown by \citet{2013A&A...551A..88C}. The most accurate posterior with a Gaussian likelihood can therefore be obtained by choosing a covariance model that adopts the best-fit parameters. This can in practice be achieved by the iterative scheme  of \citet{2017arXiv170605004V}, which observe a clear improvement on the accuracy of the cosmological constraints, however it requires either  access to a cosmology-dependent covariance estimator, or to the matrix evaluated at the best-fit cosmology. So far this was only feasible with two-point analyses, however the simulations presented in this paper, combined with our flexible emulator, facilitate incorporating the full cosmological dependence of the covariance for arbitrary statistics into the parameter estimation.

In the context of  the lensing  power spectrum in a $w$CDM universe, we verify our covariance estimation against analytical predictions based on the halo model and find a reasonable match, although not for all cosmologies.  
We study the importance of these differences with Fisher forecasts, assuming different covariance matrix scenarios and different survey configurations. Notably, we investigate whether the impact on the parameter constraints is larger for variations in the cosmology with a fixed covariance estimator, or for variations in estimators at a fixed cosmology. This question is central for determining the next steps to take in the preparation of the lensing analyses for next generation surveys. 


This document is structured as follow: we review in Section \ref{sec:th} the theoretical background and methods; in Section \ref{sec:simulations} we describe the construction and assess the accuracy of the numerical simulations; we present in Section \ref{sec:Covariance} our comparison between different covariance matrix estimation techniques, and investigate their impact on cosmological parameter measurements; we discuss our results and conclude in Section \ref{sec:conclusion}.  Further details on the simulations, the emulator and the analytical covariance matrix calculations can be found in the appendices. 

\section{Theoretical Background}
\label{sec:th}

In this Section we present an overview of the background required to carry out these investigations. We first review the modelling aspect of the two-point functions and the corresponding covariance, then describe how these quantities are measured from numerical simulations, and finally we lay down the Fisher forecast formalism that we later use as a metric to measure the effect on cosmological parameter measurements of adopting (or not) a cosmology-dependent covariance matrix. Although our main science goal is to outgrow the two-point statistics, they nevertheless remain an excellent point of comparison that most experts can easily relate to. The method described here can be straightforwardly extended to any other lensing estimator, however we leave this for future work.

\subsection{2-point weak lensing model}
\label{subsec:Cell}

The basic approach of two-point cosmic shear  is that the cosmology dependence is captured by the 
matter power spectrum, $P(k,z)$, which is therefore the fundamental quantity we attempt to measure.
Many tools exist to compute $P(k,z)$, including fit functions such as  {\sc HaloFit} \citep{Smith03, Takahashi2012}, emulators \citep{Heitmann2013, DarkEmulator}, the halo model \citep{MeadFit} or the  reaction approach \citep{2018arXiv181205594C}.
The weak lensing power spectrum $C_{\ell}^{\kappa}$ is related to the matter power spectrum  by\footnote{While $C_{\ell}$ in principle refers to full-sky calculations with $\ell$ taking on integer values, we consistently use the flat-sky approximation in this work, and hence $\ell$ should be interpreted as real-valued.}:
\begin{eqnarray}
C_{\ell}^{\kappa} = \int_{0}^{\chi_{\rm H}} \frac{{\rm d}\chi}{\chi^2} W^{2}(\chi) P\left(\frac{\ell+1/2}{\chi},z(\chi)\right),
\label{eq:C_ell}
\end{eqnarray}
where $\chi_{\rm H}$ is the comoving distance to the horizon, $\ell = k\chi$ and $W(\chi)$ is the lensing efficiency function for lenses at redshift $z(\chi)$, which depends on the source redshift distribution $n(z)$ via:
\begin{eqnarray}
W(\chi) = \frac{3 H_0^2 \Omega_{\rm m}}{2 c^2}\chi (1+z)\!\!\int_{\chi}^{\chi_{\rm H}}\!\!n(\chi')\frac{\chi' - \chi}{\chi'}{\rm d}\chi'.
\end{eqnarray}
Here $H_0$ is the value of the Hubble parameter today, $c$ is the speed of light in vacuum, and $n(\chi) = n(z){\rm d}\chi / {\rm d}z$.  
The lensing power spectrum (equation  \ref{eq:C_ell}) is directly converted into the cosmic shear correlation function $\xi_{\pm}(\vartheta)$ with:
\begin{eqnarray}
\xi_{\pm}(\vartheta) = \frac{1}{2\pi}\int_{0}^{\infty} \!\! C_{\ell}^{\kappa}\, {\rm J}_{0/4}(\vartheta \ell) \ell {\rm d}\ell,
\label{eq:C2xi}
\end{eqnarray}
where $\vartheta$ is the angular separation on the sky, and ${\rm J}_{0/4}(x)$ are Bessel functions of the first kind.
Equations (\ref{eq:C_ell} - \ref{eq:C2xi}) are quickly computed with line-of-sight integrators 
such as {\sc Nicaea}\footnote{{\sc Nicaea}: www.cosmostat.org/software/nicaea/} or {\sc cosmoSIS}\footnote{{\sc CosmoSIS}: https://bitbucket.org/joezuntz/cosmosis/wiki/Home}, and we refer to  \citet{2017MNRAS.469.2737K} and \citet{2017arXiv170205301K} for recent reviews
on the accuracy of this  lensing model.



\subsection{2-point weak lensing covariance}
\label{subsec:Cov_Cell_th}

Essential to any analysis of the cosmic shear 2-point function is an estimate of the lensing power spectrum covariance matrix, ${\rm Cov_{{\rm tot}}^{\kappa}}$, that enters in the likelihood calculation from which the best fit cosmological parameters are extracted. 
This covariance matrix consists of  three contributions, often written as:
\begin{eqnarray}
{\rm Cov_{{\rm tot}}^{\kappa}} = {\rm Cov_{{\rm G}}^{\kappa}} + {\rm Cov_{{\rm NG}}^{\kappa}} + {\rm Cov_{{\rm SSC}}^{\kappa}}.
\label{eq:Cov_ell}
\end{eqnarray}

The first term on the right-hand side is referred to as the `Gaussian covariance', which would be the only contribution if the matter field was Gaussian. It can be calculated as: 
\begin{eqnarray}
{\rm Cov}_{\rm G}^{\kappa} =  \frac{2}{N_{\ell}} \left[C_{\ell}^{\kappa} + \frac{\sigma_{\epsilon}^2}{\bar{n}}\right]^2 \delta_{\ell \ell'},
\label{eq:GaussCov}
\end{eqnarray}
where $C_{\ell}^{\kappa}$ is evaluated from equation (\ref{eq:C_ell}), $\sigma_{\epsilon}$ characterizes the intrinsic shape noise (per component) of the galaxy sample, $\bar{n}$ is the mean galaxy density of the source sample, and $N_{\ell}$ is the number of independent multipoles being measured in a bin centred on $\ell$ and with a width $\Delta \ell$. The quantity $N_{\ell}$  scales linearly with the area of the survey as $2 N_{\ell} = (2\ell +1) f_{\rm sky} \Delta \ell$, $f_{\rm sky}$ being the sky fraction defined as ${A_{\rm survey}}/(4\pi)$.  The term $\delta_{\ell \ell'}$ is the Kronecker delta function, and its role is to forbid any correlation  between different multipoles, one of the key properties of the Gaussian term.

The second term of equation (\ref{eq:Cov_ell}) is the `non-Gaussian connected term', which introduces a coupling between the measurements at multipoles $\ell$ and $\ell'$. This enhances the overall variance and further makes the off-diagonal elements non-zero, by an amount that depends on the parallel configurations of the connected trispectrum, $T^{\kappa}(\boldsymbol{\ell},-\boldsymbol{\ell},\boldsymbol{\ell}',-\boldsymbol{\ell'})$, which can be computed analytically either from a halo-model approach \citep{Takada2009a} or from perturbation theory \citep{HEPT}.
The ${\rm Cov_{{\rm NG}}^{\kappa}}$ term is then given by:
\begin{eqnarray}
{\rm Cov_{{\rm NG}}^{\kappa}} = \frac{1}{A_{\rm survey}}
\int_{|\boldsymbol{\ell}|\in \ell} 
\frac{{\rm d}{\boldsymbol{\ell}^2}}{A(\ell)} 
\int_{|\boldsymbol{\ell'}|\in \ell'} 
\frac{{\rm d}{\boldsymbol{\ell'}^2}}{A(\ell')}T^{\kappa}(\boldsymbol{\ell},\boldsymbol{-\ell}, \boldsymbol{\ell}', -\boldsymbol{\ell}'),
\end{eqnarray}
where $A(\ell)$ is the area of an annulus in multipole-space covering the bin centred on $\ell$.
The lensing trispectrum $T^{\kappa}$ is computed in the Limber approximation from the three-dimensional matter trispectrum $T_{\delta}$: 
\begin{eqnarray}
T^{\kappa}(\boldsymbol{\ell}_1,\boldsymbol{\ell}_2, \boldsymbol{\ell}_3, \boldsymbol{\ell}_4) = \int_0^{\chi_{\rm H}} \frac{{\rm d}\chi}{\chi^6} W^{4}(\chi)T^{\delta}(\boldsymbol{k}_1,\boldsymbol{k}_2, \boldsymbol{k}_3, \boldsymbol{k}_4,z(\chi)). 
\label{eq:trispectrum}
\end{eqnarray}

The last term in equation (\ref{eq:Cov_ell}) is called the `Super Sample Covariance', or SSC, which describes the coupling of survey modes to background density fluctuations $\delta_{\rm b}$ larger than the survey window $M$. It is evaluated as \citep{SSC,Takada2013a}:
\begin{eqnarray}
{\rm Cov_{{\rm SSC}}^{\kappa}} = \nonumber
\end{eqnarray}
\begin{eqnarray}
\frac{1}{A_{\rm survey}} \!\! \int_0^{\chi_{\rm H}} \!\! \frac{{\rm d}\chi}{\chi^6} W^{4}(\chi)\; \sigma^{2}_{\rm b}(\chi, {\cal M}) \left(\frac{\partial P(k,z)}{\partial \delta_{\rm b}}\right) \left(\frac{\partial P(k',z)}{\partial \delta_{\rm b}}\right),
\label{eq:Cov_SSC}
\end{eqnarray}
with $k = \ell/\chi$, $k' = \ell'/\chi$ and $z = z(\chi)$.
The term $\sigma_{\rm b}$ denotes the variance of super-survey modes for the mask ${\cal M}$, 
while the derivatives of the power spectrum can be estimated from e.g. separate universe simulations or fit functions to these results \citep{SSC, SSC_Barreira}, or from the halo model directly \citep{Takada2013a}. 
 Note that to first order, this SSC term also scales with the inverse of the survey area.

In this paper we employ the halo model to compute the matter trispectrum and the response of the power spectrum to background modes, using the same implementation that was validated with numerical simulations in \citet{KiDS450} and \citet{2017arXiv170605004V}. Details of the code are provided in Appendix \ref{sec:CovAna}. 
In order to match the simulations, we considered a survey area of 100 deg$^2$ in these calculations, and the mask ${\cal M}$ is assumed to be square. Beyond the SSC term, no survey boundary effects were incorporated in the model in this work.

\subsection{2-point measurements from simulations}
\label{subsec:Cell_sim}

Our main weak lensing simulation products consist of convergence $\kappa$-maps and galaxy catalogues that include positions, shear, convergence and redshift for every objects. 
The  lensing power spectra $\widehat{{C}_{\ell}^{\kappa}}$ were estimated directly from the Fourier transform of $\kappa$-maps (see Sec. \ref{subsec:lightcone} for details about their constructions), as:
\begin{eqnarray}
\widehat{{C}_{\ell}^{\kappa}} = \langle |\widetilde{\kappa}(\boldsymbol \ell)|^2 \rangle,
\end{eqnarray}
where the brackets refer to an angular averaging over the Fourier ring of radius $\ell$.
For both simulation measurements and model predictions, we adopted a log-space binning scheme, spanning the range $[35 \le  \ell \le 10^4]$ with 20 bins. 
 The lensing power spectrum covariance was computed from an ensemble of $N$ measurements $\widehat{{C}_{\ell}^{\kappa,i}}$, following:
\begin{eqnarray}
{\rm Cov}_{\rm sim}^{\kappa} = \frac{1}{N-1}\sum_{i=1}^{N} \left[\widehat{{C}_{\ell}^{\kappa,i}} - \langle{C}_{\ell}^{\kappa}\rangle \right] \left[\widehat{{C}_{\ell'}^{\kappa,i}} - \langle{C}_{\ell'}^{\kappa}\rangle \right].
\label{eq:Cov_sim}
\end{eqnarray}
This expression contains all at once the three terms from equation (\ref{eq:Cov_ell}) with the caveat that the SSC term may not be fully captured due to the finite simulation volume; we present in Sec. \ref{sec:Covariance} a comparison between the two approaches. The shear 2-point correlation functions $\widehat{\xi_{\pm}}(\vartheta)$ were extracted from our simulated galaxy catalogues with 
{\sc TreeCorr} \citep{TreeCorr}, which basically measures:
\begin{eqnarray}
\widehat{\xi_{\pm}}(\vartheta) = \frac{\sum_{ij} w_i w_j\left(e_{\rm t}^i e_{\rm t}^j \pm e_{\times}^i e_{\times}^j\right)\Delta_{ij}}{\sum_{ij} w_i w_j}.
\end{eqnarray}
Here $e_{\rm t/\times}^i$ are the tangential and cross components of the ellipticity measured from galaxy $i$, $w_i$ is a weight generally related to the shape quality and taken to be unity in this work, and the sums run over all galaxy pairs separated by an angle $\vartheta$ falling in the angular  bin; the binning operator $\Delta_{ ij}=1.0$ in that case, otherwise it is set to zero. Following \cite{KiDS450}, we computed the $\widehat{\xi_{\pm}}(\vartheta)$ in 9 logarithmically-spaced angular separation bins between 0.5 and 300 arcmin. 

\subsection{Fisher forecasts}
\label{subsec:Fisher_th}

Given a survey specification, a theoretical model and a covariance matrix, we can estimate the constraints on four cosmological parameters by employing the Fisher matrix formalism. 
In particular, we are interested in measuring the impact on the constraints from different changes in the covariance matrix, either switching between estimator techniques at a fixed cosmology, or varying the input cosmology for a fixed estimator. 

The Fisher matrix $\mathcal F_{\alpha\beta}$ for parameters $p_{\alpha,\beta}$ quantifies the curvature of the log-likelihood at its maximum and provides a lower bound on parameter constraints under the assumption that the posterior is well approximated by a Gaussian. We can construct our matrix $\mathcal F_{\alpha\beta}$ from the derivative of the theoretical model  $C_{\ell}^{\kappa}$ with respect to the cosmological parameter $[p_{\alpha,\beta}] = [\Omega_{\rm m}, \sigma_8, h, w_0]$, from the covariance matrix ${\mathbf C}$, and from the derivative of the  covariance matrix with respect to these cosmological parameters. Under the additional assumption that the underlying data is Gaussian distributed, we can write \citep{Tegmark:1997rp}:
\begin{eqnarray}
\label{eq:fisher}
{\mathcal F}_{\alpha \beta} = \sum_{\ell,\ell'}\frac{\partial C_{\ell}^{\kappa}}{\partial p_\alpha}\left[{\mathbf C}\right]^{-1}_{\ell \ell'}\frac{\partial C_{\ell'}^{\kappa}}{\partial p_\beta} + \frac{1}{2} {\rm Tr} \left[
{\mathbf C}^{-1} \frac{\partial{{\mathbf C}}}{\partial p_\alpha}{\mathbf C}^{-1}
\frac{\partial{{\mathbf C}}}{\partial p_\beta}   \right]\,.
\end{eqnarray}

\citet{2013A&A...551A..88C} argues that parameter-dependent covariance matrices are not suitable for Fisher forecasts, which are only accurate for Gaussian likelihoods with fixed covariance. In light of this, we neglected the second term of equation (\ref{eq:fisher}), which at the same time simplified the evaluation. 
Equipped with this tool, it is now straightforward to compare the impact of using ${\mathbf C} \equiv {\rm Cov}^{\kappa}_{\rm tot}$ (equation \ref{eq:Cov_ell}) or ${\mathbf C} \equiv {\rm Cov}^{\kappa}_{\rm sim}$ (equation \ref{eq:Cov_sim}) in our Fisher forecast, and to investigate the effect of varying the input cosmology at which the  covariance matrix is evaluated (and fixing that value, so the derivative of the covariance is still set to zero). Specifically, we monitored changes of the area of the Fisher ellipses, which we took as a metric of the global constraining power. This analysis was repeated with  different configurations of the  $\sigma_{\epsilon}$, $\bar{n}$ and $A_{\rm survey}$ parameters, which we adjusted to construct covariance matrices that emulate the KiDS-1300, DES-Y5 and LSST-Y10 surveys. Whereas the analytic calculations can evaluate the terms at any specified area and noise levels, the simulations estimates had to be area-rescaled. This introduced a small error since technically the SSC term does not exactly scale that way, but the size of this error is negligible compared to other aspects of the calculations, especially for featureless square masks.  In addition, we opted to implement the shape noise term  in the simulations simply by adding its analytic contribution, which we obtained from evaluating ${\rm Cov}_{\rm N} =  \left({\rm Cov}_{\rm G}^{\kappa} - {\rm Cov}_{\rm G, \sigma_{\epsilon}=0}^{\kappa}\right)$ with $A_{\rm survey}=100$ deg$^2$. This includes both the pure shape noise term and the mixed term, obtained from equation (\ref{eq:GaussCov}). Overall, we computed the survey covariance as:
\begin{eqnarray} 
{\rm Cov}^{\kappa}_{\rm sim}\Big|_{\rm survey} = \left({\rm Cov}^{\kappa}_{\rm sim} + {\rm Cov}_{\rm N} \right)  \times\left(\frac{A_{\rm sim}}{A_{\rm survey}}\right). 
\end{eqnarray}
Having established our methods, we now turn to the description of the {\it cosmo}-SLICS numerical simulations from which we extracted our light-cone data and evaluated ${\rm Cov}^{\kappa}_{\rm sim}$.

\section{Weak Lensing  Simulations}
\label{sec:simulations}

There exists a number of ways to construct simulated light-cones for cosmic shear studies, and we adopted here the multiple-plane prescription detailed in \citet{HDVvW12}; this method was thoroughly tested to meet the accuracy requirements of ongoing weak lensing surveys \citep[see, e.g.,][]{Heymans2012, KiDS450}. Briefly, the construction pipeline proceeds as follow: after the initial design for volume, particle number and cosmology was specified, an $N$-body code generated density snapshots at a series of  redshifts, chosen to fill the past light-cone. Under the Born approximation, the mass planes were aligned and ray-traced at a pre-selected opening angle, pixel density and source redshifts. In our implementation, this post-processing routine constructed as many mass over-density, convergence and shear maps as the number of density checkpoints in the light-cone. Finally, galaxies were assigned positions and redshifts, and their lensing quantities were obtained by interpolating from the maps. We refer the reader to HD18 for more details on the implementation of this pipeline with the SLICS simulations, and focus hereafter on the new aspects specific to the  {\it cosmo}-SLICS.

\subsection{Choosing the cosmologies}
\label{subsec:LH}

\begin{figure}
\begin{center}
\includegraphics[width=3.5in]{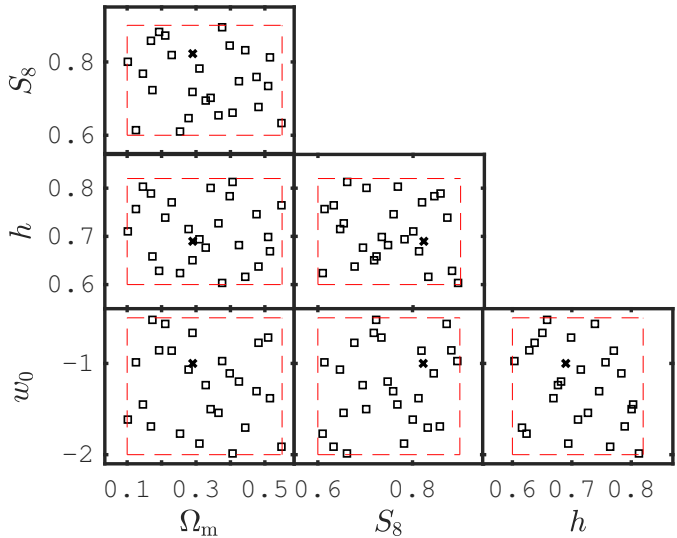}
\caption{Cosmological parameters covered by the {\it cosmo}-SLICS. Our fiducial cosmology 
is depicted here with the `$\times$' symbols.}
\label{fig:cosmo}
\end{center}
\end{figure}

The first part of the design consisted in identifying the parameter space that we wished to sample. Although a significant part of this paper focuses on power spectrum covariance matrices, the {\it cosmo}-SLICS have a broader range of applicability, and our primary science goal is, we recall,  to provide the means to carry out alternative  analyses of the current state-of-the-art weak lensing data,  paving the way for LSST and {\it Euclid}. Cosmic shear is maximally sensitive to a particular combination of $\Omega_{\rm m}$ and $\sigma_8$, often expressed as  $S_8 \equiv \sigma_8 \sqrt{\Omega_{\rm m}/0.3}$, but also varies at some level with all other parameters. In particular, tomographic lensing analyses are sensitive to  the growth of structures over cosmic time and hence probe the dark energy equation of state $w_0$, a parameter that we wish to explore. Furthermore, because of recent claims of a tension in the measurements of the Hubble parameter between CMB and direct $H_0$ probes \citep{2018ApJ...861..126R, 2017MNRAS.465.4914B,2018arXiv180706205P}, we decided to vary $h$ as well. In order to reduce the parameter space, we kept all other parameter fixed. More precisely, we fixed $n_{\rm s}$ to 0.969, $\Omega_{\rm b}$ to 0.0473 thereby matching the SLICS input values, we ignored any possible evolution of the dark energy equation of state, and we assumed that all neutrinos are massless. 
In the end, we settled for modelling variations in $[\Omega_{\rm m}, S_8, h, w_0]$.

We examined  the current $2\sigma$ constraints from the KiDS-450 and DES-Y1 cosmic shear data\footnote{Results from the first HSC cosmic shear analysis \citep{2018arXiv180909148H} were released after the completion of our simulations, and their $2\sigma$ lower limit on $\Omega_{\rm m}$ extends slightly outside of our range. If the {\it cosmo}-SLICS were used in this HSC data analysis, the error contours would likely appear truncated below $\Omega_{\rm m} =0.1$.} \citep{KiDS450,DES1_Troxel}, which are both 
well bracketed by the range $\Omega_{\rm m} \in$ [0.10, 0.55] and $S_8 \in$ [0.60, 0.90].  
This upper bound on $S_8$ falls between the upper $1\sigma$ and the $2\sigma$ constraints from {\it Planck}, but this is not expected to cause any problems since the {\it cosmo}-SLICS are designed for lensing analyses. Constraints on the dark energy equation of state parameter from these lensing surveys allow for  $w_0 \in$ [-2.5, -0.2]. This wide range of values is expected to change rapidly with the improvement of photometric redshifts, hence we restricted the sampling range to $w_0 \in$ [-2.0, -0.5]. This choice could impact the outskirts of the contours obtained from  a likelihood analysis based on the {\it cosmo}-SLICS, however this should have no effect  on the other  parameters. Constraints on $h$ from lensing alone are weak, with KiDS-450 allowing a wide range of values and hitting the prior limits, and DES-Y1 presenting no such results. We instead selected the region of $h$ informed by the Type IA supernovae measurements from \citet{2016ApJ...826...56R}. The $5\sigma$ values are close to $h \in$ [0.64, 0.82], and  we further extended the lower limit to  0.60 in order to avoid likelihood samplers from approaching the edge of the range too rapidly.  A summary of our final parameter volume is presented in Table \ref{table:prior}.

Inspired by the strategy of the {\it Cosmic Emulator}\footnote{{\sc CosmicEmu:} http://www.hep.anl.gov/cosmology/CosmicEmu/}  \citep{Heitmann2013}, we sampled this four-dimensional parameter space with a Latin hyper-cube\footnote{We used {\tt lhsdesign}, a Latin  hyper-cube generator included in the  {\sc Matlab} Statistics Function kit.}, and constructed an emulator to interpolate at any point within this range \citep[see also][for other examples relevant to cosmology]{DarkEmulator,EuclidEmulator_etal_2018, MassiveNuS}. A Latin hyper-cube is an efficient sparse sampling algorithm  designed to maximise the interpolation accuracy while minimising the node count \citep[see][and references therein for more details on the properties of these objects]{Heitmann2013}. 

Given our finite computing resources, we had to compromise on the number of nodes, which ultimately reflects on the accuracy of the interpolation. We therefore quantify the interpolation error as follow:  1- we varied the number of nodes from 250 down to 50 and 25, then generated for each case a Latin hyper-cube  that covered the parameter range summarised in Table \ref{table:prior}; 2- we evaluated the $\xi_\pm$ theoretical predictions  at these points and trained our emulator on the results (details about our emulator implementation, its accuracy and training strategy can be found in Appendix \ref{sec:emulator}); 3- we constructed a fine regular grid over the same range, and  compared at each point the predictions from our emulator with the `true' predictions computed on the grid points; 4- we examined the fractional error and decided on whether our accuracy benchmark was reached, demanding an uncertainty no larger than 3\%, which is smaller but comparable in size to the accuracy of the {\sc HaloFit} model itself. We also recall that the current uncertainty caused by photometric redshifts significantly exceeds this 3\% threshold, and that the smaller scales are further affected by uncertainty about baryon feedback mechanisms, hence this interpolation error should be sub-dominant.

We present the fractional error in Fig. \ref{fig:AccGrids} for the  25 nodes case; we achieve a 1-2\% accuracy over most of the parameter range, which meets our accuracy requirement, and which we report as our fiducial interpolation error. We emphasise that this error size is not strictly applicable to all types of measurements,  for instance the $\xi_+$ interpolation becomes less accurate than that for angular scales larger than two degrees. Instead, this should be viewed as a representative error given an arbitrary lensing signal that varies in cosmology with similar strength as the $\xi_+$ observable over the range $0.5 < \vartheta < 72$ arcmin. 

Increasing the node counts from 25 to 50 significantly reduces the size of the regions in parameter space where the accuracy exceeds 2\%, which are now pushed to  small pockets on the outskirts. Further inflating to 250 nodes moves  the bulk of the accuracy below the 1\% level. Since our current accuracy target is less strict, we therefore developed the {\it cosmo}-SLICS on 25 $w$CDM plus one $\Lambda$CDM nodes, but may complete the Latin hyper-cube with more nodes as in \citet{Rogers19} in the future;  the exact parameter values are listed in Table \ref{table:cosmo}, and their two-dimensional projections are presented in Fig. \ref{fig:cosmo}. 



\begin{table}
   \centering
   \caption{Ranges of the cosmological parameters varied in the {\it cosmo}-SLICS, compared to those of the {\it MassiveNuS},  the DH10 and the {\it DarkMatter} simulation  suites. Also listed are some of the properties relevant to their use in cosmic shear analyses, including the box size ($L_{\rm box}$, in $h^{-1}$Mpc), the number of particles $N_{\rm p}$ and the highest redshift available. Neutrino masses are listed in eV.}
   \tabcolsep=0.11cm
      \begin{tabular}{@{} cccccccc @{}} 
      \hline
      & {\it cosmo}-SLICS & {\it MassiveNuS} & DH10 & {\it DarkMatter} \\
      \hline
         $\Omega_{\rm m}$ & $[0.10, 0.55]$ & $[0.18, 0.42]$ & $[0.07, 0.62]$ & $[0.15, 0.70]$\\
          S$_{8}$ & $[0.60, 0.90]$ & $[0.38, 1.20]$ & $[0.38, 1.03]$ & $[0.40, 1.35]$\\
          $h$ & $[0.60, 0.82]$ & $0.70$ & $0.70$ & $0.72$\\
          $w_0$ & $[-2.0, -0.5]$  & $-1.0$ & $-1.0$ & $-1.0$\\
          $M_\nu$& $0.0$ & $[0.0, 0.62]$ & $0.0$  &$0.0$ \\
\hline
$L_{\rm box}$ &  $505$ &$512$ &$140$ &$240$\\
$N_{\rm p}$   &  $1536^3$ & $1024^3$& $256^3$&$512^3$ \\
$z_{\rm max}$ &  $3.0$ & $45.0$ &$2.0$ & $45.0$\\
\hline
      \end{tabular}
    \label{table:prior}
    \end{table}

\begin{table}
   \centering
   \caption{Cosmological parameters in the 25+1 {\it cosmo}-SLICS models, with S$_{8}$ is defined as $\sigma_8\!\!\sqrt{\Omega_{\rm m}/0.3}$. In all runs, the baryon density, primordial tilt and neutrino density have been fixed to $\Omega_{\rm b}=0.0473$, $n_{\rm s} = 0.969$ and $\Omega_{\nu} = 0$. Two matched-seed $N$-body simulations are evolved at each of these nodes, as detailed in Sec. \ref{subsec:Nbody}.}
   \tabcolsep=0.11cm
      \begin{tabular}{@{} cccccccc @{}} 
      \hline
      ID &   $\Omega_{\rm m}$ &   S$_{8}$ &  $h$ & $w_0$  & $\sigma_8$ & $\Omega_{\rm c}$ & $\Omega_{\Lambda}$\\
\hline
FID & 0.2905 & 0.8231 & 0.6898 & -1.0000 & 0.8364 & 0.2432 & 0.7095\\
00 & 0.3282 & 0.6984 & 0.6766 & -1.2376 & 0.6677 & 0.2809 & 0.6718\\
01 & 0.1019 &  0.7826 & 0.7104 & -1.6154 & 1.3428 & 0.0546 & 0.8981\\
02 & 0.2536 &  0.6133& 0.6238 & -1.7698 & 0.6670 & 0.2063 & 0.7464\\
03 & 0.1734 &  0.7284& 0.6584 & -0.5223 & 0.9581 & 0.1261 & 0.8266\\
04 & 0.3759 &  0.8986& 0.6034 & -0.9741 & 0.8028 & 0.3286 & 0.6241\\
05 & 0.4758 &  0.7618& 0.7459 & -1.3046 & 0.6049 & 0.4285 & 0.5242\\
06 & 0.1458 &  0.7680& 0.8031 & -1.4498 & 1.1017 & 0.0985 & 0.8542\\
07 & 0.3099 &  0.7861& 0.6940 & -1.8784 &  0.7734 & 0.2626 & 0.6901\\
08 & 0.4815 &  0.6804& 0.6374 & -0.7737 &  0.5371 & 0.4342 & 0.5185\\
09 & 0.3425 &  0.7054& 0.8006 & -1.5010 &  0.6602  & 0.2952 & 0.6575\\
10 & 0.5482 &  0.6375& 0.7645 & -1.9127 &  0.4716 & 0.5009 & 0.4518\\
11 & 0.2898 &  0.7218& 0.6505 & -0.6649 &  0.7344 & 0.2425 & 0.7102\\
12 & 0.4247 &  0.7511& 0.6819 & -1.1986 & 0.6313 & 0.3774 & 0.5753\\
13 & 0.3979 &  0.8476& 0.7833 & -1.1088 & 0.7360 & 0.3506 & 0.6021\\
14 & 0.1691 & 0.8618 & 0.7890 & -1.6903 &  1.1479& 0.1218 & 0.8309\\
15 & 0.1255 &  0.6131& 0.7567 & -0.9878 & 0.9479 & 0.0782 & 0.8745\\
16 & 0.5148 &  0.8178 & 0.6691 & -1.3812 & 0.6243 & 0.4675 & 0.4852\\
17 & 0.1928 & 0.8862 & 0.6285 & -0.8564 & 1.1055 & 0.1455 & 0.8072\\
18 & 0.2784 &   0.6500 & 0.7151 & -1.0673 &  0.6747& 0.2311 & 0.7216\\
19 & 0.2106 & 0.8759 & 0.7388 & -0.5667 & 1.0454 & 0.1633 & 0.7894\\
20 & 0.4430 &  0.8356& 0.6161 & -1.7037 &  0.6876 & 0.3957 & 0.5570\\
21 & 0.4062 & 0.6620 & 0.8129 & -1.9866 &  0.5689 & 0.3589 & 0.5938\\
22 & 0.2294 &  0.8226& 0.7706 & -0.8602 &  0.9407& 0.1821 & 0.7706\\
23 & 0.5095 &   0.7366& 0.6988 & -0.7164 & 0.5652 & 0.4622 & 0.4905\\
24 & 0.3652 &   0.6574& 0.7271 & -1.5414 & 0.5958 & 0.3179 & 0.6348\\
\hline
       \end{tabular}
    \label{table:cosmo}
    \end{table}

\subsection{Preparing the light-cones}
\label{subsec:geometry}

Prior to running the $N$-body code, we needed to specify the box size, the particle count and redshift dumps of the projected mass maps, which  must form contiguous light-cones along the line of sight. Following HD18, we fixed the simulation volume to $L_{\rm box} = 505 ~h^{-1}{\rm Mpc}$ on the side (note that $h$ varies between models) and the particle count to $N_{\rm p} = 1536^3$, offering an excellent compromise between large scales coverage and small scales resolution. This set-up allows to estimate cosmic shear correlation functions beyond a degree and under the arc minute without significant impact from the two limitations above-mentioned, thereby covering most of the angular range that enter the KiDS analyses. By fixing the box size however, the number of redshift dumps  up to $z_{\rm max}$ varies with cosmology due to differences in  the redshift-distance conversion. We further split these volumes in halves along one of the Cartesian axis and randomly chose one of the six possibilities (three directions for the projections axis times two half-volume options) at every redshift dump. 
We finally aligned the resulting cuboids to form a long pencil, we  worked out the comoving distance to the mid-plane of each of these cuboids, converted\footnote{The distance-to-redshift relations are obtained from the public  {\tt w0waCDM} module within {\sc python}  {\tt astropy.cosmology} numerical package.} distances to redshift in the specified cosmology, and proceeded from redshift $z=0$ until the back side of the last cuboid exceeds $z_{\rm max}$, with  $z_{\rm max} = 3.0$. The list of redshifts found that way were then passed to the main $N$-body code which set out to produce particle dumps and mass sheets  for each entry. The total number of redshift dumps ranges from 15 (for models-08 and -23) to 28 (for model-01).

\subsection{Cosmological simulations with matched pairs}

\label{subsec:Nbody}

\begin{figure}
\begin{center}
\includegraphics[width=3.5in]{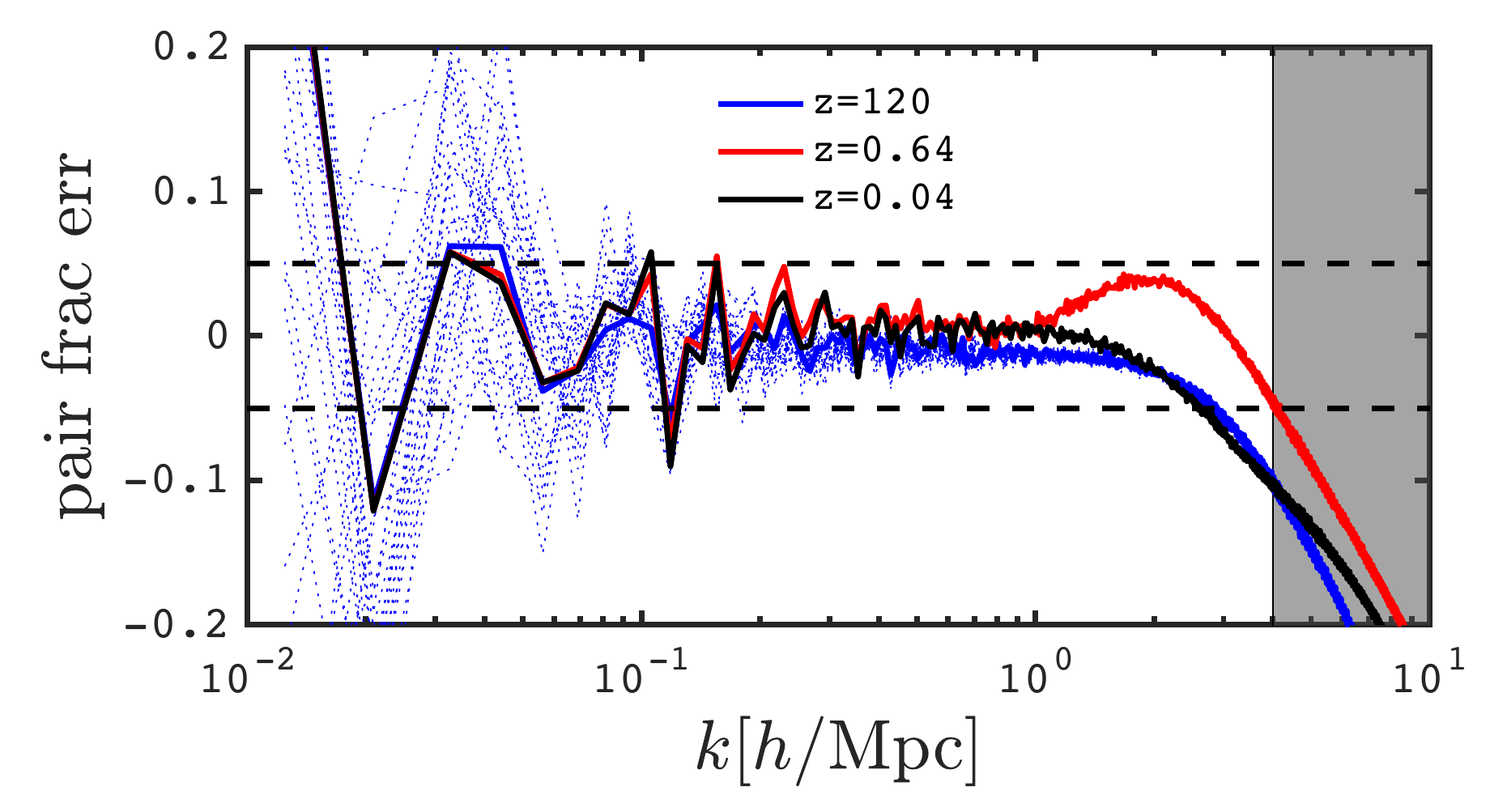}
\caption{Fractional difference between the mean of simulation pairs at the fiducial cosmology (i.e. model-FID) and the input theoretical model $P(k)$, obtained with {\sc HaloFit}. Faint blue dotted lines show the results for a number of random pairs at the initial redshift $z_i=120$, while the thick blue line highlights the best pair. The sampling variance cancels to better than 5\% also at $z = 0.64$ and $0.04$, as demonstrated respectively by the red and black lines. The grey zone indicates the regime where the discrepancy exceeds 10\%.
}
\label{fig:Pk_pair}
\end{center}
\end{figure}

The $N$-body calculations were carried out with the gravity solver CUBEP$^3$M \citep{2013MNRAS.436..540H}
in a setup similar to that described in HD18, except for key modifications due to the $w$CDM nature 
of our runs. 
Dark matter particles were initially placed on a regular grid, then displaced using linear perturbation theory given  an initial input power spectrum $P(k,z_{\rm i})$ and a Gaussian noise map, with $z_{\rm i}=120$.
Different cosmological models required distinct transfer functions $T(k)$, obtained from running the Boltzmann code {\sc camb} \citep{CAMB} with the  parameters values taken from Table \ref{table:cosmo}.
The initial power spectrum was then computed as  $P(k,z_{\rm i}) = A_{\sigma_8} D^2(z_{\rm i}) T(k) k^{n_{\rm s}}$, where $D(z_{\rm i})$ is the linear growth factor, and the normalisation  parameter $A_{\sigma_8}$ is defined such that  $P(k, z=0)$ has the $\sigma_8$ value given by the model. The initial condition generator included with the public CUBEP$^3$M release can only compute growth factors in $\Lambda$CDM cosmologies, hence we computed $D(z_{\rm i}, \Omega_{\rm m}, \Omega_{\Lambda}, w_0)$ with {\sc Nicaea} instead, then manually input the results in the  generator.
 
Since the central goal of the {\it cosmo}-SLICS is to model the cosmological signal of novel weak lensing methods, it is important to ensure that the simulation sampling variance does not lead to mis-calibrations. Extra-large volume simulations can achieve this through spatial averaging, however these are expensive to run. Instead, we produced a pair of noise maps in which the sampling variance cancels almost completely, such that the mean of any estimator extracted from the pair will be very close to the true ensemble mean. We achieved this in a relatively simple way: 

\begin{enumerate}
\item We generated a large number of initial conditions at our fiducial cosmology and extracted their power spectra $P(k,z_{\rm i})$; 
\item We computed the mean power spectrum for all possible pair combinations and  selected the pair whose mean was the closest to the theoretical predictions, allowing a maximum of 5\% residuals; 
\item We further demanded that neither of the members of a given pair is a noise outlier. What we mean by this is that the fluctuations in $P(k,z_{\rm i})$ must behave as expected from a Gaussian noise map and scatter evenly across the input power spectrum. Quantitatively, we required the fluctuations to cross the mean at almost every $k$-mode. This last requirement further prevented power leakage from large to small scales, which otherwise affects the late-time structure formation. 
\end{enumerate}

Fig. \ref{fig:Pk_pair} shows the fractional difference between the {\sc HaloFit} predictions (set to the horizontal line with zero $y$-intercept) and the mean initial $P(k, z_{\rm i})$ measured from our best pair (solid blue); other random pairs are also shown (thin dotted blue lines) and exhibit much larger variance. The drop at high $k$ is caused by the finite mass resolution of our simulations;
the grey zone indicates the scales where the departure is greater than 10\% at redshift $z=0.0$, which occurs at $k = 4.0 ~h {\rm Mpc}^{-1}$. We used the same pair of noise maps in the initial conditions for our 25 $w$CDM cosmologies, further ensuring that the sample variance in $P(k, z_{\rm i})$ is exactly the same across models, and that differences are attributed solely to changes in the input cosmological parameters.

After this initialisation step, the gravity solver evolved the particles until redshift zero, writing to disk the particles' phase space and the projected densities at each snapshot. The  background expansion subroutine of CUBEP$^3$M has been adapted to allow for $w_0 \neq -1$ cosmologies by Taylor-expanding the FRW equation to third order in the time coordinate. The exact value of the particle mass depends on the volume and on the matter density, hence varies with $h$ and $\Omega_{\rm m}$, spanning the range $[1.42, 7.63] \times 10^9 M_{\odot}$. The $N$-body computations were carried out on 256 compute nodes on the Cedar super computer hosted by Compute Canada, divided  between  64 {\sc mpi} tasks and further parallelised with 8 {\sc openmp} threads; they ran for 30-70 hours depending on the cosmology.  After completion of every simulation, we computed the matter power spectra at every snapshot then erased the particle data to free up space for other runs\footnote{Dark matter halo catalogues were stored, with properties and format fully described in HD18; the halo mass function is presented in Appendix \ref{sec:SLICS_vs_model}.}.  
The red and black lines in Fig. \ref{fig:Pk_pair} show the fractional difference between the non-linear predictions from \citet{Takahashi2012} and the mean $P(k)$ measured from the matched pair at lower redshifts. They demonstrate that the phase cancellation survives well the non-linear evolution.

One potential catch in our matched-pair method is that it is only calibrated against the two-point function, and there is no formal mathematical proof that the sampling variance cancels at the same level for higher order statistics.
Evidence points in that direction however: in the initial conditions, the density fields follow Gaussian statistics, hence all the information is captured by the matter power spectrum. Minimising the variance about $P(k)$ is thereby equivalent to minimising the variance about the cosmological information,  irrespective of the measurement technique. The results of \citet {Pontzen2} are encouraging and demonstrate that the matched-pair technique of \citet{AnguloPontzen} introduces no noticeable bias on the matter-matter, matter-halo and halo-halo power spectra, nor on the halo mass function, void mass function and matter PDF. 
Additionally, some estimators reconnect with the two-point functions on large scales \citep[e.g. shear clipping, as in][]{Giblin18}, and for these we expect a significant noise  cancellation as well.




\subsection{Ray-tracing the light-cone}
\label{subsec:lightcone}


Closely following the methods of HD18, we constructed mass over-density, convergence and shear maps from the output of the $N$-body runs. Every light-cone map subtends 100 deg$^2$ on the sky and is divided in 7745$^2$ pixels. 
For each redshift dump $z_{\rm l}$, we randomly chose one of the six projected density fields, we shifted its origin, then interpolated the result onto the light-cone  grid to create a mass over-density map $\delta_{\rm 2D}({\boldsymbol \theta},z_{\rm l})$.  We needed here to minimise a second source of sampling variance that arises from the choice of our observer's position, and which we refer to as the  `light-cone sampling variance'. This is distinct from the `Gaussian sampling variance' caused by drawing Fourier modes from a noise map in the initial condition generator. Since the number of mass planes required to reach a given redshift varies across cosmology models, there is an inevitable amount of residual light-cone sampling variance introduced in the $\delta_{\rm 2D}({\boldsymbol \theta},z_{\rm l})$ maps. We nevertheless reduced this by matching the origin-shift vectors and the choice of projection planes at the low-redshift end in our construction. 

We computed convergence maps  from a weighted sum over the mass planes:
\begin{eqnarray}
 \kappa( {\boldsymbol \theta},z_{\rm s})\! =\! \frac{3 H_{0}^{2} \Omega_{\rm m}}{2 c^2}\!\! \sum_{\chi_{\rm l}=0}^{\chi_{\rm H}}\! \delta_{\rm 2D}({\boldsymbol \theta},\chi_{\rm l}) (1 + z_{\rm l})  \chi_{\rm l} \bigg[\!\sum_{\chi_{\rm s} = \chi_{\rm l}}^{\chi_{\rm H}}\!\! n(\chi_{\rm s})\frac{\chi_{\rm s} - \chi_{\rm l}}{\chi_{\rm s}} {\Delta}\chi_{\rm s}  \bigg] \Delta \chi_{\rm l}, \hspace{-10mm} \nonumber \\
          \label{eq:kappa_disc}
\end{eqnarray}
where  
$\Delta \chi_{\rm l} =  L_{\rm box}/{\rm nc}$,  nc = 3072 being our grid size.
We used equation (\ref{eq:kappa_disc})  to construct a series of $\kappa({\boldsymbol \theta},z_{\rm s})$ maps for which the source redshift distribution is given by $n(z) = \delta(z - z_{\rm s})$,  where $z_{\rm s}$  corresponds to the redshift of the back plane of every projected sub-volume  that make up the light-cone. Shear maps, $\gamma_{1,2}({\boldsymbol \theta},z_{\rm s})$, were obtained by filtering the convergence fields in Fourier space as described by \citet{1993ApJ...404..441K}. Our specific implementation of this transform makes use of the periodicity of the full simulation volume to eliminate the boundary effects into the light-cone, as detailed in \citet{HDVvW12}. Thereafter, any quantity ($\delta_{\rm 2D}, \kappa, \gamma_{1,2}$) required at an intermediate redshift (e.g. for a galaxy at coordinate ${\boldsymbol \theta}$ and redshift $z_{\rm gal}$) can be interpolated from these series of maps. For both members of the matched pair and for every cosmological models, we repeated this ray-tracing algorithm with 400 different random shifts and rotations, thereby probing each {\it cosmo}-SLICS node  800 times, or total area of 80,000 deg$^2$. We stored the maps for only 50 of these given their significant sizes, but provide galaxy catalogues for all others. These {\it pseudo}-independent light-cone maps and catalogues are the main {\it cosmo}-SLICS simulation products that we make available to the community.


\subsection{Accuracy}
\label{subsec:accuracy}

\subsubsection{Matter power spectrum}

As we mentioned before, the calibration of a weak lensing signal can be affected by limitations in the simulations, more specifically by the accuracy of the non-linear evolution, by the finite resolution and by the finite box size. These  systematic effects impact every estimator in a different way, and generally exhibit a scale and redshift dependence  \citep[see][for such a study on $\xi_{\pm}$ from the SLICS]{2015MNRAS.450.2857H}. In many cases however, one can estimate roughly the range of $k$-modes (or the $\vartheta$ values) that enters a given measurement, as in figure A1 of \citet{2017arXiv170605004V}, hence it is possible to construct an unbiased calibration by choosing only the data points for which the {\it cosmo}-SLICS are clean of these systematics. We observe from Fig. \ref{fig:Pk_pair} that our fiducial cosmology run recovers the non-linear model to better than 2\% up to $k=1.0 ~h {\rm Mpc}^{-1}$ at all redshifts, then the agreement slowly degrades with increasing $k$-modes, crossing 5\%  at  $k=2-3 ~h {\rm Mpc}^{-1}$ and 10\%  at $4-6 ~h {\rm Mpc}^{-1}$, depending on redshift. This comparison is not necessary representative of the true resolution of the {\it cosmo}-SLICS, since the {\sc HaloFit} predictions themselves have an associated error. It is shown in \citet{2015MNRAS.450.2857H} that the CUBEP$^3$M simulations agree better with the {\it Cosmic Emulator}, extending the agreement up to  higher $k$-modes. Unfortunately we cannot use this emulator as our baseline comparison since all of our $w$CDM nodes lie outside the allowed parameter range.

\begin{figure}
\begin{center}
\includegraphics[width=3.5in]{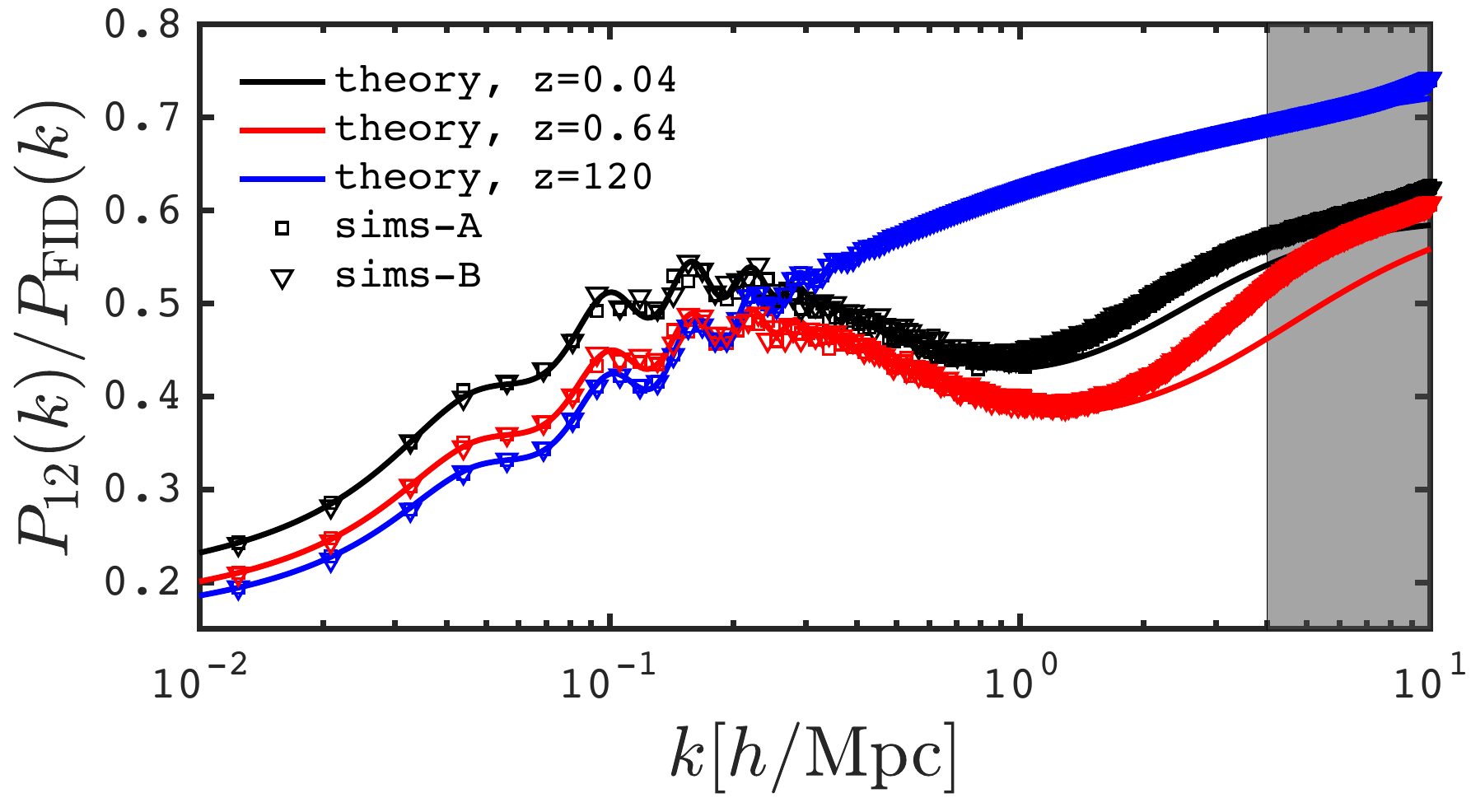}
\caption{Ratio between the power spectrum $P(k,z)$ in model-12 and in model-FID  (see Table \ref{table:cosmo}).
The lines show the predictions from {\sc HaloFit}, while the square and triangle symbols are measured from the pair of {\it cosmo}-SLICS $N$-body simulations. Upper (black), middle (red) and lower (blue) lines correspond to redshifts $z=0$, $0.6$ and $120$, respectively. Other cosmologies are shown in Appendix \ref{sec:SLICS_vs_model}.}
\label{fig:deriv_pk}
\end{center}
\end{figure}

With regards to the growth of non-linear structure across redshifts and cosmologies, the accuracy of the simulations is cleanly inspected with ratios of power spectra, where the small residual sampling variance cancels exactly, owing to the fact that all pairs of $N$-body calculations originate from the same two noise maps. A comparison between the {\it cosmo}-SLICS measurements and the  {\sc HaloFit} calculations therefore reveals the degree of agreement in a noise-free manner. We show in Fig. \ref{fig:deriv_pk} a representative example, the ratio between the model-12 and model-FID power spectra, $P_{12}(k)/P_{\rm FID}(k)$.  The different colours represent three redshifts, and the vertical offset is caused by differences in the linear growth factor. We observe an excellent match over a large range of scales for the two runs (labelled `sims-A' and `sims-B' in the figure). Some discrepancy is seen at small scales  where {\sc HaloFit} and the {\it cosmo}-SLICS are only 5-8\% accurate anyway. 
 A more detailed comparison can be found  in Appendix \ref{sec:SLICS_vs_model}, where for example we measure that beyond $k=2.0 ~h{\rm Mpc}^{-1}$, this ratio agrees to within 10\% at $z\sim 0.6$, and 5\% at $z\sim0.0$. In summary, ratios from simulations are mostly within a few percent of the ratios from the predictions, but some larger departures are observed at low redshift in dark energy models where $w_0 \ll-1.0$, which we attribute to inaccuracies in the  calibration 
of the \citet{Takahashi2012} predictions in that parameter space.

\begin{figure}
\begin{center}
\includegraphics[width=3.5in]{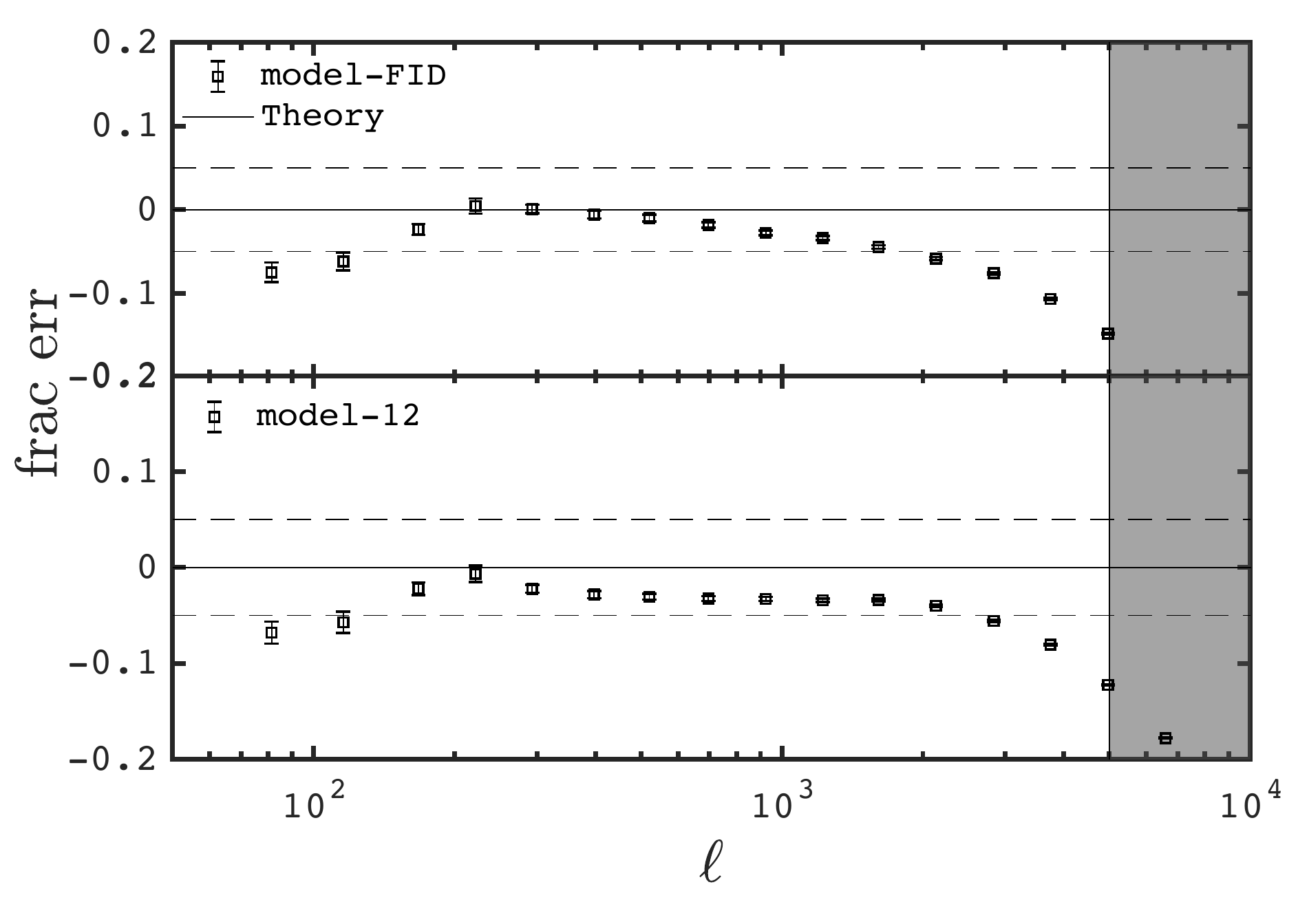}
\caption{Fractional difference between the $C_{\ell}^{\kappa}$ estimated from the simulation pairs and the input theoretical model, for sources at $z_{\rm s} = 1.0$.  The fiducial  and model-12  cosmologies are shown in the upper and lower panels, respectively. The mean and error bars are calculated from resampling every simulation 400 times; we show here the error on the mean.}
\label{fig:Cell_pair}
\end{center}
\end{figure}

\begin{figure}
\begin{center}
\includegraphics[width=3.5in]{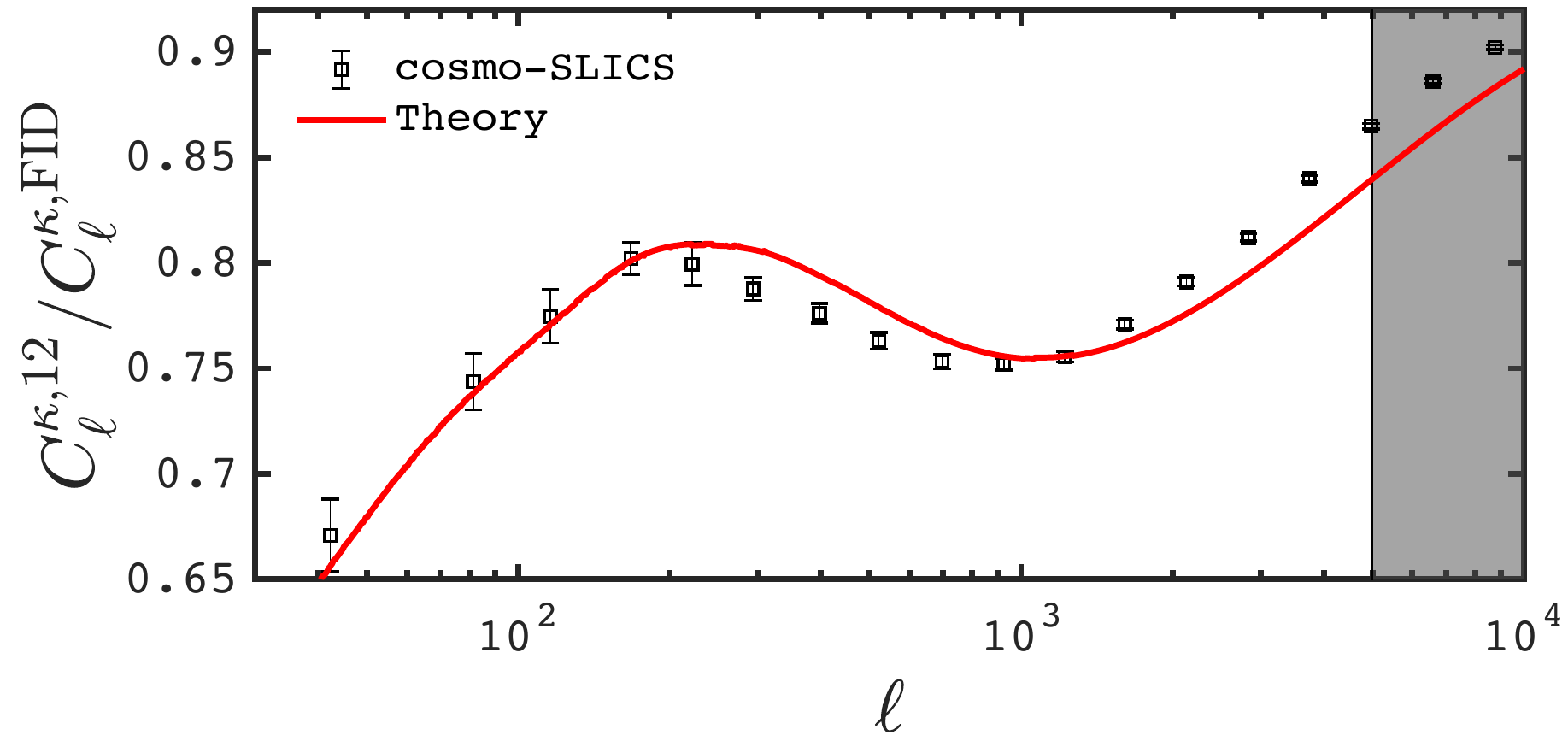}
\caption{Ratio between the convergence power spectrum $C_{\ell}^{\kappa}$ from model-12 and model-FID. 
Other models are presented in Appendix \ref{sec:SLICS_vs_model}.}
\label{fig:deriv_Cell}
\end{center}
\end{figure}

\begin{figure}
\centering
\includegraphics[width=0.49\textwidth]{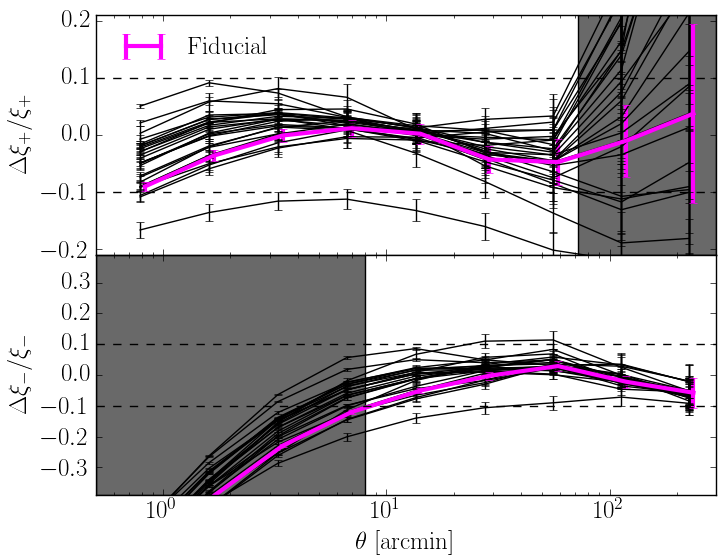} 
\caption{The fractional differences between the {\it cosmo}-SLICS measurements of $\xi_\pm$ for all models, averaged here across the 50 light-cones, and the corresponding theoretical predictions from {\sc Nicaea}  \citep[with the {\sc HaloFit} calibration from][]{Takahashi2012}. The magenta line corresponds to the measurements from the fiducial cosmology, and the grey bands indicate angular scales we recommend to exclude from an emulator training on these simulations.  Simulations and predictions are both constructed with the KV450 $n(z)$ here, and we plot the error on the mean.} \label{fig:AccCS}
\end{figure}

 %
 %

\subsubsection{Lensing 2-point functions}
\label{subsubsec:2pt_lensing}

For the particular goal of testing the accuracy of the light-cone products, we examined the lensing power spectrum for each of the 800 {\it pseudo}-independent realisations described in Sec. \ref{subsec:lightcone}, assuming a single source plane at  $z_{\rm s} \sim 1.0$. We present the $C_{\ell}^{\kappa}$ measurements from model-FID and model-12 in Fig. \ref{fig:Cell_pair}, compared to the predictions from {\sc Nicaea}. The grey band identifies  a relatively ambitious cut on the lensing data at $\ell = 5000$; most forecasts \citep[e.g][]{LSST_SRD} are more conservative and reject the $\ell > 3000$ multipoles.  The agreement between simulations and theory is of the order of a few percent over most of the multipole  range for these two cosmologies; the drop at high-$\ell$ is once again caused both by limitations in the simulation's resolution and by inaccuracies in the  non-linear predictions.   Fig. \ref{fig:deriv_Cell} next presents the ratio between these  two models, and is therefore the light-cone equivalent of  Fig. \ref{fig:deriv_pk}. The same trends are recovered, namely a generally good agreement at large scales, followed by an overshooting of a  few percent compared to the theoretical models at smaller scales. This disagreement is a known source of uncertainty in the non-linear evolution of the matter power spectrum and hence must be included in the error budget in data analyses that include these scales. It is however sub-dominant compared the uncertainty on baryonic feedback over these same scales, which reaches up to 40\%, depending on the hydrodynamical simulations \citep{Semboloni11, HWVH15, MeadFit, 2018arXiv180108559C}, and hence is not worrisome for lensing analyses that marginalise over the baryon effects. Ratios computed from other models are presented in Appendix \ref{sec:SLICS_vs_model}.

The accuracy of the shear 2-point correlation functions $\xi_{\pm}(\vartheta)$ was next investigated, this time in a more realistic application of the {\it cosmo}-SLICS: we populated the simulated light cones with mock galaxies following a $N(z)$ described by the KiDS+VIKING-450 lensing data \citep[][KV450 hereafter]{KV450} and compared the mean value from each cosmological model with the theoretical predictions. The fractional difference, presented in Fig. \ref{fig:AccCS}, shows that for many models we recover an agreement of a few percent over most of the scales included in the KiDS-450 cosmic shear analysis (the other angular scales are in the grey regions). Some models exceed the 10\% agreement marks, highlighting once again limitations in the {\sc HaloFit} calibration. This is discussed in greater detail in Appendix \ref{sec:emulator}.

\section{Covariance matrices}
\label{sec:Covariance}

As a first application of the {\it cosmo}-SLICS, we investigated the  accuracy of the covariance matrix of the convergence power spectra constructed from the 800 light-cones (see Sec. \ref{subsubsec:2pt_lensing}). This enquiry was motivated by a recent study from \citet{Petri16}, where it is shown that a lensing covariance matrix estimated with  {\it pseudo}-independent realisations could be as accurate as one estimated from truly independent simulations, leading to negligible biases on cosmological parameters constraints. Their results are based on a smaller simulation suite with degraded properties compared to the {\it cosmo}-SLICS or the SLICS: they use 200 independent $N$-body simulations with $L_{\rm box} = 240 ~h^{-1}{\rm Mpc}$ and $N_{\rm p} = 512^3$, which they ray-trace up to 200 times each. The authors  warn that their findings have to be revisited with better mocks before claiming that the method is robust, a verification we carry out in Sec. \ref{subsec:recyclings}. We further validate the two estimators with the analytical calculations described in Sec. \ref{subsec:Cov_Cell_th}, then explore in Sec. \ref{subsec:CovCosmo} the impact of variations in cosmology on the covariance, and propagate the effect onto error contours about four cosmological parameters. Lastly,  we demonstrate in Sec. \ref{subsec:emu_cov} how our Gaussian process emulator can learn the cosmology dependence of these matrices and hence be used in an iterative algorithm similar to the analytical model strategy, but now based exclusively on numerical simulations.

\subsection{Simulation-based vs. analytical model: a comparison }
\label{subsec:recyclings}

\begin{table}
   \centering
   \caption{Survey characteristics used in the analytical covariance calculations. All include a Gaussian distributed shape noise with standard deviation $\sigma_{\epsilon}=0.29$ per component.}
   \tabcolsep=0.11cm
      \begin{tabular}{@{} cccccccc @{}} 
      \hline
      Survey & Area (deg$^{2}$)& $n_{\rm gal} (\rm arcmin^{-2})$  \\
    \hline 
    KiDS   &  1300   & 7.54  \\
    DES-Y5 &  5000  & 5.07  \\
    LSST   &  15000 & 26.00 \\
    \hline
    \end{tabular}
    \label{table:surveys}
\end{table}

In this comparative study, we considered four lensing covariance matrix estimators:
\begin{enumerate}
\item {Our `baseline' was constructed from  800 truly independent measurements of $C_{\ell}^{\kappa}$ extracted from the SLICS, with galaxy sources placed at $z_{\rm s} = 1.0$. 
We additionally estimated the uncertainty on that covariance from bootstrap resampling these 800 measurements 1000 times;} 
\item{We identified 14 pairs of simulations within the SLICS whose initial $P(k,z_i)$ also satisfy the matched-pair criteria described in Section \ref{subsec:Nbody} (e.g. their mean closely follows the solid blue line in  Fig. \ref{fig:Pk_pair}). We resampled the underlying $N$-body simulations to produce 800 {\it pseudo}-independent $C_{\ell}^{\kappa}$ measurements  and an associated covariance matrix for each of these 14 pairs. We refer to this method as the `matched SLICS' estimate, and treated the variance between the 14 matrices as the uncertainty on the technique;}
\item{We estimated the covariance matrix from the 800 {\it pseudo}-independent power spectra extracted from the {\it cosmo}-SLICS. We assigned the same uncertainty on that method as on  the matched-SLICS method (item 2 above), both being equivalent in their nature. In the fiducial cosmology, we refer to this method as the `model-FID' covariance estimate. We also estimated a matrix for the other 25 cosmological points,  which we label  `model-00', `model-01' and so on;}
\item{At each of the 25+1 cosmologies sampled, we computed the analytic covariance model presented in equations (\ref{eq:Cov_ell}-\ref{eq:Cov_SSC}), keeping distinct the Gaussian, non-Gaussian and SSC terms.}
\end{enumerate}

We first examined for these four estimators the diagonal and the off-diagonal parts separately, then investigated the overall impact of their residual differences with a Fisher forecast about $\Omega_{\rm m}$, $S_8$, $w_{0}$ and $h$. We began with an inspection of the noise-free case before including survey-specific shape noises, galaxy densities and sky coverage. Aside from assuming a global square footprint, we did not apply survey masks in this comparison. This would introduce an extra level of complexity in the comparison, which we would rather keep at a more fundamental level.

\begin{figure}
\begin{center}
\includegraphics[width=3.5in]{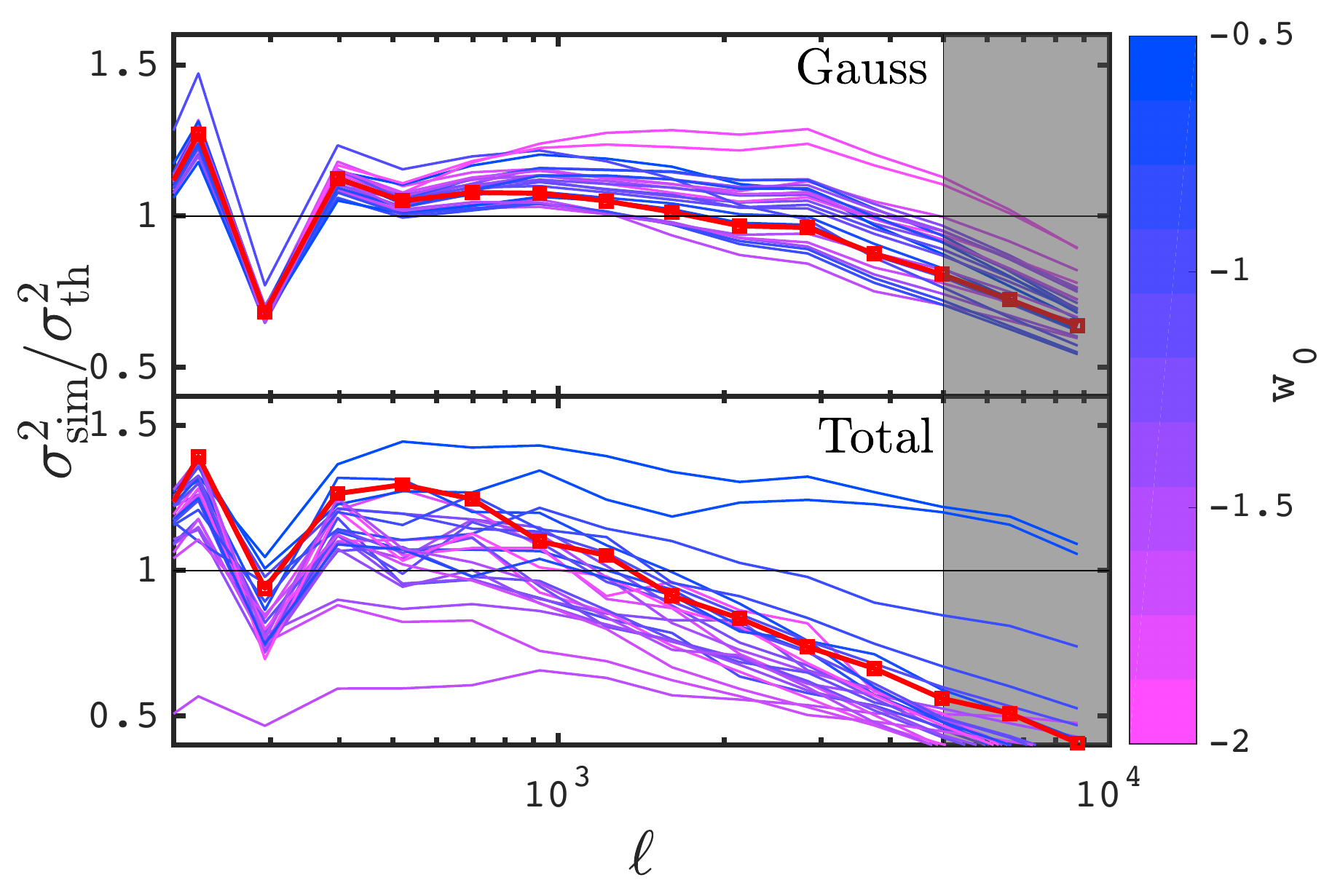}
\caption{Ratio between the variance of the shape noise-free lensing power spectrum estimated from the {\it cosmo}-SLICS simulations and that obtained from the analytical calculations. The upper panel is for the Gaussian ${\rm Cov}_{\rm G}^{\kappa}$ term only, while the lower panel shows our results for the full ${\rm Cov}_{\rm tot}^{\kappa}$ estimates. The lines are colour-coded as a function of $w_0$, ranging from magenta ($w_0\sim -2$) to blue ($w_0 \sim -0.5$), with the fiducial model shown in red squares. Models with high (low) $w_0$  exhibit larger (smaller) ratios. }
\label{fig:CovCompZ}
\end{center}
\end{figure}

\begin{figure*}
\begin{center}
\includegraphics[width=7.5in]{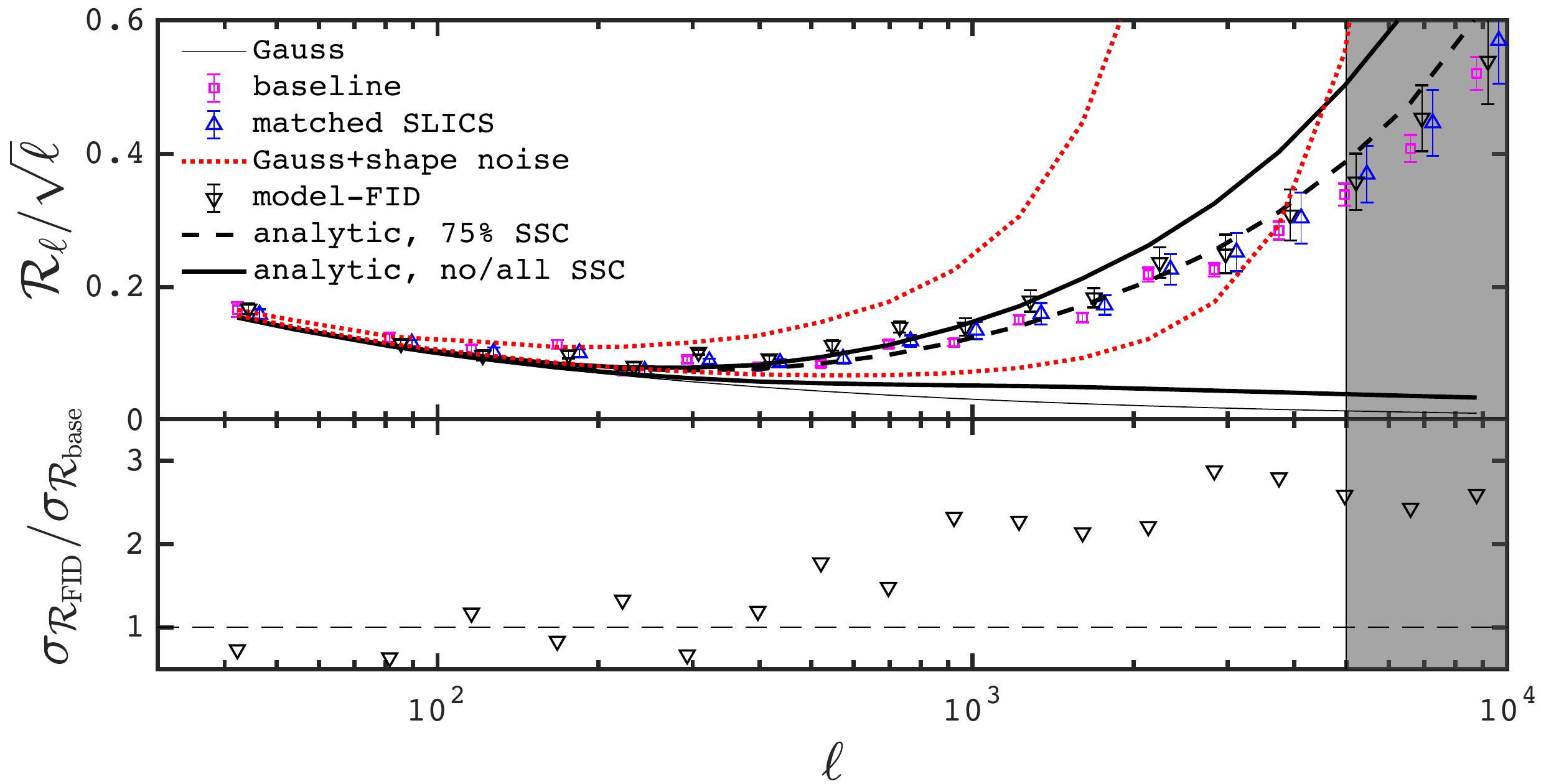}
\vspace{-5mm}
\caption{(upper:) Ratio between the diagonal of the  lensing power spectrum covariance matrices and the noise-free Gaussian term (i.e. equation \ref{eq:F}). We further divide this ratio by $\sqrt{\ell}$ to increase the readability of the low-$\ell$ part. The magenta squares correspond to the `baseline' measurement estimated from 800 independent light-cones with error bars  from bootstrap resamplings. The blue upward pointing triangles show the results from multiple ray-tracing the 14 matched-pairs found in the SLICS, while the black downward triangles are from the {\it cosmo}-SLICS (see main text in Section \ref{subsec:recyclings} for more details). The error bars on the two sets of triangles are estimated from the scatter over the 14 matched SLICS pairs.
Horizontal positions are offset  for clarity.  The thick solid and dashed lines represent the analytic calculations with 0, 75 and 100\% of the SSC term (see equation \ref{eq:Cov_ell}). The  red dotted-lines show the Gaussian term only, but this time with shot noise included assuming either KiDS (left) or LSST (right)  survey configuration described in Table \ref{table:surveys}. (lower:) Ratio between the error on ${\mathcal R}_{\ell}$ estimated from the {\it cosmo}-SLICS and from the baseline methods.}
\label{fig:NG_G_ratio_FID}
\end{center}
\end{figure*}

\subsubsection{Diagonal elements}

Even though the diagonal part of the covariance is generally the easiest to capture, we do not expect a perfect match between the simulation-based and the analytic methods since differences are already clear at the power spectrum level (see Fig. \ref{fig:Cell_pair}). 
We show in Fig. \ref{fig:CovCompZ} the ratio between the variance estimated from the {\it cosmo}-SLICS  and the analytical estimate, for all cosmologies and in the shape noise-free case, again assuming $z_{\rm s}=1$. The baseline and matched SLICS methods closely follow the {\it cosmo}-SLICS hence are not shown here for clarity. We examined both the ratio between the Gaussian terms (upper panel, computed from equation \ref{eq:GaussCov}) and between the diagonal of the full covariance (lower panel), colour-coding the results as a function of $w_0$. Departure from unity in this figure are caused by: 1-residual sampling variance (especially at low $\ell$-modes); 2- pixelization of the simulations and slight differences in the $\ell$-binning that impact the mode-count  3-  resolution limits in the simulations and 4- potential inaccuracies in the theoretical models. We further observe that the high-$\ell$ mismatch is higher in ${\rm Cov}_{\rm tot}^{\kappa}$ than in ${\rm Cov}_{\rm G}^{\kappa}$, which likely follows from the fact that the Gaussian term is only quadratic in $C_{\ell}^{\kappa}$, whereas it is raised to a higher power inside the trispectrum, (to the third power, within first order perturbation theory); consequently the discrepancies observed in the $C_{\ell}^{\kappa}$ are expected to  scale more rapidly in the latter case.  Models with high and low $w_0$ are shown with blue and magenta lines, respectively. While the Gaussian terms show no colour trend, there is a clear split in the full covariance ratios (lower panel), where blue lines are generally higher than magenta lines. Given that order 50\% discrepancies are seen at almost all scales in some models, this points to major differences in the SSC terms, which consequently suggests differences in the halo-mass function. We confirmed this conclusion in Appendix \ref{sec:SLICS_vs_model}, where we show that the match in halo mass function degrades for cosmologies with dark energy $w_0$ significantly different from $-1.0$.

Finally, when repeating the above comparison for different redshifts in the model-FID cosmology, we note that the agreement in the full variance improves at higher redshift, where non-linear evolution is less important.


We next investigated the relative departure from pure Gaussian statistics on the diagonal by dividing the full  matrix by the Gaussian term. It is therefore convenient to define:
\begin{eqnarray}
{\mathcal R}_{\ell} \equiv {\rm diag}\left[\frac{{\rm Cov_{\rm tot}^{\kappa}}}{{\rm Cov_{{\rm G}}^{\kappa}}}\right],
\label{eq:F}
\end{eqnarray}
which we evaluated separately for the four methods described at the beginning of this section.
The baseline measurement of ${\mathcal R}_{\ell}$ is reported as the magenta squares in Fig. \ref{fig:NG_G_ratio_FID}, and clearly captures the non-Gaussian features reported before \citep[e.g.][see their figure 1]{Takahashi2009a}. In comparison, the purely Gaussian term ${\rm Cov_{{\rm G}}^{\kappa}}$ is shown with the thin solid line, which significantly underestimates the simulated variance for $\ell$-modes larger than a few hundreds. The matched SLICS are shown with the blue upward triangles, and the {\it cosmo}-SLICS model-FID with the black downward triangles.  At all scales, we recover an excellent match between these three simulation-based approaches. More precisely, the baseline  and the model-FID agree  to within 20\%, corresponding to a 10\% difference on the non-Gaussian part of the error bar about $C_{\ell}^{\kappa}$. We further examined the agreement with the analytical calculations of ${\mathcal R}_{\ell}$ for three cases: ${\rm Cov_{{\rm G}}^{\kappa}}+ {\rm Cov_{{\rm NG}}^{\kappa}}$ + 0\% SSC contribution, shown on Fig. \ref{fig:NG_G_ratio_FID} as the lower thick solid line;  +75\% SSC, shown with the thick dashed line;  +100\% SSC, shown with the upper thick solid line. All simulation-based estimates are bracketed by the two solid lines (except at a few noisy points, e.g. $\ell$ = 190), consistent with capturing most but not all of the SSC contribution. The $k$-modes smaller than $2\pi/L_{\rm box}$ are absent from the simulations and hence do not contribute to the measured SSC, which instead comes from the  simulated volume that is not part of the light-cones  \citep[this conclusion was also reported in][for the baseline estimate]{2017arXiv170605004V}. The bottom panel of Fig. \ref{fig:NG_G_ratio_FID} compares the error on $\mathcal{R}_{\ell}$ between the baseline and the model-FID methods, showing that our gain of a factor 400 in computation resources incurs a  degradation in precision about ${\mathcal R}_{\ell}$ by a factor of $\sim2-3$.

To frame this comparison in a broader context, we further add to the figure two cases where the shape noise has been included in the Gaussian term, following a KiDS-like (upper/left dotted red curve) and a LSST-like (lower/right) survey configuration (see Table \ref{table:surveys} for the numerical specifics of these surveys). In the KiDS-like case, the diagonal is dominated by this noise component, which means that differences of order 10-20\% in the non-Gaussian terms are negligible in the total error. In the LSST-like survey however, the shape noise is massively reduced and becomes mostly sub-dominant, meaning that differences between the covariance estimators are expected to have a larger impact.

\subsubsection{Off-diagonal elements}

\begin{figure}
\begin{center}
\includegraphics[width=3.5in]{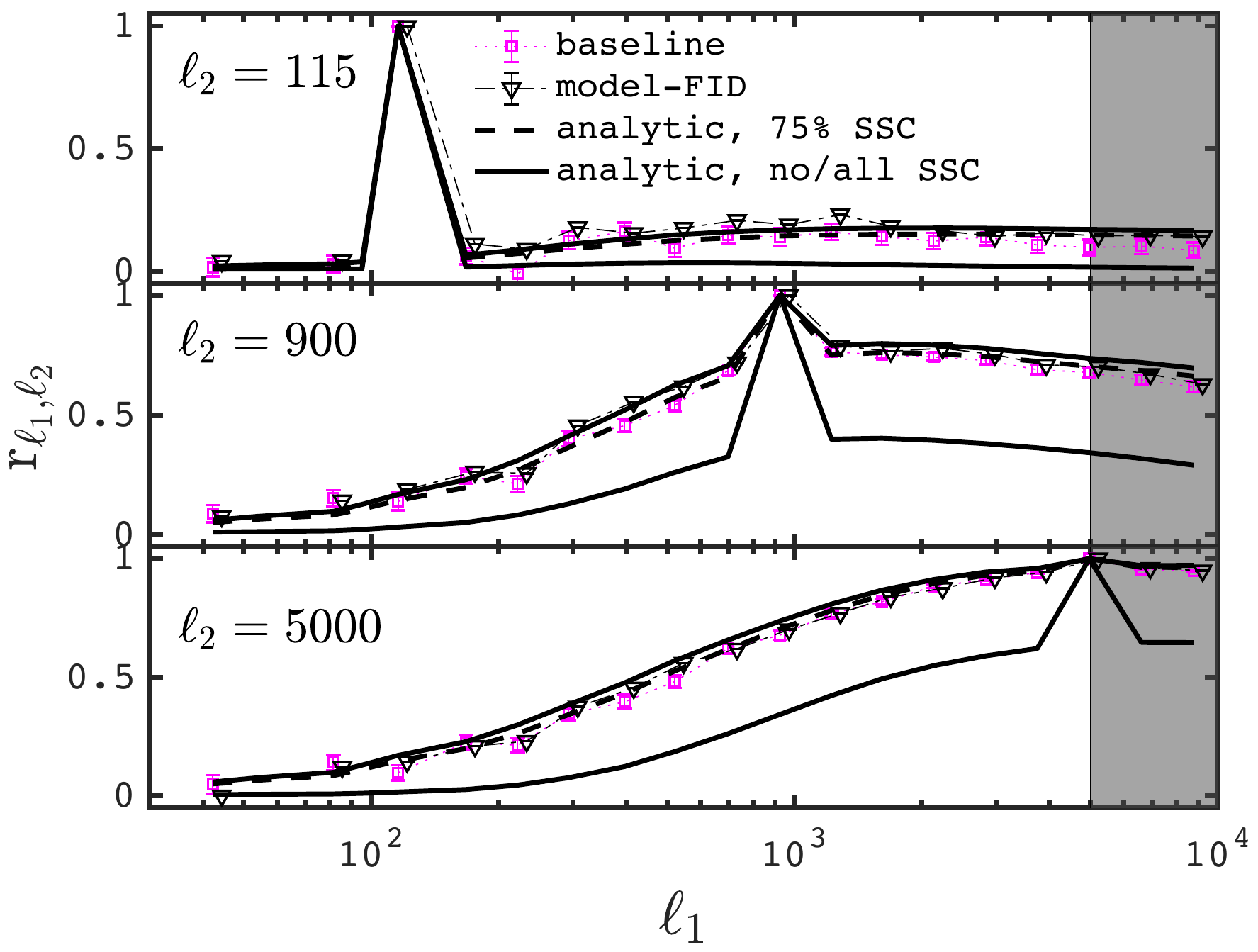}
\caption{Comparison between the cross-correlation coefficients measured from the baseline method (magenta squares), from the {\it cosmo}-SLICS (triangles) and from the analytic model with different amounts of SSC (thick and dashed lines). The spikes seen in these panels indicate the point of crossing with the diagonal, where $r_{\ell\ell'} \equiv 1.0$  for $\ell = \ell'$.}
\label{fig:r}
\end{center}
\end{figure}

We next constructed and compared the four cross-correlation coefficient matrices, defined as $r_{\ell \ell'} = {\rm Cov}^{\kappa}_{\ell \ell'}/\sqrt{{\rm Cov}^{\kappa}_{\ell \ell}{\rm Cov}^{\kappa}_{\ell' \ell'}}$, which highlight the amplitude of the mode-coupling. The results  are presented in Fig. \ref{fig:r}, where we show slices through the matrices while holding one of the components fixed ($\ell'=115, 900$ and $5000$). From the upper to the lower panel, we present $r_{\ell,115}$, $r_{\ell,900}$ and $r_{\ell,5000}$, using the symbol convention of Fig. \ref{fig:NG_G_ratio_FID}. We observe an excellent agreement between the simulation-based methods, which both appear to be consistent with capturing about 75\% of the SSC contribution once compared with the analytic methods. These results correspond to the shape noise-free case and thereby provide the upper limit on the importance of these off-diagonal terms; the inclusion of shape noise significantly down-weights their overall contributions, further diluting the small differences  between the estimators observed in Figs. \ref{fig:NG_G_ratio_FID} and \ref{fig:r}. 

\subsubsection{Fisher forecast}
\label{subsubsec:fisher}

\begin{figure*}
\begin{center}
\includegraphics[width=3.5in]{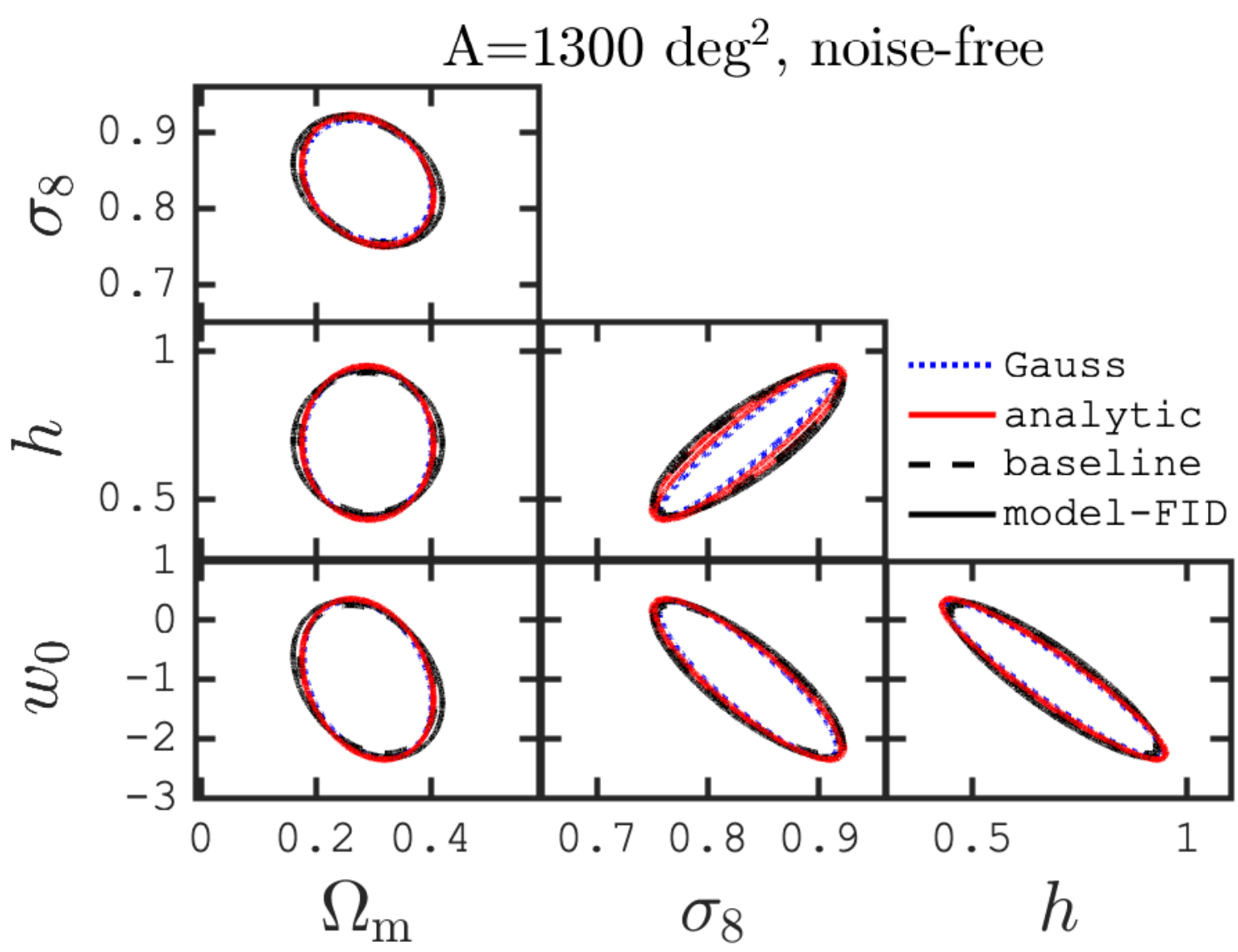}
\includegraphics[width=3.5in]{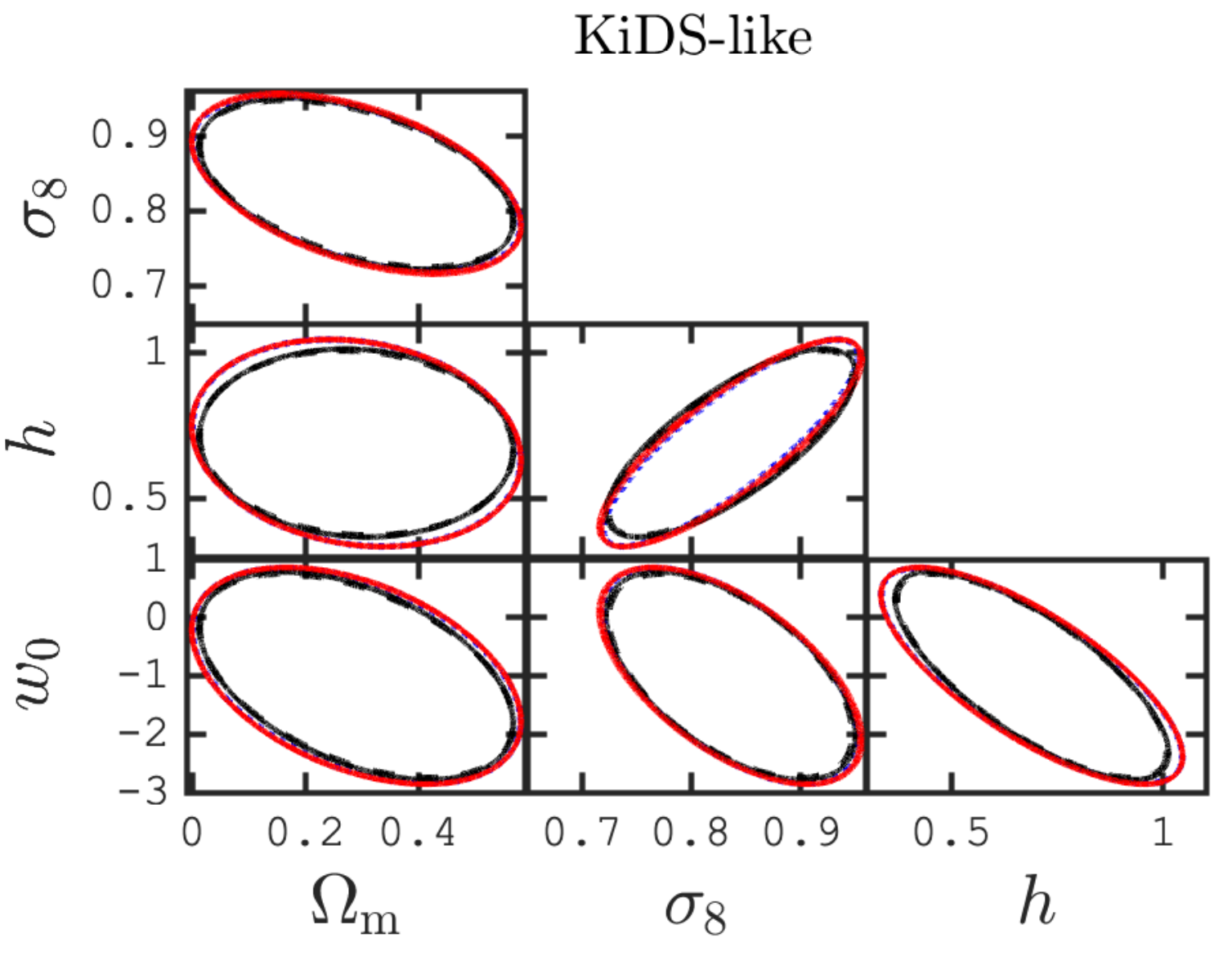}
\includegraphics[width=3.5in]{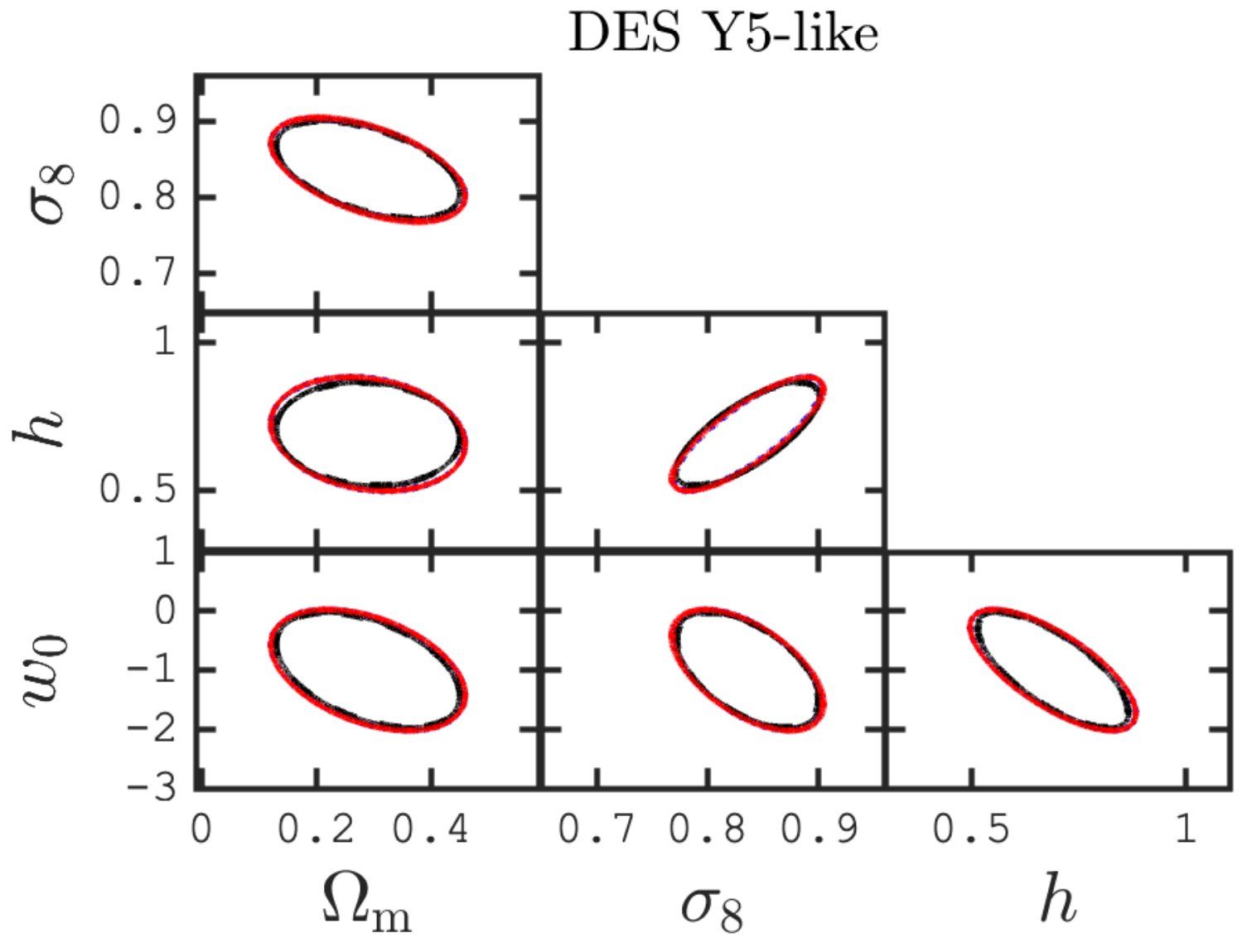}
\includegraphics[width=3.5in]{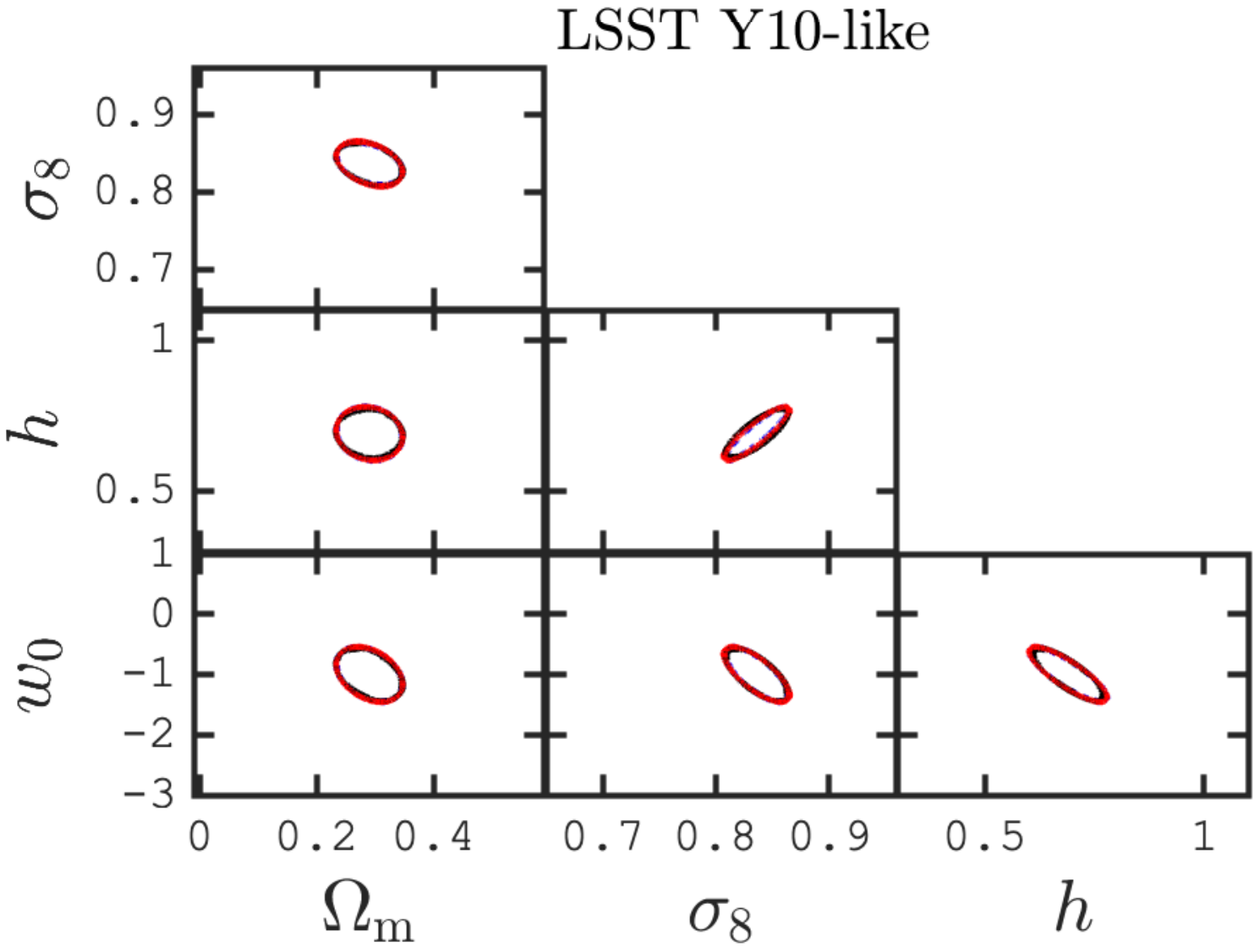}
\caption{Measurement forecasts on  cosmological parameters obtained with different estimates for the covariance matrix (shown with the different lines in the sub-panels), and for different survey properties. Curves show the 95.4\% confidence intervals. In our LSST-Y10 configuration, and cycling through the panels starting from the uppermost, the ${\rm Cov}^{\kappa}_{\rm G}$ term accounts for (92, 98, 72, 96, 91, 94)\% of the area. }
\label{fig:Fisher_contours}
\end{center}
\end{figure*}

The four different methods agree qualitatively on most properties of the full covariance matrix, but differ in the details, exhibiting various noise levels and converging on coupling strengths that are at times slightly offset. Given that it is unclear which of these covariance estimates is the best,  we sought to find out whether   these differences matter for weak lensing data analyses. To answer this, we carried out a series of Fisher forecast analyses based on equation (\ref{eq:fisher}) in which we cycled through three of our four covariance matrix options (baseline, model-FID and analytic, but we dropped the matched SLICS for redundancy reasons) and examined the differences in the  constraints on $\Omega_{\rm m}, \sigma_8, w_0$ and $h$. 
We additionally fragmented the analytical case in its three components to further our insight on the relative importance of each term.
We included multipoles  in the range $35 < \ell < 3000$, inspired by the fiducial angular scale selection of the LSST {\it Science Requirement Document} \citep{LSST_SRD}. 

Starting with the analytic methods, the forecasted constraints from the Gaussian-only matrix are shown  in Fig. \ref{fig:Fisher_contours} with the dashed-blue lines, the Gaussian+non-Gaussian case with the inner solid red lines, and the total covariance with the outer solid red line (these three lines are plotted in every panel, but overlap in most cases). In the first survey configuration (upper-left triangle plot), we assumed an area of 1300 deg$^2$ with no shape noise. Our results are consistent with the findings of \citet{2018JCAP...10..053B}, where it is demonstrated that the Gaussian and the SSC terms together capture most of the uncertainty about the cosmological parameters, whereas  ${\rm Cov}_{\rm NG}^{\kappa}$ contributes minimally. Adopting the area of the Fisher ellipses as a metric, neglecting the non-Gaussian term amounts to underestimating the areas by 5-7\% only, except for the $[\sigma_8 - h]$ join contour where the change reaches 18\%. Differences in survey geometry and data vectors can explain why we observe a sensitivity in this particular parameter plane while \citet{2018JCAP...10..053B} do not: their measurements, made with fine tomographic sampling, are more sensitive to the growth of structure, which translates into tighter constraints in general. The degeneracy direction of the $[w_0 - \Omega_{\rm m}]$ is also flipped for the same reason. These conclusions about the relative non-importance of ${\rm Cov}_{\rm NG}^{\kappa}$  cannot be generalised to all weak lensing measurement techniques, since some alternatives (e.g. peak statistics) may be more sensitive than $C_{\ell}^{\kappa}$ to the non-Gaussian signal, and therefore might receive a larger contribution from the ${\rm Cov}_{\rm NG}^{\kappa}$ term.

The simulation-based methods are also shown on these plots; the baseline with the  dashed black lines and the {\it cosmo}-SLICS results with the solid black lines.  Although it is difficult to observe in the figure, the Fisher ellipses from these two methods differ by 10-15\% in area;  the baseline and the analytic estimates  (assuming 100\% SSC) differ by less than 7\%, while the model-FID and the analytic method by less than 11\%. 
Whether these apparently slight differences matter or not depends on the overall error budget of the measurement. In the KiDS-450 cosmic shear analysis for example, these changes were shown to be sub-dominant compared to the uncertainty associated with the photometric redshift estimation or with the baryon feedback models \citep{KiDS450}. 
This is bound to change as the statistical power of weak lensing surveys increases, and for this reason we repeated the forecasts with three survey configurations (summarised in Table \ref{table:surveys}). 
%
%

First, we included shape noise and sky coverage in amounts that mimic the KiDS survey configuration defined in Table \ref{table:surveys} (upper right triangle plot). In this case, the two simulation-based methods provide areas that differ by less than 6\%, and by at most 15\% with the analytical estimate. Second, we lowered the galaxy density but increased the area to emulate a DES-Y5 survey (lower left triangle). In that case, the baseline and the {\it cosmo}-SLICS methods agree to better than 4\%, with a 10-16\% match in area with the analytic method. We finally increased both the area and the density to generate a LSST Y10-like survey (lower right), in which case the match in areas between the two simulation estimates decreases to the 10\% level, while preserving the agreement with the analytic model seen in the DES-Y5 set-up. In summary, when propagated into a Fisher forecast, the three covariance matrices predict cosmological constraints that agree well given their radically different estimation methods. One could then possibly interpret the scatter in area as an uncertainty on the error contours, sourced by systematic error on the covariance. 

Once we move away from the two-point statistics however, the simulation-based methods are often the only option left. If we further wish to evaluate the covariance matrix at an arbitrary point in parameter space (i.e. at the best-fit cosmology given by the data), then {\it cosmo}-SLICS could be a prime estimation method, which we present next.

\subsection{Dependence on cosmology}
\label{subsec:CovCosmo}

\begin{figure*}
\begin{center}
\includegraphics[width=7.5in]{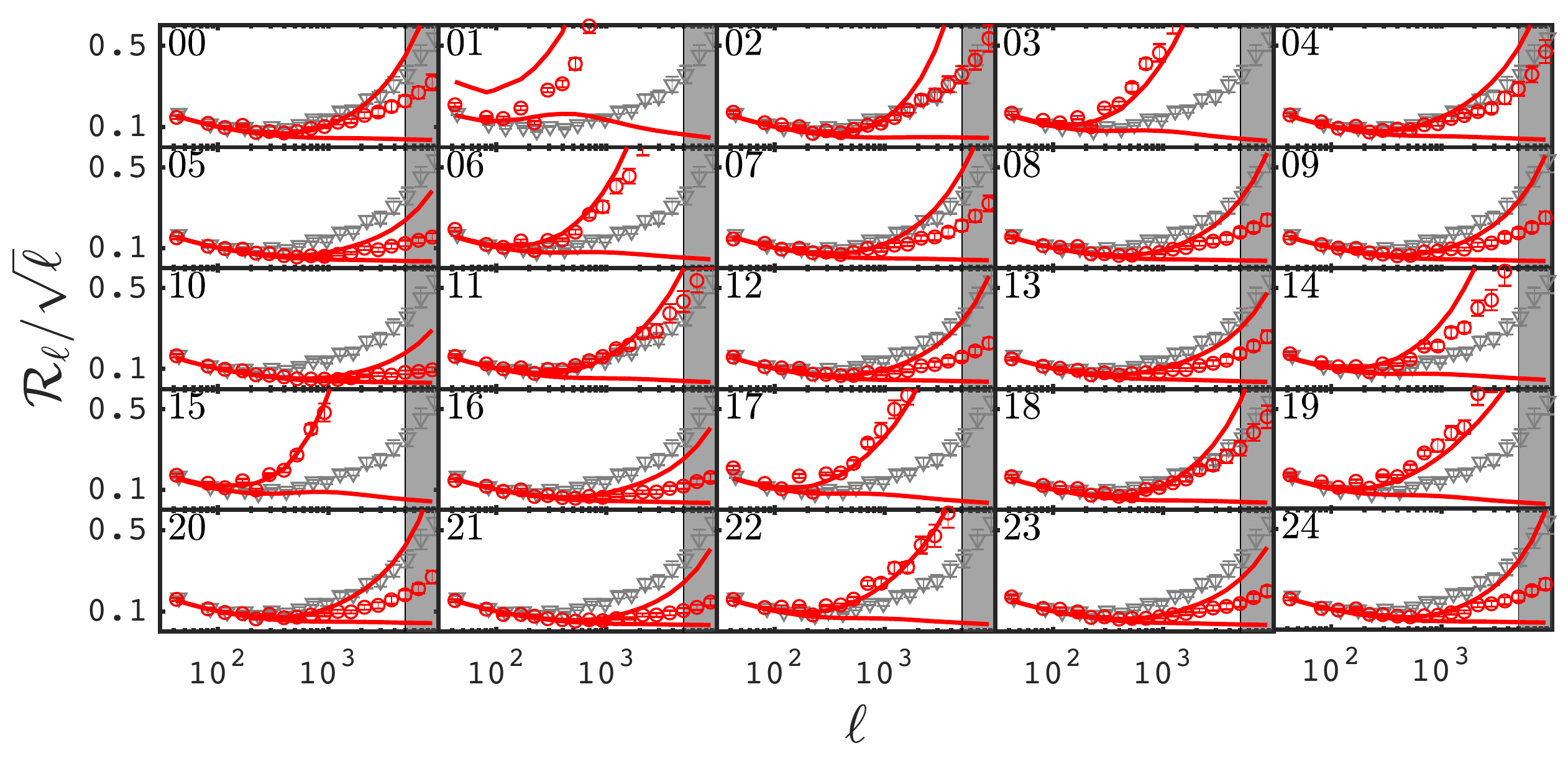}
\vspace{-5mm}
\caption{Similar to the upper panel of Fig. \ref{fig:NG_G_ratio_FID}, but now showing with red circles the results from all different {\it cosmo}-SLICS models, and with red lines the corresponding analytical predictions with none and all of the SSC contribution. For reference, we also overplot with grey triangles the model-FID in each of the panels. }
\label{fig:NG_G_ratio_cosmo}
\end{center}
\end{figure*}

\begin{figure}
\begin{center}
\includegraphics[width=3.5in]{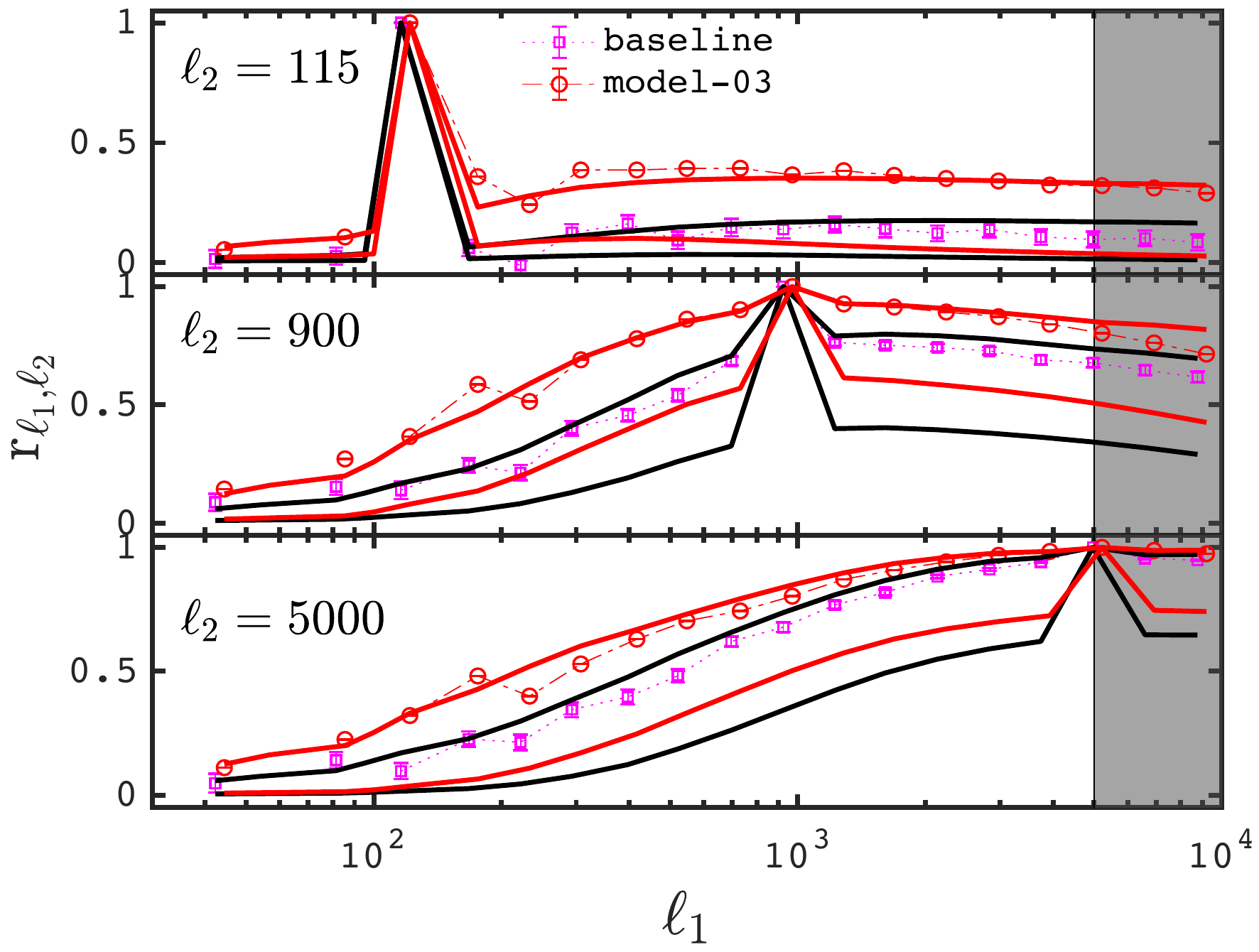}
\includegraphics[width=3.5in]{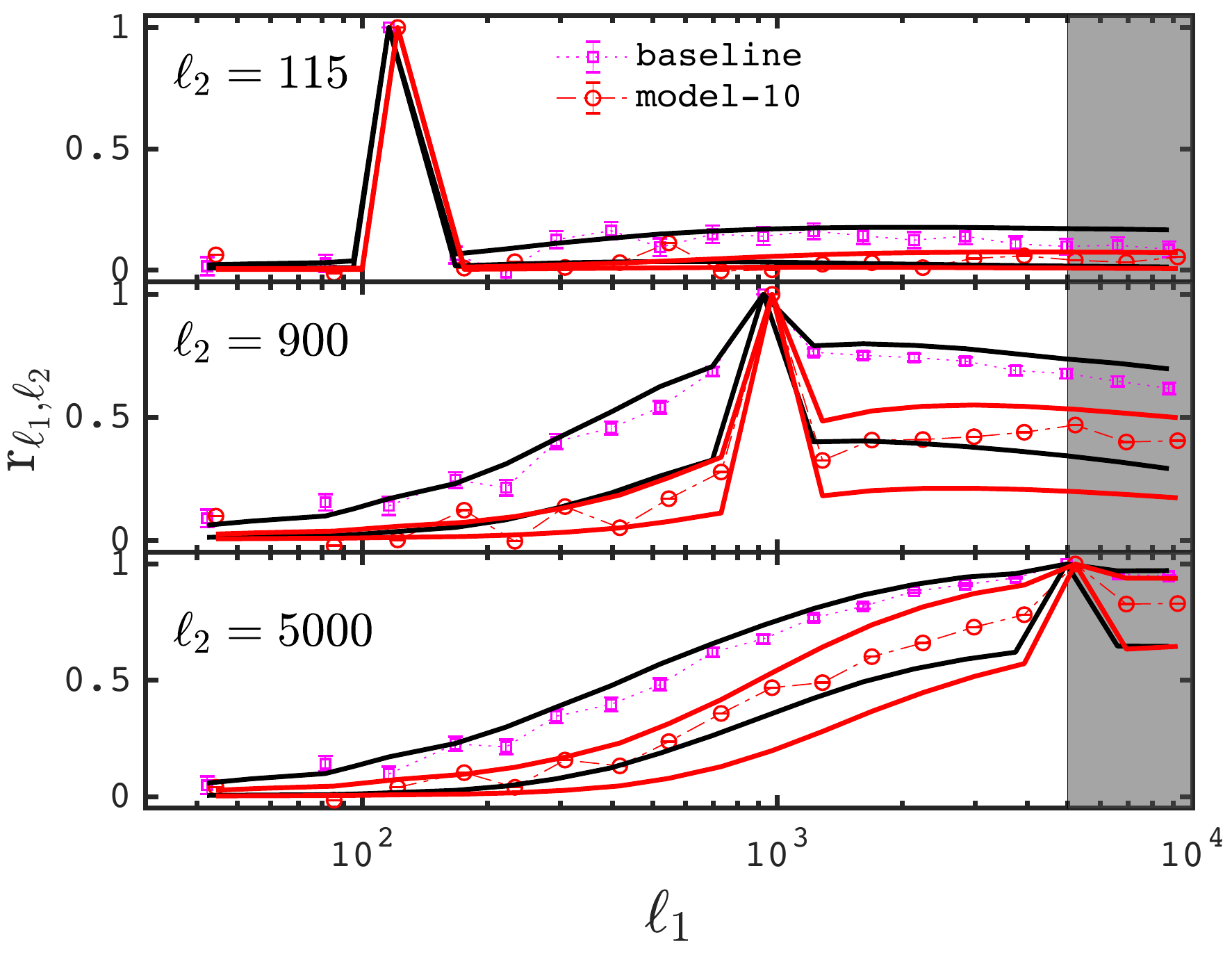}
\caption{Same as Fig. \ref{fig:r}, but for different cosmologies. The magenta squares and black lines are taken from Fig. \ref{fig:r} and show the baseline estimator and the analytic model at the fiducial cosmology. The red circles and red lines are from the {\it cosmo}-SLICS and the analytic predictions respectively, for model-03 (upper) and model-10 (lower). Results from all other models are similar to these.}
\label{fig:r_12}
\end{center}
\end{figure}

\begin{figure*}
\begin{center}
\includegraphics[width=3.5in]{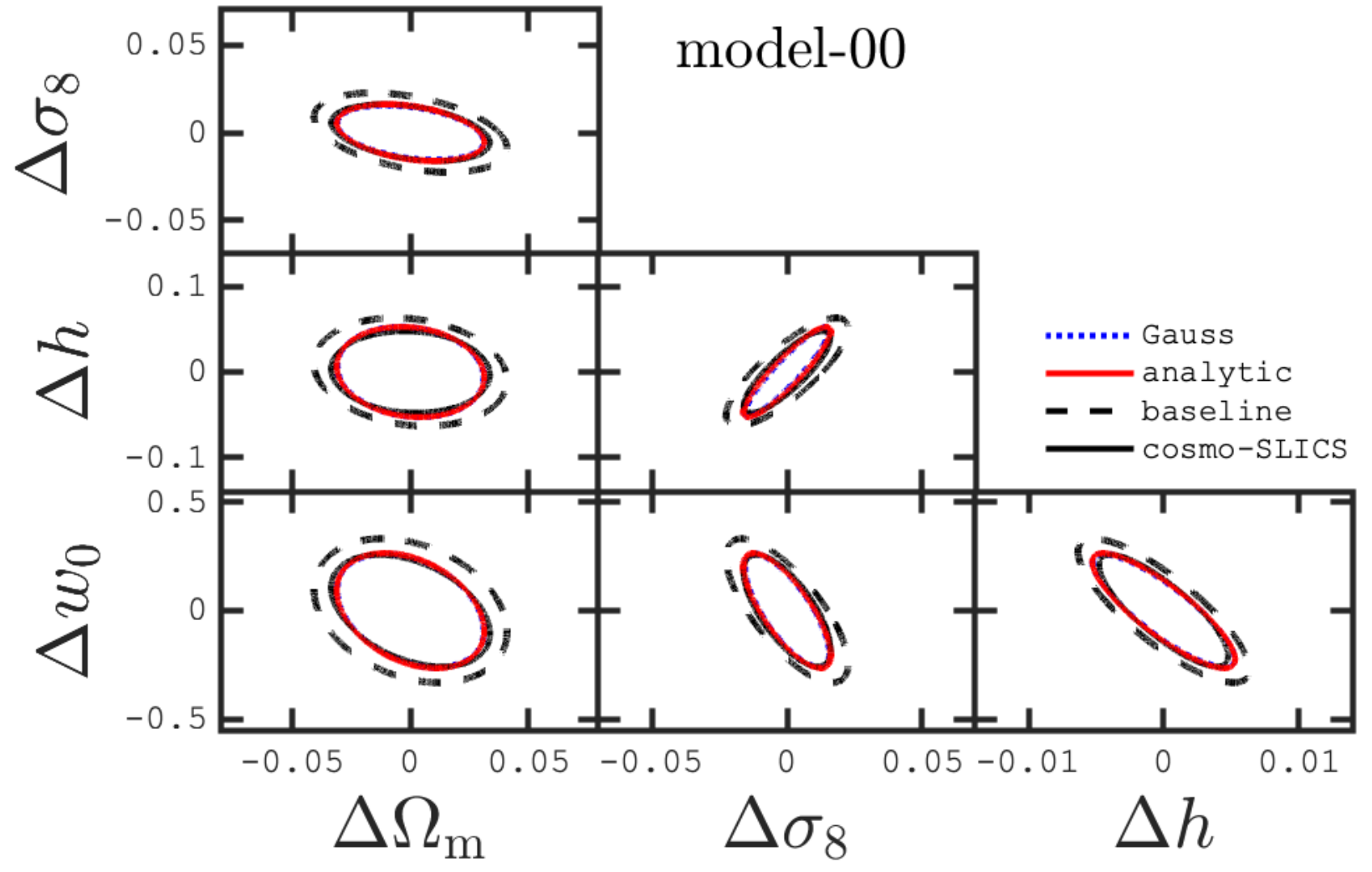}
\includegraphics[width=3.5in]{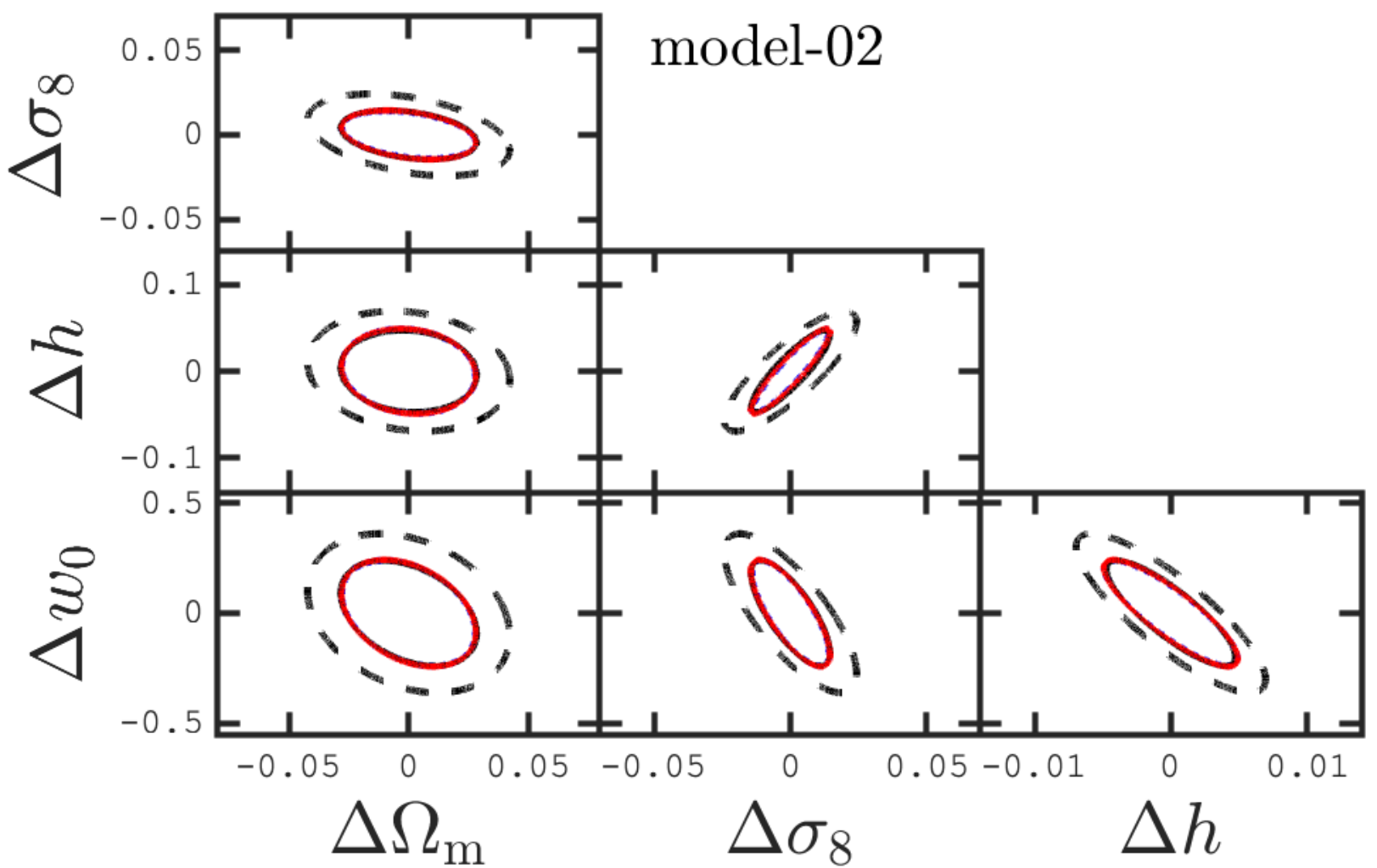}
\includegraphics[width=3.5in]{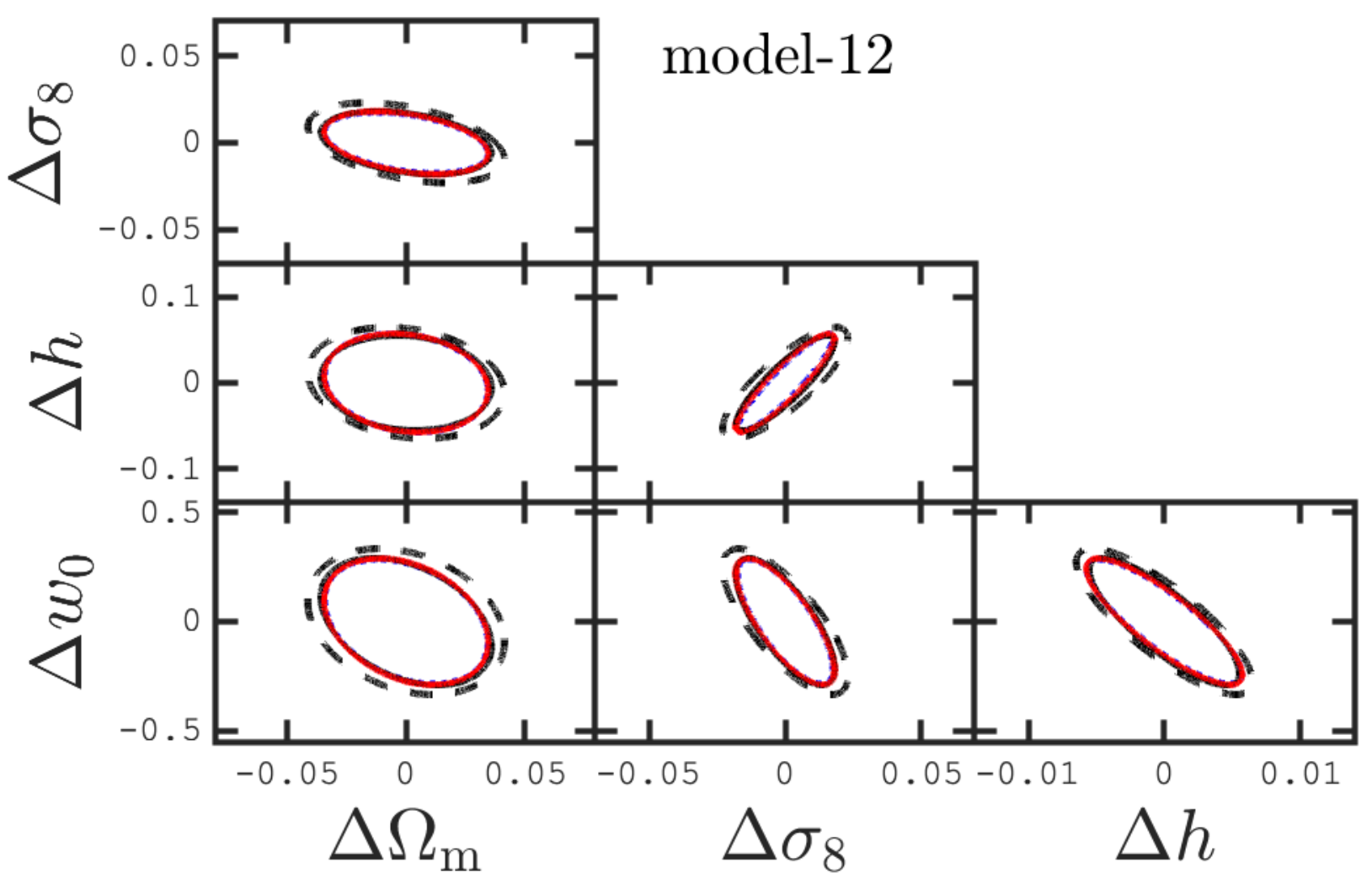}
\includegraphics[width=3.5in]{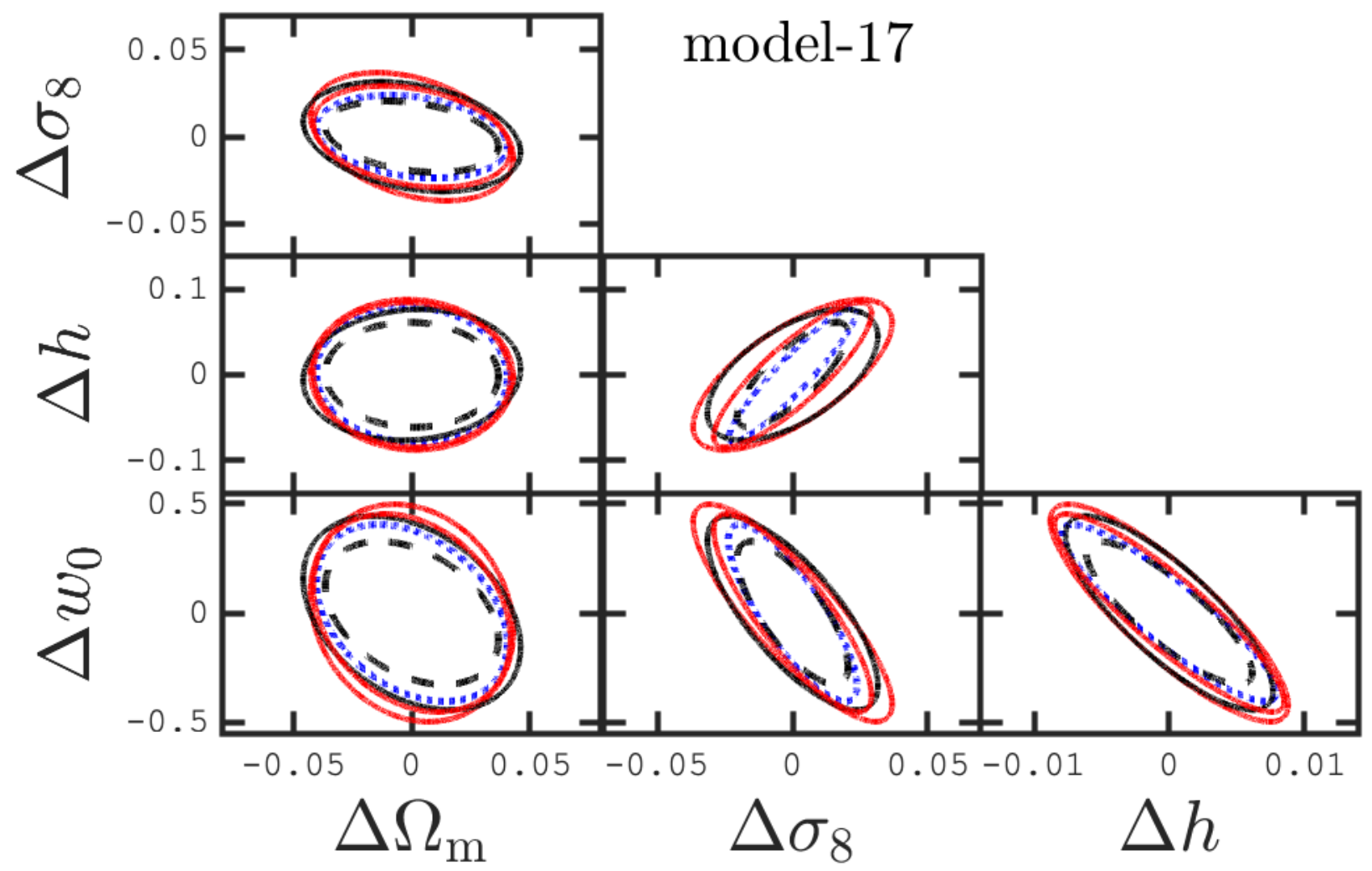}
\caption{Measurement forecasts on cosmological parameters from an LSST Y10-like survey, obtained with different estimates for the covariance matrix, and for different input cosmology. Curves show the 95.4\% confidence intervals. Measurement are shown relative to  the input value (hence the `$\Delta$' in the axis labels) in order to align the different cosmologies to the origin and highlight the change in size of the error contours caused by variations in cosmology.} 
\label{fig:Fisher_contours_cosmo}
\end{center}
\end{figure*}

We have established in the last section that the lensing covariance matrix estimated from the model-FID is well suited for current $C_{\ell}^{\kappa}$-based lensing analyses\footnote{Analyses based on correlation functions $\xi_\pm$ further need to account for the finite box effects inherent to the SLICS simulations.}, and possibly for upcoming experiments as well. Achieving this accuracy with only two independent $N$-body simulations opens up a new path to study the impact that variations in cosmology have on the lensing covariance and on the parameter constraints, regardless of the choice of weak lensing estimator.
%
%
The matched-pair strategy presented in this work could play a key role, as there are no large ensembles required anymore: one simply needs to resample the {\it cosmo}-SLICS nodes (or other simulation pairs produced in a similar way) and to interpolate between the nodes to the desired cosmology, as suggested by \citet{2008PhRvD..78f3529S}. 

That being said, multiple studies suggest that varying the covariance matrix in a multivariate Gaussian likelihood is neither mathematically correct \citep[e.g.][]{2013A&A...551A..88C} nor necessary \citep{2018arXiv181111584K}, and that instead one should evaluate the matrix at the best fit cosmology and keep it fixed in the likelihood. 
This approach was adopted by \citet{2017arXiv170605004V} who use the same analytic covariance model as ours in their analysis of the combined KiDS-450$\times$GAMA data. At the parameter inference stage, they first guess an initial cosmology at which the covariance matrix is  evaluated, they next solve for the best fit cosmology given the data and that initial covariance matrix, they then update the covariance  with these new parameters and recalculate a new best fit cosmology;  convergence on the posterior distributions of the parameters is achieved after 2-3 iterations.

It seems however that a consensus on the subject has not been reached, considering that cosmology-dependent covariance matrices are utilised as a cross-check in the angular power spectrum analysis of the BOSS-DR12 data \citep[][see their figure 10]{2019MNRAS.485..326L}, in the HSC-Y1 cosmic shear analysis \citep{2018arXiv180909148H}, or in the hybrid\footnote{The covariance matrix used by \citet{Kilbinger2013} consists of a non-Gaussian term estimated from an ensemble of mocks at a fixed cosmology, and a Gaussian term that varies with cosmology in the likelihood.}  approach of the CFHTLenS cosmic shear analysis \citep{Kilbinger2013}. We do not intend to settle the issue here, but rather wish to enable this type of inquiries with simulation-based covariance estimators. 

Besides deciding on whether to fix the covariance or let it vary within the likelihood sampling, anchoring the matrix (or converging) to different points in cosmology will have consequences on the parameter constraints, by an amount we need to quantify. We therefore examined in this Section what happens to the Fisher forecast contours when we varied the cosmology at which the covariance matrix is fixed. We adopted the same data vector as in Sec. \ref{subsubsec:fisher}, and present the results at the 25 $w$CDM cosmologies from both the analytic model and the {\it cosmo}-SLICS estimator. 


The diagonal terms are plotted in Fig. \ref{fig:NG_G_ratio_cosmo} for all models (in red circles), compared to the  model-FID estimate (grey triangles) and the analytic model with and without the SSC term (red solid). We first observe that the simulation-based estimates fall between the two analytic cases for all cosmologies except models-03 and -19, two models for which $w_0$ is close to $-0.5$  and hence their SSC term is not well calibrated (we examine the halo mass function of model-03 in Appendix \ref{sec:SLICS_vs_model}). Since other components are known to be uncertain as well, we conclude that this bracket adequately bounds the simulation results most of the time. 

Our second observation is that although rarely in agreement, the {\it cosmo}-SLICS and analytic estimates are highly correlated: the red curves and symbols move up or down with respect to the model-FID in the same way, although not  by the same amount, suggesting that at a fundamental level, variations in cosmology push the mode-coupling term in the right direction. In fact, this aligns with some of the tests carried out in \citet{2017MNRAS.465.4016R}, where the consistency in the $\Omega_{\rm m}$ and $\sigma_8$ scalings is established between a tree-level perturbation theory trispectrum and a small number (50) of numerical simulations. 
Although a direct comparison is unfortunately not possible, our results appear to follow their scaling relations.
For example, they find that decreasing $S_8$  from 0.82 to 0.7 reduces the trace of the lensing covariance matrix by about 50\%, while increasing $S_8$ to 0.9 augments it by 50\%. The {\it cosmo}-SLICS models-00, -08 and -11 feature a similar decrease in $S_8$ with respect to the model-FID, and also display a reduction in their traces  by 49\%, 72\% and 63\%, respectively\footnote{For this calculation only we employ a similar $\ell$-binning scheme and reject bins with centres outside the range $\ell \in [115 - 2900]$; \citet{2017MNRAS.465.4016R} carried out their analysis over the range  $\ell \in [100 - 2500]$. Further differences exist in our redshift distributions: ours  consist of a single plane at $z_{\rm s} = 1.0$, whereas theirs follows a broad {\it Euclid}-like $n(z)$ peaking at $z=0.9$.}.  When increasing the lensing signal to $S_8 \sim 0.9$ with models-04, -17 and -19, we find that the traces vary by +9\%, +25\% and -22\%, respectively. 
The scatter in scaling values is caused by the variations in the other parameters, which in the end contribute to the covariance and further complicate this comparison. In their study,  \citet{2017MNRAS.465.4016R}  compute the scaling of the Frobenius norm with $\Omega_{\rm m}$ and $\sigma_8$, but are unable to validate the trispectrum scaling on an element-by-element basis. Given the large size of their error bars, the numerical convergence that they recognise is not achieved, and the important role of other cosmological parameters such as $h$ and $w_0$, we conclude that despite a broad agreement with their results, it is currently impossible to assert the accuracy of analytical trispectrum calculation outside $\Lambda$CDM, up to and beyond $\ell = 3000$. In this context, the {\it cosmo}-SLICS offer an avenue to push our understanding of the lensing covariance one step further, exploring new cosmologies without being restricted to two-point statistics.

The off-diagonal components of these matrices are next presented in Fig. \ref{fig:r_12} for two representative cosmologies (models-12 and -20). The agreement with the analytic models is similar to the fiducial scenario shown in Fig. \ref{fig:r}, being mostly bracketed by the two solid curves in both cases.  We overplot on this figure the previous baseline (in magenta squares) and the predictions at the fiducial cosmology (in black solid line) to illustrate that the cosmology scaling of $r_{\ell\ell'}$ is well captured by both methods. We have verified that this holds for all other models as well, which we therefore decided not to show.

We finally present in Fig. \ref{fig:Fisher_contours_cosmo} our Fisher forecasts in the LSST Y10-like case (i.e. equivalent to the bottom-right triangle plots of Fig. \ref{fig:Fisher_contours}),  but now varying the input cosmology of the covariance matrix. We show here representative results from four models out of 25 to illustrate our point, comparing in each case the constraints from the analytic model and from the {\it cosmo}-SLICS; we also include the baseline model  as a reference. The impact of cosmology on these ellipses is striking, especially between models-02 and -17, with changes in area that sometimes almost reach a factor 6. The simulations and theoretical models trace each other generally well across many of these scenarios, matching on average the ellipses' area at the 15-25\% level, even though they exhibited major differences in $\cal R_{\ell}$. The worst agreement occurs for models-03, -17 and -19, in which the areas of simulation-based ellipses are up to 16\% smaller than for the analytic method. These models all have extreme values of $w_0$, for which the halo mass function is not well calibrated (see Appendix \ref{sec:SLICS_vs_model}).

Also obvious from Fig. \ref{fig:Fisher_contours_cosmo} is that changing the cosmology has a much larger effect than changing estimator at a fixed cosmology (e.g. switching from the model-FID to the analytical estimates or the baseline in the top-left triangle plots of Fig. \ref{fig:Fisher_contours}). In other words, it is more important to estimate the lensing covariance matrix at the correct cosmology than to fine-tune the estimator, especially if computed at the wrong cosmology. In light of this it becomes clear that the ability to evaluate the covariance matrix at a flexible cosmology is critical, and in order to achieve this for an arbitrary weak lensing signal, we propose to train an emulator on the 25 {\it cosmo}-SLICS covariance matrices and interpolate at the desired cosmology. The next section presents a toy example that illustrates how this can be achieved in an actual lensing data analysis.

\subsection{Emulation of the cosmic shear covariance}
\label{subsec:emu_cov}

\begin{figure}
\begin{center}
\includegraphics[width=3.6in]{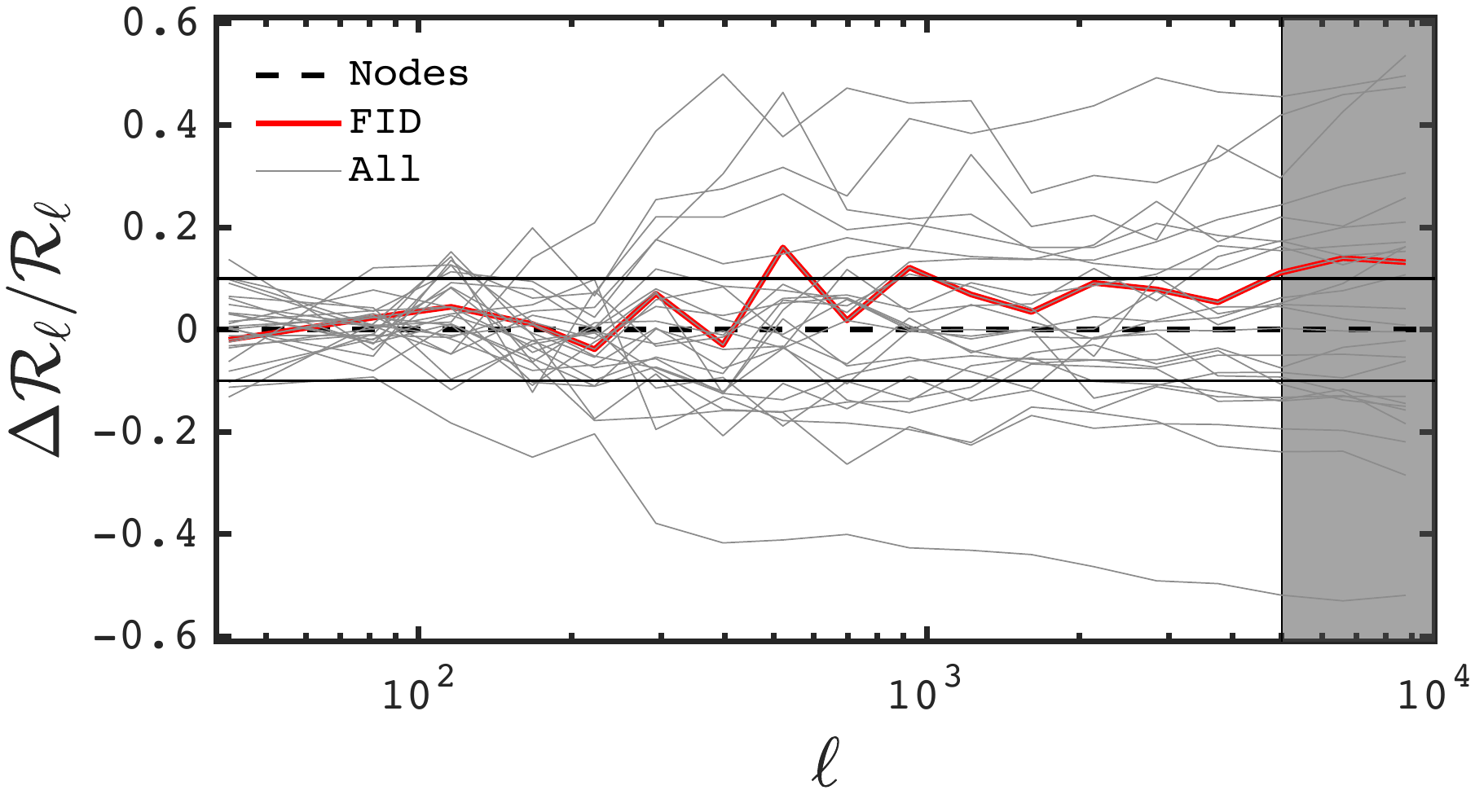}
\caption{Fractional difference on ${\mathcal R}_{\ell}$ between the measurements at the 25+1 {\it cosmo}-SLICS nodes and the interpolated predictions from our GP emulator, obtained in our `leave-one-out' cross-validation test. The thick red line represents the $\Lambda$CDM prediction after training on the $w$CDM models, and the thin horizontal lines indicate the $\pm 10\%$ range.}
\label{fig:GP_F_ell}
\end{center}
\end{figure}

In this section we present how well our Gaussian process (GP) emulator can learn the cosmology dependence of the covariance matrices from the 25 {\it cosmo}-SLICS nodes. More precisely, we trained the emulator on the ${\mathcal R}_{\ell}$ measurements presented in Fig. \ref{fig:NG_G_ratio_cosmo} and defined in equation (\ref{eq:F}). In this setup, we imagine that we have confidence in the analytical Gaussian term only, but would prefer to use the ${\rm Cov}_{\rm NG}^{\kappa}$ and ${\rm Cov}_{\rm SSC}^{\kappa}$ terms from the simulations; ${\rm Cov}_{\rm G}^{\kappa}$ and the {\it cosmo}-SLICS estimate of ${\mathcal R}_{\ell}$ can therefore be combined to compute the full variance about the cosmic shear signal at any cosmology.  

Following a similar approach to  \citet{MiraTitan} and \citet{EuclidEmulator_etal_2018}, we emulated the principal components of log${\mathcal R}_{\ell}$, which varies over a reduced dynamical range (we refer the reader to Appendix \ref{sec:emulator} for more details about our GP emulator). We assessed the accuracy of our method with a `leave-one-out' cross-validation test, in which we trained the emulator on all but one of the nodes, then compared at that cosmology the emulated prediction with the left-out measurement. Our results, presented in Fig. \ref{fig:GP_F_ell}, indicate an accuracy of better than 20\% for most of the models, with some outliers that perform less well in this test. Notably, removing (extreme) models-01, -02, -10 or -14 resulted in a particularly poor interpolation. We recall that by construction, cross-validation provides a lower limit on the accuracy, since it requires the emulator to interpolate to cosmologies at the outer edges of the training set range, and from an incomplete set of training nodes. The only representative case occurs when leaving out the $\Lambda$CDM model-FID, as it resides outside the Latin hyper-cube. For this reason, we consider this special case as the benchmark accuracy of our covariance emulator. 

The thick red line in Fig. \ref{fig:GP_F_ell} represents the comparison between our $\Lambda$CDM ${\mathcal R}_{\ell}$ prediction after training on the 25 $w$CDM models, and the test value measured from the model-FID. This test reveals that our GP emulator matches the test case to better than 10\%, a promising result that can likely be generalized to other lensing statistics provided the reasonable assumption that the variation of the covariance with cosmology is of similar amplitude.  The exact accuracy of the covariance emulator based on the {\it cosmo}-SLICS of course needs to be assessed for every lensing method, but the tests presented in this section should serve as guidelines, and provide an order-of-magnitude estimation of the accuracy one can achieve that way.

\section{Discussion}
\label{sec:conclusion}

\begin{figure}
\begin{center}
\includegraphics[width=3.6in]{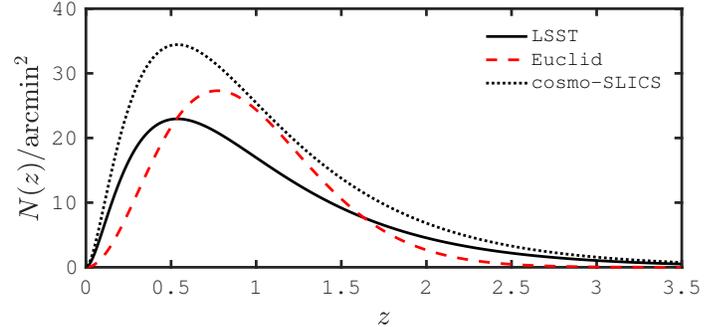}
\caption{Galaxy redshift distribution  from the LSST and {\it Euclid} forecasts, compared to the {\it cosmo}-SLICS  catalogues.}
\label{fig:nz}
\end{center}
\end{figure}

As mentioned earlier in this paper, the fundamental motivation behind the production of the {\it cosmo}-SLICS simulations is to provide a public training set with which new weak lensing observables can be developed. One can then wonder why we have focused on Fisher analyses of two-point statistics, with no more mention of these alternative techniques. The reason behind this choice is sound however: we needed to assess the accuracy of our simulated data, which is straightforward in the case of two-point statistics given that analytical predictions are readily available. And although we have not established the performance of all possible weak lensing estimators, the fact that both the mean and the covariance of the lensing power spectra are 
in overall agreement with the analytical predictions provides compelling evidence that other higher-order moments are correctly captured as well. Of course this has to be demonstrated in every case, but not necessarily for all cosmologies. 

We provide shear, convergence and mass over-density maps for 25 light-cones per seed, per node, for a total of 5000 deg$^2$ per cosmology, and 130,000 deg$^2$ in total. The lensing maps can then be ray-traced  to construct a series of mock galaxy catalogues with a user's defined $N_{\rm s}(z)$ and shape noise, while the mass maps can be populated with foreground `lens' galaxies of a given $N_{\rm l}(z)$ and a controlled linear bias \citep[as in, e.g.][]{2017arXiv170605004V}. 
The storage footprint of these maps is significant, ranging from 14.4 to 26.9 Gb per light-cone per cosmology for the set of maps. We are unfortunately not equipped to host 800 light-cones per cosmology in that form, so instead we opted for the more compact option of storing mock galaxy catalogues. Even with a density as large as 45 gal/arcmin$^2$, keeping 800 copies per cosmology with 6 entries per object (RA, Dec, $z_{\rm spec}$, $\gamma_1$, $\gamma_2$, $\kappa$) requires just over 8Tb. We selected a redshift distribution that exceeds at all redshift the forecasts from LSST and {\it Euclid}, such that the {\it cosmo}-SLICS catalogues can be down-sampled to match either data sets. In all cases, the source redshift distributions assume a functional form given by:
\begin{eqnarray}
n(z) \propto z^2 {\rm exp}\left[-\left(\frac{z}{z_0}\right)^\beta \right]
\end{eqnarray}
and are normalized such that $\sum n(z) dz = n_{\rm gal}$ (see Fig. \ref{fig:nz}).
In their Science Requirement Document,  \citet{LSST_SRD} use $n_{\rm gal}=$ 30 gal/arcmin$^2$, $\beta = 0.68$ and $z_0 = 0.11$ (see their figure F4); the {\it Euclid Theory Working Group}  instead quote  $n_{\rm gal} = 30$, $\beta = 1.5$ and $z_0 = 0.637$ \citep[][see their equation 1.212]{Euclid_th}; in our simulations, we opted to use the LSST $n(z)$, augmented to reach $n_{\rm gal}=$ 45.0 gal/arcmin$^2$.

With these catalogues, a lensing covariance matrix can be evaluated at each of the 25+1 nodes, then interpolated at any given cosmology inside the parameter range with our GP emulator. One must remember that this still provides a noisy estimate of the full matrix, and that the inversion introduces extra errors that must be accounted for \citep{Hartlap2007, Dodelson2013, 2014MNRAS.442.2728T, 2016MNRAS.456L.132S}. One could eventually push the envelope further and resample the volume even more \citep[][for example, ray-traced the simulations $10^4$ times]{Petri16} potentially suppressing the noise down to negligible values, however this would likely hit the residual  noise inherent to our matched-pair technique. A robust verification of this idea is required, which we defer to future work. Another approach that may be worth exploring consists in working directly with the precision matrix (the inverse of the covariance matrix) without first estimating the covariance matrix, as suggested by e.g. \citet{2016MNRAS.460.1567P} and \citet{2018MNRAS.473.4150F}.

When calibrating an estimator on controlled mock data, one has to bear in mind that the numerical simulations themselves are subject to three basic limitations\footnote{For the sake of simplicity, we are factoring out from this discussion the effect of baryonic feedback, secondary signals and the detailed implementation of observational effects.}, namely their finite box sizes, their finite small-scales (or mass) resolution, and residual inaccuracies in the non-linear evolution segment of the $N$-body code. 
Given a novel measurement method, all of these aspects must be carefully considered. We recommend to assess the accuracy range of the {\it cosmo}-SLICS by training on lensing simulations with higher mass resolution   \citep[such as the  SLICS-HR introduced in][]{2015MNRAS.450.2857H} and larger volume  such as the HSC mocks \citep[][]{HSCmocks} or the MICE-GC described in \citet[][]{Fosalba2013}. That way, it becomes possible to identify the part of the {\it cosmo}-SLICS data vector that can be fully trusted.

Additionally, the parameter space can be expanded by combining our simulations with external suites. For example, sensitivity to variations in the neutrino mass $M_{\nu}$ can be probed with the {\it MassiveNuS} simulations\footnote{http://columbialensing.org/\#massivenus}, which simultaneously vary $\Omega_{\rm m}$, $\sigma_8$ and $M_{\nu}$ \citep{MassiveNuS}. Among the suites of existing simulations, we also point out the {\it Mira-Titan} simulations \citep{MiraTitan}, the {\it Aemulus} simulations \citep{Aemulus} and those from the {\it DarkEmulator} collaboration \citep{DarkEmulator}, which could also serve this purpose, however their light-cone data has not been released to the public yet.

We also acknowledge the fact that the area (100 deg$^2$) of our lines of sight prevents us from measuring structures at very large angular separations in the simulations. Although a clear limitation to some measurement techniques, the information contained at such large (linear) scales is well captured with the two-point correlation functions, and well described by the Gaussian term of the covariance matrix, for which numerical simulations are not required. 

One question remains open throughout our work on covariance, which concerns the exact amount of SSC that is actually contained in our simulation suites. Figs. \ref{fig:NG_G_ratio_FID}, \ref{fig:r}, \ref{fig:NG_G_ratio_cosmo} and \ref{fig:r_12}
provide compelling indicators that the two simulation-based covariance estimates include a large fraction, but the exact amount is difficult to measure. Some SSC contribution is expected to be captured due to the cosmological volume that is unused in the light-cone. This quantity varies with the source redshift, which therefore introduces a redshift dependence on the simulated SSC term. Additionally, the contribution from density fluctuations with modes larger that the simulation box is completely missing. A lower bound on the missing SSC term could be estimated by imposing a mask in $k$-space instead of a survey footprint in equation (\ref{eq:Cov_SSC}) and carrying out the rest of the SSC calculation to find out the difference on the end product. However our current implementation does not allow us to perform this calculation.

Another approach would consist of validating the matter trispectrum calculations separately. \citet{2017MNRAS.465.4016R} have started to address this validation in the $[\Omega_{\rm m} - \sigma_8]$ plane, but much of the $w$CDM space remains unverified as of yet. If we could establish a range of scales for which the simulations and the theory agree on $P(k)$ and $T^{\delta}(k,k')$, then we could compare the $\cal R_{\ell}$ measurements, excluding the $\ell$-modes that are contaminated by the unresolved scales, and any differences could be solely attributed to the difference in the SSC term. The latter could further be improved in $w$CDM cosmologies with a proper calibration of the halo mass function, as discussed in Appendix \ref{sec:SLICS_vs_model}. We could then possibly down-scale the analytical ${\rm Cov}_{\rm SSC}^{\kappa}$  term until a match with the mock data is achieved.  Again, 
changes to $\cal R_{\ell}$ caused by trispectrum modelling errors and resolution limits will be wrongly interpreted as variations in the total SSC contribution captured by the simulations. When we performed this test with the {\it cosmo}-SLICS excluding the $\ell$-modes in the grey zone of Fig. \ref{fig:NG_G_ratio_FID}, we estimated that our simulations contain about 75\% of the SSC at $z_{\rm s}=1$. This is also what we found in the cross-correlation coefficient terms (Fig. \ref{fig:r}), although this number varies from model to model. It is nevertheless reassuring that the global impact of these differences on the cosmological constraints is rather limited, as demonstrated by our Fisher forecasts. 
\\

\section{Conclusions}
\label{subsec:conclusion}

We  introduced in this paper the {\it cosmo}-SLICS, a new suite of $w$CDM  weak lensing simulations covering a wide parameter space. The range was chosen such as to enclose most of the posterior distributions about $\Omega_{\rm m}$, $\sigma_8$, $w_0$ and $h$  measured from the KiDS-450 and DES-Y1 cosmic shear data analyses \citep{KiDS450, DES1_Troxel}. We sampled this 4-dimensional volume at 25 points with a Latin hyper-cube and trained a GP emulator on these nodes, achieving an interpolation accuracy  of 1-2\% over most of the volume on $\xi_\pm$ in the noise-free case. At each of the 25 nodes, we evolved a pair of $N$-body simulations in which the large scale fluctuations mostly cancel, originating from specific constraints on the initial conditions. This allowed us to rapidly approach the ensemble mean with only a fraction of the  computational cost. Our method is largely inspired by the work of \citet{AnguloPontzen}, which we simplified in order to preserve  Gaussianity  in the matter density field, at the cost of losing the exactitude of the cancellation: we  instead engineered a sample variance suppression.

We further ray-traced these simulations up to 400 times each, and showed that the lensing covariance matrix about these {\it pseudo}-independent light-cones was in close agreement with the exact brute force ensemble approach, based on truly independent realizations from the SLICS suite introduced in  \citet{SLICS}. When pushed through a Fisher parameter forecast, we reached a conclusion similar to that of \citet{Petri16}, namely that re-sampling one of our matched-pair of independent simulations yields accurate constraints on dark matter and dark energy parameters.  More specifically, the area of the  $2\sigma$ confidence region varies by less than 6\% between both methods, a result that we verified holds for areas and galaxy densities that emulate the final  KiDS, DES and LSST surveys.

Having shown that our matched-pair simulation setup led to robust estimates of the lensing covariance matrix, we repeated the measurement at each of the 25+1 cosmological nodes, and compared our results with an analytical covariance calculation based on the halo model  \citep[and implemented in many KiDS cosmic shear analyses, e.g.][]{KiDS450,2017arXiv170605004V,KV450}. We found an excellent agreement on the parameter uncertainty contour between the simulation-based and the theoretical approaches, with a response to cosmology variations that by far exceeds the 6\% effect observed between our two fixed-cosmology estimates. This led us to conclude that evaluating the covariance at the correct cosmology should be prioritised  over improving the accuracy of a covariance matrix estimator at a fixed but offset cosmology, at least for the two-point functions.  The analytical methods naturally allow for this type of calculation,  where one can first evaluate the matrix at a guessed cosmology, then solve for the best fit parameters, update the matrix and iterate; the shortfall of this approach however is that the internal accuracy of the analytical covariance matrix has not been fully verified. Simulation-based covariance matrices are potentially more  flexible in terms of weak lensing measurement method, but it is now clear that biases on the parameter constraints will occur if they are evaluated at the wrong cosmology. The {\it cosmo}-SLICS offer for the first time a way to vary the cosmology in the covariance matrix that is fully simulation-based, and that can therefore be generalised to any weak lensing estimator.

Our primary goal is to facilitate the development of novel lensing techniques beyond the current two-point statistics, and for this reason we make the GP emulator\footnote{https://github.com/benjamingiblin/GPR\_Emulator} public and the simulated light-cone data available upon request. The emulator is flexible enough to train on a variety of input data vectors, and we presented two examples in this paper, the cosmic shear $\xi_{\pm}$ signal (presented in Appendix \ref{sec:emulator}) and the diagonal of the covariance matrices of the lensing power spectrum, ${\rm Cov}^{\kappa}(\ell,\ell)$ (presented in Sec. \ref{subsec:emu_cov}). We introduced various tests to assess the performance of the emulator,  and concluded that the weak lensing signal and variance can be interpolated with an accuracy of 1-2\% and $10\%$, respectively. 

We  envision that interested users will download the mock light-cone  data for their own science case, with the {\it cosmo}-SLICS supporting and accelerating the development of novel, more optimal, weak lensing measurement techniques, besides the two-point statistics. Peak statistics, shear clipping, density-split lensing statistics and Minkowski functionals are examples of promising avenues, and their full deployment relies on the availability of dedicated well controlled calibration samples such as the simulations presented herein. With its extended parameter range, the {\it cosmo}-SLICS probe far outside the target domain of many fit functions, notably for the mass power spectrum \citep[e.g the {\sc HaloFit} calibration by][]{Takahashi2012} and the halo mass function \citep{Tinker2010a}, and hence can serve to re-calibrate these tools.

A larger dimensionality in the cosmology parameter space can be achieved by combining the {\it cosmo}-SLICS with  external simulation suites in which other parameters are varied, and where lensing maps and catalogues are also made available. There is a large gain in cosmological information within reach, and its extraction will require a sustained effort within the community of weak lensing data analysts and simulation specialists. Upcoming lensing surveys such as the LSST\footnote{lsst.org}, {\it Euclid}\footnote{http://sci.esa.int/euclid/} and WFIRST\footnote{https://www.nasa.gov/wfirst} will map dark matter with a billion galaxies, and we must gear up to exploit these exquisite data sets at their maximal capacity.
\\

\begin{acknowledgements}
We would like to thank Martin Kilbinger for his assistance with dissecting {\sc Nicaea}, Alex Barreira for useful discussions on the topic of super sample covariance and for carefully reading the manuscript, Katrin Heitmann, Salman Habib, Jia Liu and Dan Foreman-Mackey for their advice on Gaussian process emulation, Vasiliy Demchenko for his help in cleaning up some of the {\it cosmo}-SLICS products, and Raul Angulo and Catherine Heymans for their suggestions on the methods and on the manuscript, respectively. JHD and BG acknowledge support from the European Research Council under grant number 647112. BJ acknowledges support by the UCL CosmoParticle Initiative.
Computations for the $N$-body simulations were enabled in part by support provided by Compute Ontario (www.computeontario.ca), Westgrid (www.westgrid.ca) and Compute Canada (www.computecanada.ca).
\\

All authors contributed to the development and writing of this paper.
JHD led the simulation effort and the analysis; BG implemented and tested the Gaussian process emulator; BJ led the modelling of the analytical covariance matrix.
\end{acknowledgements}

\bibliographystyle{aa}
\bibliography{cosmoSLICS}


\begin{appendix}
\section{The {\it cosmo}-SLICS Emulator}
\label{sec:emulator}

\subsection{Emulation strategy}

In this Section, we describe the basics of employing a Gaussian process regression emulator to train on the {\it cosmo}-SLICS suite and thus predict weak lensing statistics for $w$CDM cosmologies. We present the accuracy of the emulator's predictions of the shear correlation functions, $\xi_\pm$, as a function of the galaxy angular separation and cosmological parameters, by comparing to theoretical predictions from {\sc Nicaea}, ran with the recalibrated {\sc HaloFit} model \citep{Takahashi2012}, and assume these results representative of those which would be obtained for an arbitrary cosmological statistic measured from these simulations. We calculated the shear correlation functions from our simulations using the public  {\sc TreeCorr} software in 9 bins of angular separation, $\vartheta$, logarithmically spaced between 0.5 and 300 arcmin. We further show to what extent the accuracy of the emulator depends on the distribution of the cosmological parameters, $\boldsymbol{\pi}=\{ \Omega_{\rm{m}}, S_8, h, w \}$, rather than the noise on the training set predictions, by  replacing the simulated $\xi_\pm$ from {\it cosmo}-SLICS with the noise-free theoretical $\xi_\pm$. We used the public {\sc Scikit learn} Gaussian process regression code\footnote{\url{https://scikit-learn.org/stable/modules/gaussian_process.html}}  and the KV450 $n(z)$ for all analyses in this Section.     

 \medskip

The mathematics behind GP regression emulators have been covered extensively in previous work; we refer the interested reader to \citet{Rasmussen_Williams_2006} for a general discussion of GP and to \citet{Habib_etal_2007} and \citet{2008PhRvD..78f3529S} for its applications in cosmology. Here we summarise only the key details of this methodology. 

\medskip

GP regression is a non-parametric Bayesian machine learning algorithm for constraining the distribution of functions which are consistent with observed data. Typically, we have a training data set, $\mathcal{D}$, consisting of $n$ measurements of an observable, $\boldsymbol{y}$, corresponding to different input parameters $\boldsymbol{\pi}$, i.e. $\mathcal{D}=\{ (\boldsymbol{\pi}_j,y_j)|j=1,...,n \}$. The {\it cosmo}-SLICS $\xi_\pm$ predictions can be regarded as 9 such data sets corresponding to the 9 $\vartheta$ bins, with each set consisting of the measurements from the $n=26$ different $d$-dimensional cosmological parameter vectors, $\boldsymbol{\pi}$, where $d=4$. Based on this training set, the task of the GP emulator is to learn the distribution of functions, $f(\boldsymbol{\pi})$, which are consistent with the mapping between the training set input parameters - the `nodes' - and output, via

\begin{eqnarray}
y(\boldsymbol{\pi}) = f(\boldsymbol{\pi}) + \epsilon_{\rm n}(\boldsymbol{\pi}) \,, 
\label{eqn:yModel}
\end{eqnarray}

\noindent where $\epsilon_{\rm n}(\boldsymbol{\pi})$ is a noise term sampled from a mean-zero Gaussian distribution with a standard deviation given by the error on $y(\boldsymbol{\pi})$, the training set observable. The prediction, $y^*$, corresponding to an arbitrary coordinate $\boldsymbol{\pi^*}$, is then sampled from a generalisation of a Gaussian posterior probability distribution over the range of consistent functions. In other words, the GP emulator interpolates the observables from the input coordinates of the training set to trial coordinates across a $d$-dimensional parameter space.


\medskip

A key ingredient of our posterior is the Gaussian prior distribution of functions deemed to reasonably map between input and output. The prior is determined by a mean, conventionally taken to be zero, and a covariance function, known as the `kernel'. The kernel can take various functional forms, each described by a vector of hyperparameters, $\boldsymbol{h}$, governing the kernel's behaviour. Following \cite{Heitmann_etal_2009}, in this work we adopted the squared exponential form, which has $\boldsymbol{h} = \{ A, p_1, \cdots, p_d\}$ and specifies the covariance between the functions $f(\boldsymbol{\pi})$ and $f(\boldsymbol{\pi^*})$ as

\begin{eqnarray}
K(f,f^*; \boldsymbol{h}) \equiv  {\rm{cov}}\left( f(\boldsymbol{\pi}),f(\boldsymbol{\pi^*}) ; \boldsymbol{h} \right) = A \prod_{l=1}^d {\exp}\left[ \frac{(\pi_l - \pi^*_l) ^2}{p_l^2} \right] . \label{eqn:kernel}
\end{eqnarray}

\noindent This kernel has the following properties: (1) the covariance varies smoothly within the parameter space; (2) it depends only on the Euclidean distance between points, such that $K(f,f^*; \boldsymbol{h}) = K(f^*,f; \boldsymbol{h})$; (3) predictions become maximally correlated when $\boldsymbol{\pi} = \boldsymbol{\pi^*}$; (4) the correlation is large for points in relative proximity and small for largely separated points; (5) each $p_l$ corresponds to the functions' characteristic length-scale of variation in each of the $d$ dimensions, while $A$ is the kernel amplitude.

\medskip

The emulator is generally trained by finding values for the hyperparameters which define a distribution of functions that are optimally consistent with all realisations in the training set. In this work, we fit for these using the method built-in to {\sc{Scikit learn}}, which employs a gradient ascent optimisation of the marginal likelihood conditioned on the training set. Emulator accuracy is also strongly affected by the shape of the observable being predicted, performing best for smooth monotonic functions with narrow dynamic ranges. Since the $\xi_\pm(\vartheta)$ statistics vary over orders of magnitude, ${\ln} \xi_\pm(\vartheta)$ presents a wiser choice of quantity to emulate. We found that emulation performance is further improved by decomposing the ${\ln} \xi_\pm(\vartheta)$ observable into a linear sum of $n_\Phi$ orthogonal basis vectors, $\phi^i_\pm(\vartheta)$ where $i \in [1,n_\Phi]$, using a principal component analysis (PCA),  

\begin{eqnarray}
{\ln} \xi_\pm(\vartheta; \boldsymbol{\pi}) = \mu_\pm(\vartheta) + \sum_{i=1}^{n_{\phi}} \phi^i_\pm(\vartheta) w_\pm^i(\boldsymbol{\pi}) + \epsilon_\pm^i(\boldsymbol{\pi})  + \epsilon_\pm^{\rm{PCA}}(\boldsymbol{\pi})  \,,
\end{eqnarray} \label{eqn:Logxipm_PCAmodel}

\noindent where $\mu_\pm(\vartheta)$ is the mean across the training set ${\ln} \xi_\pm(\vartheta; \boldsymbol{\pi})$ predictions, and the orthogonal basis functions, $\phi^i_\pm(\vartheta)$, are calculated from a PCA of the mean-subtracted training set. In this formulism, the weight parameters, $w_\pm^i(\boldsymbol{\pi})$, specifying how much each basis function contributes to the ${\ln} \xi_\pm(\vartheta; \boldsymbol{\pi})$ recipe for a given $\boldsymbol{\pi}$, now become the target of our emulator's predictions, taking the place of $y(\boldsymbol{\pi})$ in equation (\ref{eqn:yModel}), rather than ${\ln} \xi_\pm(\vartheta; \boldsymbol{\pi})$ itself. The $\epsilon_\pm^{\rm{PCA}}$ and $\epsilon_\pm^i$ are terms arising from two different sources of error, that vary slightly between the {\it cosmo}-SLICS cosmologies.

\medskip
$\epsilon_\pm^{\rm{PCA}}$ arises if one uses an insufficient number of basis functions to reconstruct the emulated statistic. PCA decomposition is a standard procedure \citep[see for example][]{Habib_etal_2007,2008PhRvD..78f3529S, MiraTitan}, facilitating improvements in emulation time where $n_\Phi$ is less than the length of the statistic of interest, in this case determined by the number of $\vartheta$ bins. Computational expense is not a problem for our $\xi_\pm(\vartheta)$ measured from {\it cosmo}-SLICS however, consisting of only 9 bins in angular separation. Hence we simply set $n_\phi=9$, for perfect PCA reconstruction of the ${\ln} \xi_\pm(\vartheta; \boldsymbol{\pi})$. We verified however that this number is sufficient to reconstruct more than 99.99\% of the variance in theoretical ${\ln}\xi_\pm$ sampled in 70 bins and that using more basis functions has minimal effect on the emulator accuracy. Hence, with 9 basis functions the error induced from the PCA reconstruction is negligible.  

\medskip

The remaining error term, $\epsilon_\pm^i(\boldsymbol{\pi})$, comes from the Gaussian noise, denoted by $\epsilon_{\rm n}(\boldsymbol{\pi})$ in equation (\ref{eqn:yModel}), arising from uncertainties on the training set. To inform the emulator of the error on the {\it cosmo}-SLICS predictions, we first calculated the standard deviation of the $\ln \xi_\pm(\vartheta;\boldsymbol{\pi})$ across the 25 light-cones and 2 seeds for each cosmology, $\sigma_\pm(\vartheta;\boldsymbol{\pi})$. We translated this into uncertainties on the PCA weights by computing the upper and lower bounds, given by

\begin{eqnarray}
w_\pm^{i,\rm{upper}} = \sum_{m=1}^{9} \Phi^i_\pm(\vartheta_m) \left[ \ln \xi_\pm(\vartheta_m) +  \left(\sigma_\pm(\vartheta_m)/\sqrt{50}\right) \right] \,, \nonumber
\end{eqnarray}
\vspace{-3mm}
\begin{eqnarray}
w_\pm^{i,\rm{lower}} = \sum_{m=1}^{9} \Phi^i_\pm(\vartheta_m) \left[ \ln \xi_\pm(\vartheta_m) -  \left(\sigma_\pm(\vartheta_m)/\sqrt{50}\right) \right] .
\end{eqnarray}

\noindent Here, the $\xi_\pm$ is the average of the measurements for the different light-cones and seeds per cosmology, the factor $\sqrt{50}$ is included to scale the standard deviation to an error on the mean, and for simplicity we have dropped the dependence on the cosmological parameters. The error on the PCA weight, approximated as

\begin{eqnarray}
\epsilon_\pm^i = \frac{1}{2} \left( w_\pm^{i,\rm{upper}} - w_\pm^{i,\rm{lower}} \right) \,,
\end{eqnarray}

 \noindent serves as the standard deviation of the Gaussian distribution from which the $\epsilon_{\rm n}(\boldsymbol{\pi})$ is sampled. In this work we also emulated noise-free {\sc{HaloFit}} predictions; in these cases we set the $\epsilon_{\rm n}$ for all $\boldsymbol{\pi}$ to the arbitrarily-small constant default value in {\sc{Scikit learn}}\footnote{One cannot set $\epsilon_{\rm n}=0$ or the marginal likelihood, entering into the posterior from which predictions are sampled, becomes singular.}.

\medskip

All results presented in this work demonstrating the emulator performance correspond to accuracies in the inferred $\xi_\pm$, and not the logarithmic transforms of these statistics nor the weight vectors, $\boldsymbol{w}_{\pm}(\boldsymbol{\pi})$.

\medskip


\medskip

\subsection{Emulator results}

Having established our emulation strategy, we then sought to test how accurately we can predict the $\xi_\pm(\vartheta;\boldsymbol{\pi^*})$ corresponding to an ensemble of trial cosmologies, $\boldsymbol{\pi^*}$. It is too computationally expensive to produce a fine grid of trial predictions covering the entire 4D parameter space, against which emulator accuracy can be tested. Instead we generated two separate ensembles of trial coordinates. The first, which we refer to as the ``grid" ensemble, $\boldsymbol{\pi^*_{\rm{g}}}$, seeks to illuminate how accurately we can reproduce the predictions for different regions of the emulation space. This ensemble consists of six cosmological parameter grids, with dimensions $50 \times 50$, for the six different 2D projections of the 4D space. For each grid in which two parameters vary, the remaining two are fixed to the corresponding fiducial values from $\{\Omega_{\rm{m}}=0.3251, S_8=0.75245, h=0.7082, w_0=-1.254\}$, selected on account of being the centre of the {\it cosmo}-SLICS training set. This ensemble is useful for identifying for which combinations of cosmological parameters our emulator will perform best and where there is room for improvement. The second, ``bulk", ensemble, $\boldsymbol{\pi^*_{\rm{b}}}$, consists of 300 cosmologies which probe the bulk accuracy of the emulator throughout the emulation space by varying in all 4 parameters simultaneously. We sampled these cosmologies from an independent  4-dimensional Latin hyper-cube with dimensions equal to that of the {\it cosmo}-SLICS training set. 

\medskip

\begin{figure*}
\centering
\includegraphics[width=0.49\textwidth]{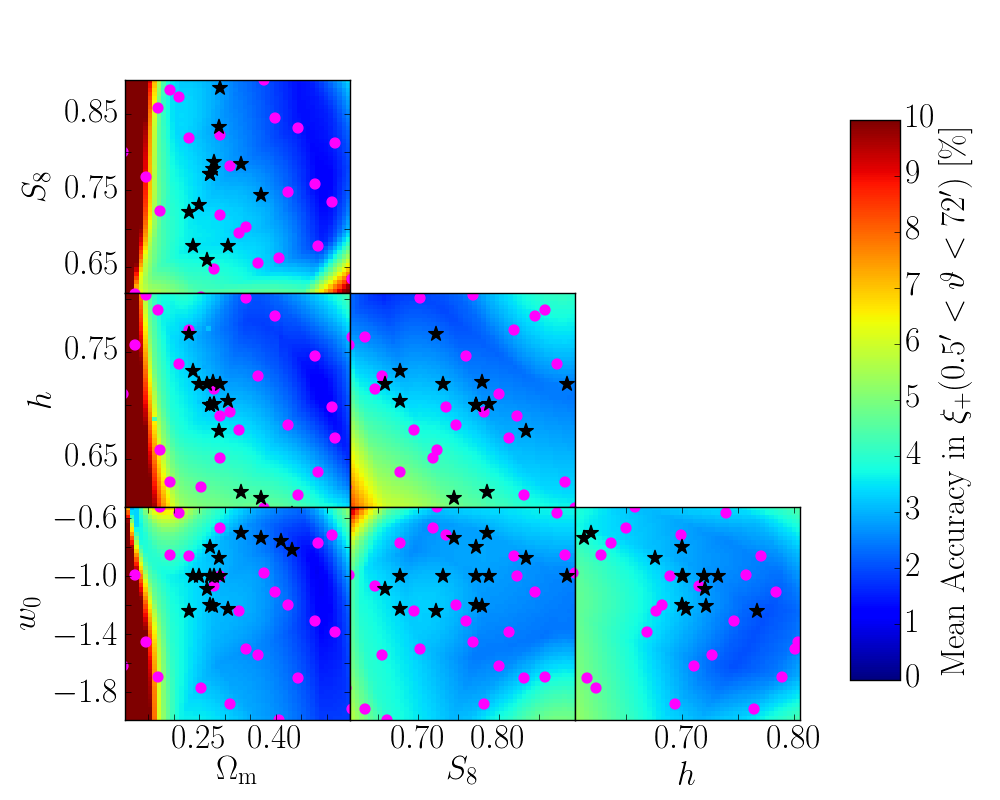} 
\includegraphics[width=0.49\textwidth]{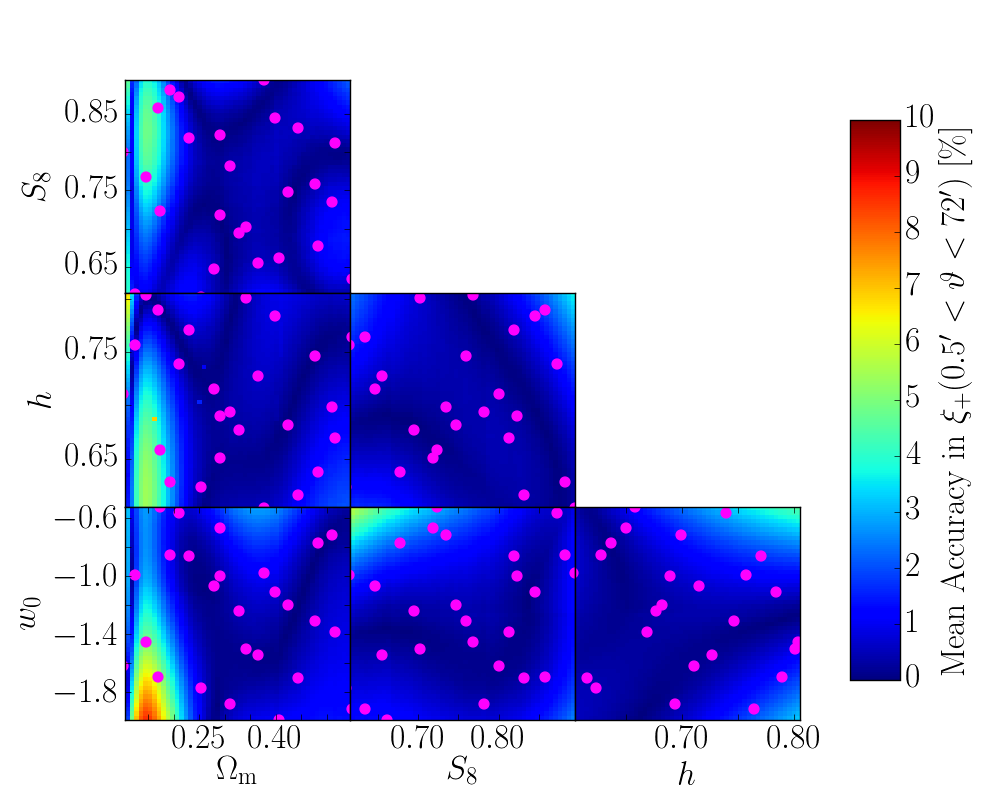} \\
\includegraphics[width=0.49\textwidth]{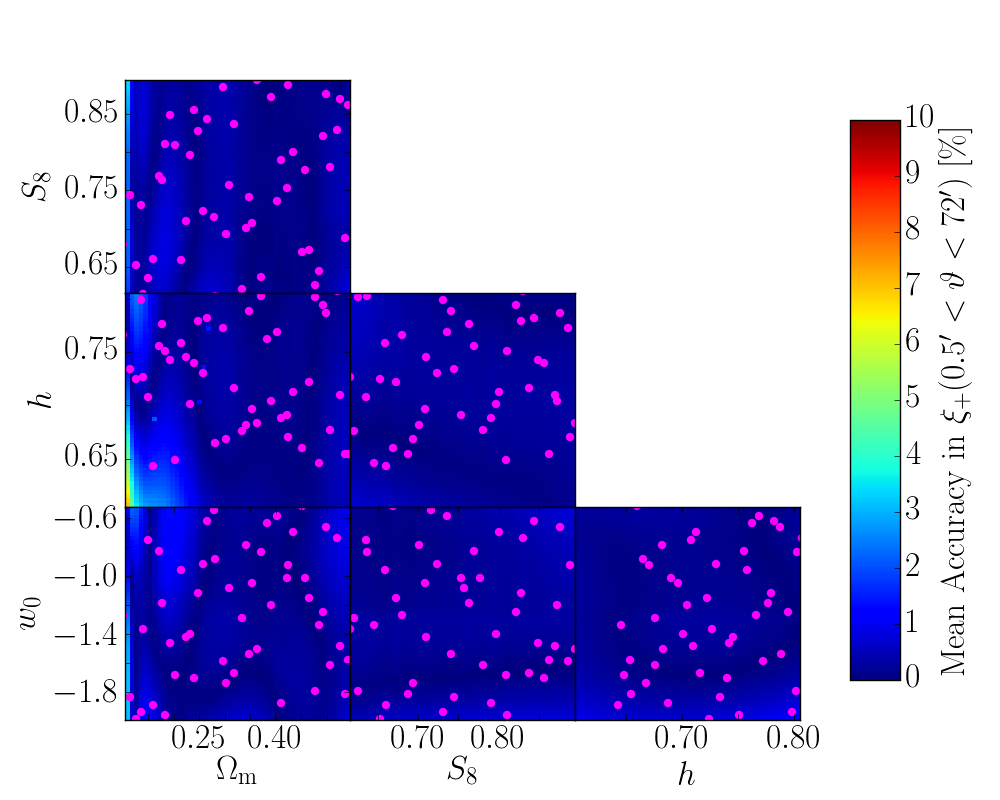} 
\includegraphics[width=0.49\textwidth]{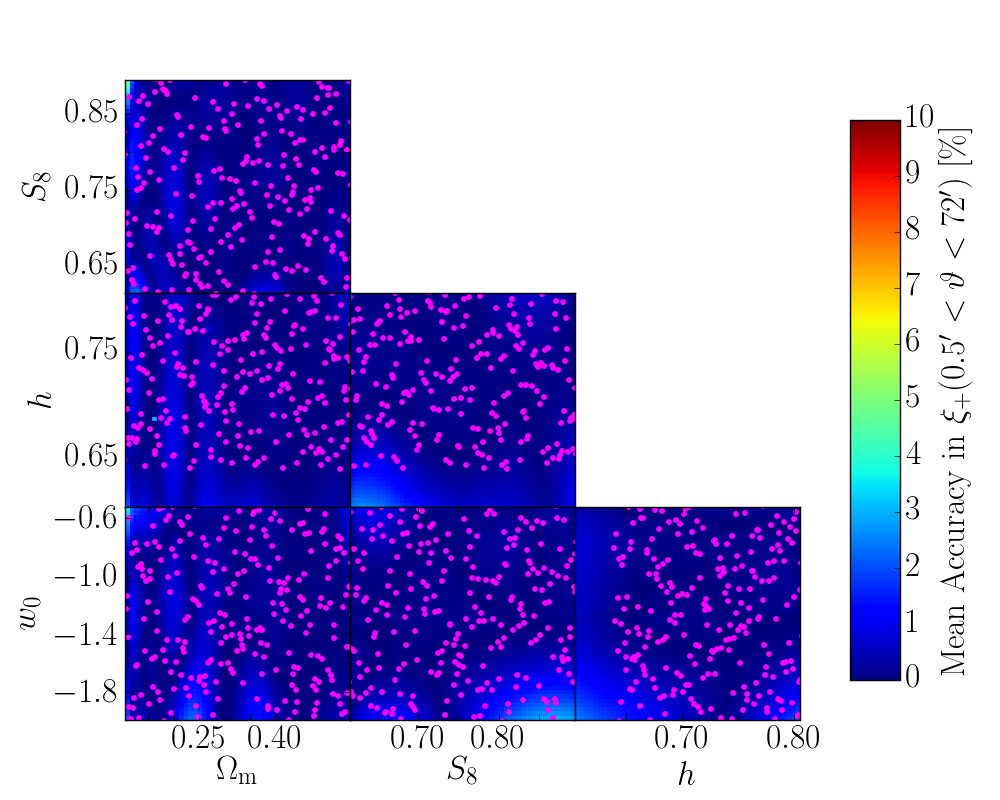} 
\caption{The observed emulator accuracies for $\xi_+$, averaged between 0.5 and 72 arcmin, with the grid ensemble of trial cosmologies, $\boldsymbol{\pi^*_{\rm{g}}}$, shown by the colour maps, when trained on the 26 {\it cosmo}-SLICS predictions (upper-left) and 26, 50 and 250 noise-free {\sc Nicaea} predictions (upper-right, lower-left, lower-right respectively). The training nodes are shown by the magenta circles. The black stars in the upper-left panel show the input parameters of the \protect\cite{Takahashi2012} simulations over our parameters volume (their two highest $\Omega_{\rm m}$ nodes have  $h$ and $S_8$ values that exceed our boundaries). For each grid in which two cosmological parameters vary, the remaining two are fixed to the corresponding fiducial values from $\{\Omega_{\rm{m}}=0.3251, S_8=0.75245, h=0.7082, w_0=-1.254\}$. The contrast between the upper panels, for which the training cosmologies are the same, indicates the extent to which simulation noise and inaccuracies in both the simulations and theoretical predictions degrade the apparent emulation accuracy.}
\label{fig:AccGrids}
\end{figure*}

\medskip

\begin{figure*}
\centering
\includegraphics[width=0.49\textwidth]{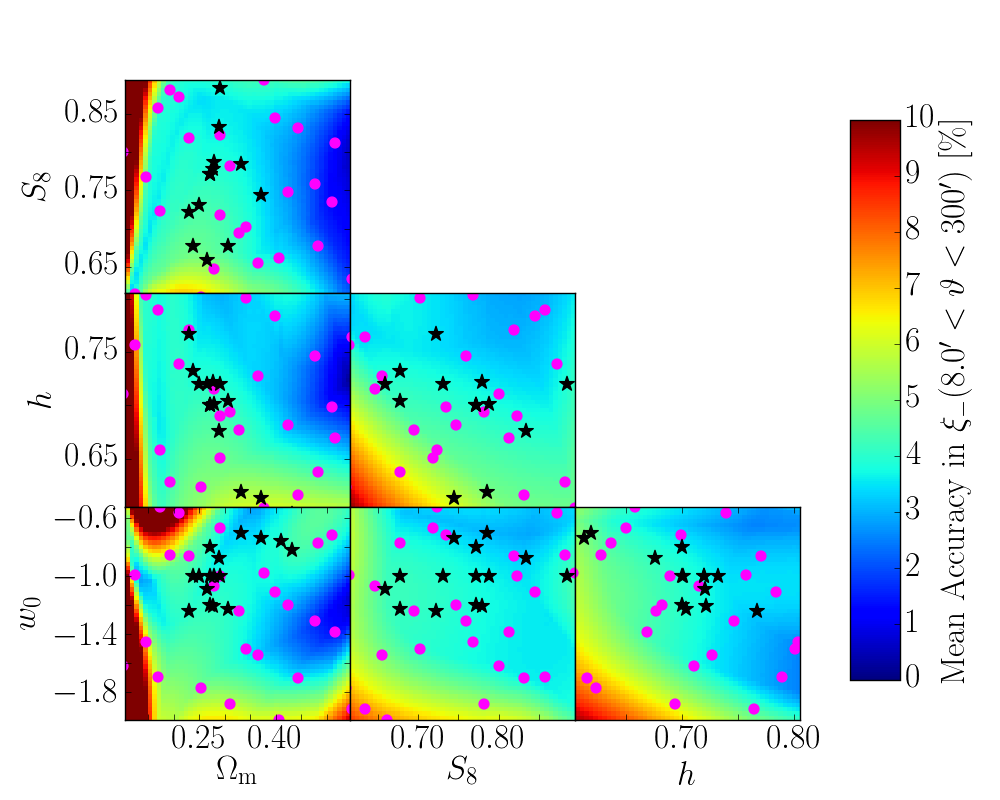} 
\includegraphics[width=0.49\textwidth]{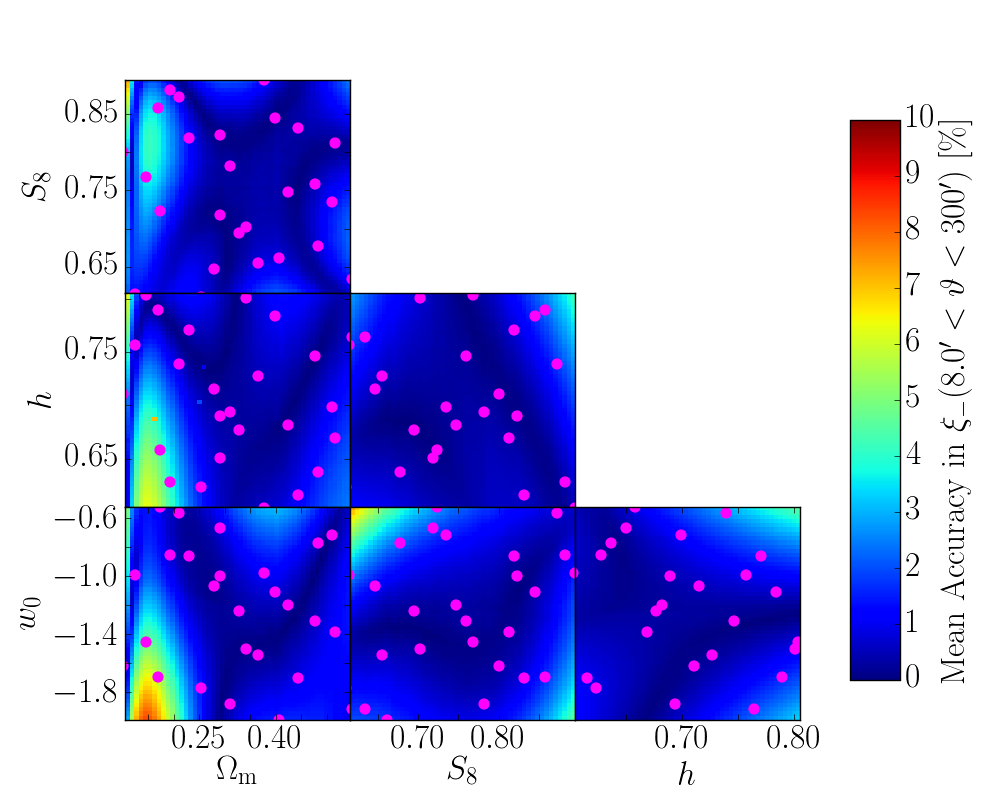} \\
\includegraphics[width=0.49\textwidth]{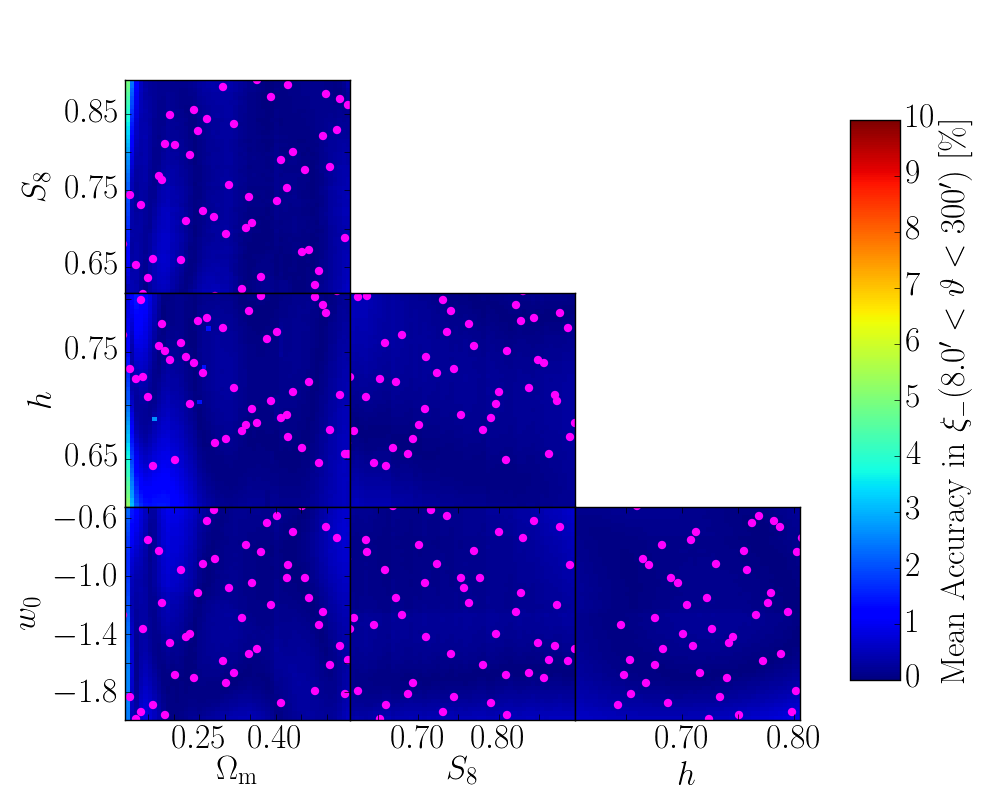} 
\includegraphics[width=0.49\textwidth]{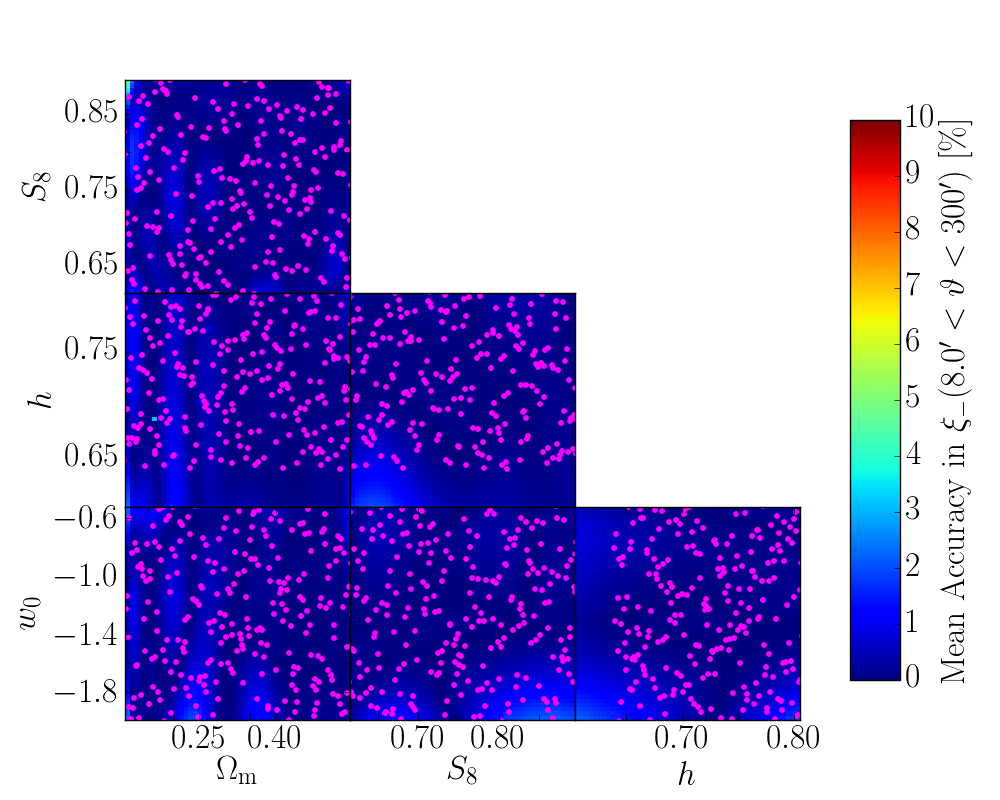} 
\caption{The same as Fig. \ref{fig:AccGrids} but for $\xi_-$ with accuracies averaged between 8 and 300 arcmin.} 
\label{fig:AccGrids2}
\end{figure*}

\medskip

A crucial ingredient in evaluating the emulator's accuracy is a theoretical prediction with which to compare the emulator's. However, the fact that the {\it cosmo}-SLICS $\xi_\pm(\vartheta;\boldsymbol{\pi})$ differ from the corresponding theoretical predictions, as shown by Fig. \ref{fig:AccCS}, means that the emulator will not recover the theoretical predictions used to gauge accuracy, even at the nodes. The disagreement between the two arises not only because of residual noise and small, non-linear angular scales that are not fully resolved in {\it cosmo}-SLICS, but also because of inaccuracies in the {\sc HaloFit} model prescription. These are caused by resolution limitations also present in the simulations used to calibrate the \citet{Takahashi2012} fitting function methodology mentioned earlier, and also the fact that the range of input cosmologies for these mocks does not cover the full range of the {\it cosmo}-SLICS input parameters, especially in the $w_0$ dimension. This is shown by the distribution of black stars (\citealp{Takahashi2012} simulation nodes) relative to the magenta circles ({\it cosmo}-SLICS nodes) in the upper-left panel of Figs. \ref{fig:AccGrids} and \ref{fig:AccGrids2}. The effect of the imperfections in the {\it cosmo}-SLICS (training) and {\sc HaloFit} (trial) predictions on the emulator performance cannot be \textit{completely} disentangled. Therefore, our results for the accuracy of the {\it cosmo}-SLICS emulator should be regarded as a conservative, ``worst case scenario"; performance would likely improve with perfect trial predictions to compare with.  

\medskip

\begin{figure}
\centering
\includegraphics[width=0.4\textwidth]{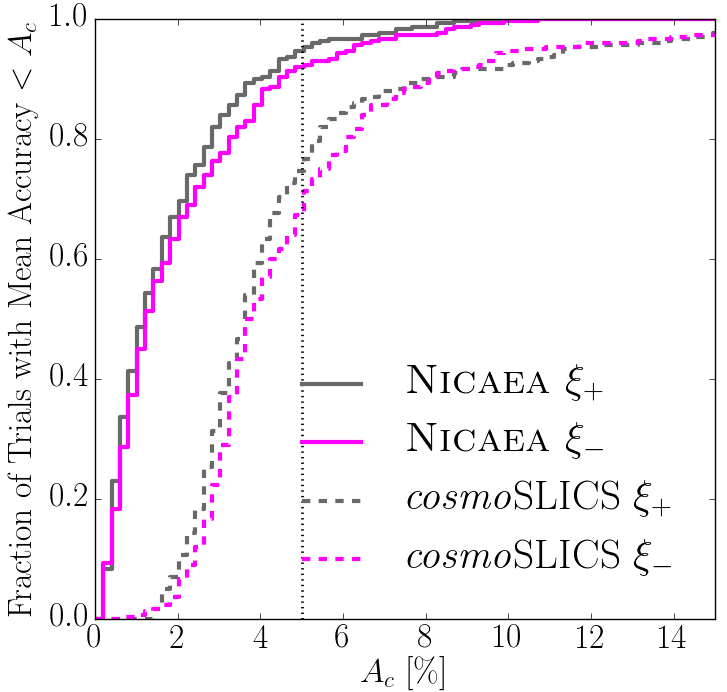} 
\caption{The fraction of the trial cosmologies from the bulk ensemble, $\boldsymbol{\pi^*_{\rm{b}}}$, with accuracies, averaged over a range of angular scales (0.5--72 arcmin for $\xi_+$, 8.0--300 arcmin for $\xi_-$), better than the value, $A_c$, plotted on the horizontal axis. The grey curves correspond to $\xi_+$ predictions, magenta to $\xi_-$. The solid curves result from training the emulator on the noise-free theoretical predictions from {\sc Nicaea}, whereas the dashed result from training on {\it cosmo}-SLICS itself. The decrement in performance when training on {\it cosmo}-SLICS is expected due to the added noise in the training set and inaccuracies in the theoretical predictions.} \label{fig:AccBulk}
\end{figure}

\medskip

To suppress the contribution of inaccuracies on non-linear scales, we considered only the $0.5<\vartheta<72$ arcmin angular range for $\xi_+$ and $8.0<\vartheta<300$ arcmin for $\xi_-$ in evaluating the emulator accuracy. This roughly corresponds to the scales used in the \citet{KiDS450} cosmic shear analysis, but with a slightly higher lower limit for $\xi_-$, to select an angular range with good agreement between {\it cosmo}-SLICS and {\sc Nicaea} predictions for this statistic (see Fig. \ref{fig:AccCS}). In addition to testing the emulator with the {\it cosmo}-SLICS training set, we also tested with noise-free {\sc Nicaea} $\xi_\pm(\vartheta;\boldsymbol{\pi})$ training sets of various sizes. Whereas training with {\it cosmo}-SLICS probes how emulator accuracy is affected by the limitations of both our simulations and the trial {\sc HaloFit} predictions, the latter isolates how well we are able to interpolate $\xi_\pm$ statistics from finite distributions of points. 

\medskip

The accuracies for the emulated $\xi_+$ and $\xi_-$, averaged across the aforementioned $\vartheta$ ranges, for the grid ensemble are shown in Figs. \ref{fig:AccGrids} and \ref{fig:AccGrids2} respectively. The upper-left panel in either figure shows the accuracies when training on {\it cosmo}-SLICS. The remaining panels correspond to the noise-free {\sc Nicaea} sets, increasing in size from that of our simulation suite, to 50 and finally 250 training predictions.

\medskip  

When training on the {\it cosmo}-SLICS mocks themselves, we observe emulation accuracies $\leq 5\%$ in both $\xi_+$ and $\xi_-$ across much of the emulation space, suggesting that the {\it cosmo}-SLICS nodes are well-placed to sample the cosmological dependence on these parameters. Noticeably worse accuracies of 5--10\% manifest at low $\Omega_{\rm{m}}$ values however. Features such as this are expected at the edges of the training set, where there is a lower concentration of nodes from which to interpolate. We also note that this region is not sampled at all by the {\sc HaloFit} training set, hence the predictions completely rely on extrapolation. Similarly, we see edge-effects at some corners in the other projections, but again most of these were not part of the model calibration. The high dependence of the $\xi_\pm$ statistics on $\Omega_{\rm{m}}$ is perhaps the reason why the feature is strongest in the 2D planes with this parameter. Comparison of the upper-left panel to the upper-right, where the training predictions are replaced by noise-free theoretical $\xi_\pm$, reveals how much of the inaccuracy seen when training on {\sc cosmo}-SLICS can be attributed to noise in the simulations and differences between {\it cosmo}-SLICS and the {\sc HaloFit} prescription. The average observed accuracy reduces to $\leq 2\%$ although worse performance continues to be observed at $\Omega_{\rm{m}}<0.2$.

\medskip

The lower two panels of Figs. \ref{fig:AccGrids}--\ref{fig:AccGrids2} show the emulation accuracy when the training sets consist of 50 and 250 noise-free theoretical predictions respectively, with nodes indicated by the magenta points\footnote{The $h$-range for these training nodes, $\in [0.65,0.8]$, reflects that of a previous experimental design for the {\it cosmo}-SLICS suite, before the lower limit of $h=0.6$ was chosen to better represent observational constraints. The cosmologies of the grid ensemble were selected to cover the range of the \textit{present} {\it cosmo}-SLICS suite, hence why the 50 and 250 magenta points do not cover the full grid size in projections featuring $h$. It is not necessary to adjust the distribution of 50 and 250 training points however, since these training sets already permit very accurate extrapolation to these low $h$ values.}. We found that these numbers of training points are sufficient to achieve accuracies around the level of 1\% across all of the explored parameter space, and that the improvement between 50 and 250 nodes is negligible, suggesting the former already samples the cosmological dependence of the $\xi_\pm$ very well. The noticeable improvement increasing from the 26 to 50 training nodes could be considered argument for running {\it cosmo}-SLICS simulations at 50 distinct cosmologies. However, we remind the reader that given an amount of computing resources fixed to 50 runs, opting for running all different cosmologies would lack the benefits of our matched-pair simulation strategy,  which facilitate an unbiased estimate of the true $P(k)$ and $\xi_\pm(\vartheta)$ with a small amount of noise (see Section \ref{subsec:Nbody}). 
We interpret these results instead as evidence that augmenting {\it cosmo}-SLICS with an additional 24 cosmologies each having the matched-pair simulations, would be quite beneficial to emulation performance, especially at low $\Omega_{\rm{m}}$ values, but going beyond this sized suite is unnecessary. Also worth considering is that in this parameter space, baryons contribute to up to 50\% of the total matter density, hence will likely have a different and stronger feedback on the lensing signal. 

\medskip

The results of exploring the bulk accuracy of the emulator, where all 4 cosmological parameters were varied simultaneously in the 300 trial ensemble, is plotted in Fig. \ref{fig:AccBulk}. Here we show the fraction of trial cosmologies for which the mean accuracy across the fiducial angular separation range is better than the threshold, $A_c$, plotted on the horizontal axis. We see that when training on the $N=26$ noise-free  theoretical $\xi_\pm$, our emulator recovers more than 90\% of the trial predictions to better than 5\% accuracy (solid magenta and grey curves). Further inspection reveals that the trial cosmologies with mean accuracies worse than 5\% all reside on the edges of the hyper-cube defined by the training set, where emulation is expected to perform less well. In particular, we see cosmologies with $\Omega_{\rm{m}}<0.2$  over-represented, by factors of 3 (considering $\xi_-$ predictions) and 5 (considering $\xi_+$), in the set of trials which failed to achieve this mean accuracy. This is consistent with our accuracy tests involving the grid ensemble, further pointing to a necessity for extra training nodes to improve the emulation for this part of the parameter space.       

\medskip

The dashed lines in Fig. \ref{fig:AccBulk} demonstrate the cumulative mean accuracy when we instead trained on the {\it cosmo}-SLICS predictions. We observe a decrement in performance relative to the noise-free training set results as expected; for  25\%(33\%) of the trial cosmologies, the mean emulator accuracies for the $\xi_+$($\xi_-$) statistics are worse than 5\%. The slight assymetry in performance for these two statistics is also consistent with grid ensemble tests, where accuracy for emulating $\xi_+$ (Fig. \ref{fig:AccGrids}) when training on the {\it cosmo}-SLICS predictions was slightly better than emulations of $\xi_-$ (Fig. \ref{fig:AccGrids2}). We emphasise once again that these results represent a conservative view of emulation accuracy given {\it cosmo}-SLICS as a training set, owing to the imperfections of the theoretical predictions used for comparison. We hence conclude that our simulation suite permits emulated predictions with accuracies at the level of $\simeq 5\%$ or better. It is possible that accuracy would improve further given an alternative interpolation strategy, such as sparse polynomial chaos expansion, as exercised by \cite{EuclidEmulator_etal_2018}. We leave investigation of this for future work.


\section{Comparison with Theory}
\label{sec:SLICS_vs_model}

\begin{figure*}
\begin{center}
\includegraphics[width=6.0in]{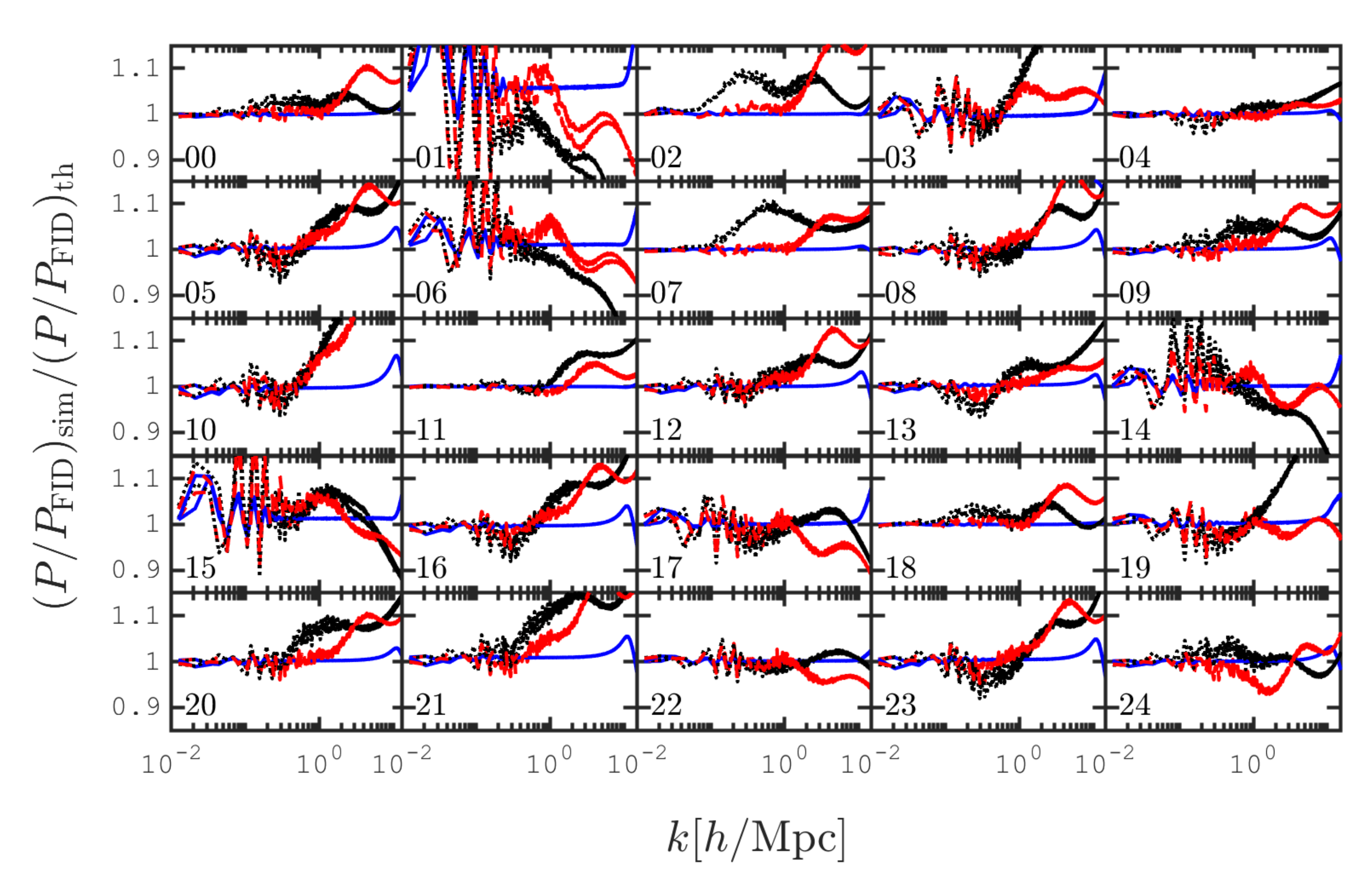}
\caption{The sampling variance cancels when computing ratio between simulated power spectra, which eases the comparison with theoretical predictions. This figure shows a comparison between these ratios, when computed from the {\it cosmo}-SLICS (denoted with subscript `$\rm{sim}$') or from {\sc HaloFit} (subscript `$\rm{th}$'). More precisely, we compute $P_{\rm model}(k)/P_{\rm FID}(k)$ for both cases and for all 25 cosmological models, and examine the ratio between the two estimates at $z=120$ (blue), $z\sim0.6$ (red) and $z\sim0.0$ (black). }
\label{fig:ratio_ratio}
\end{center}
\end{figure*}

\begin{figure*}
\begin{center}
\includegraphics[width=7.0in]{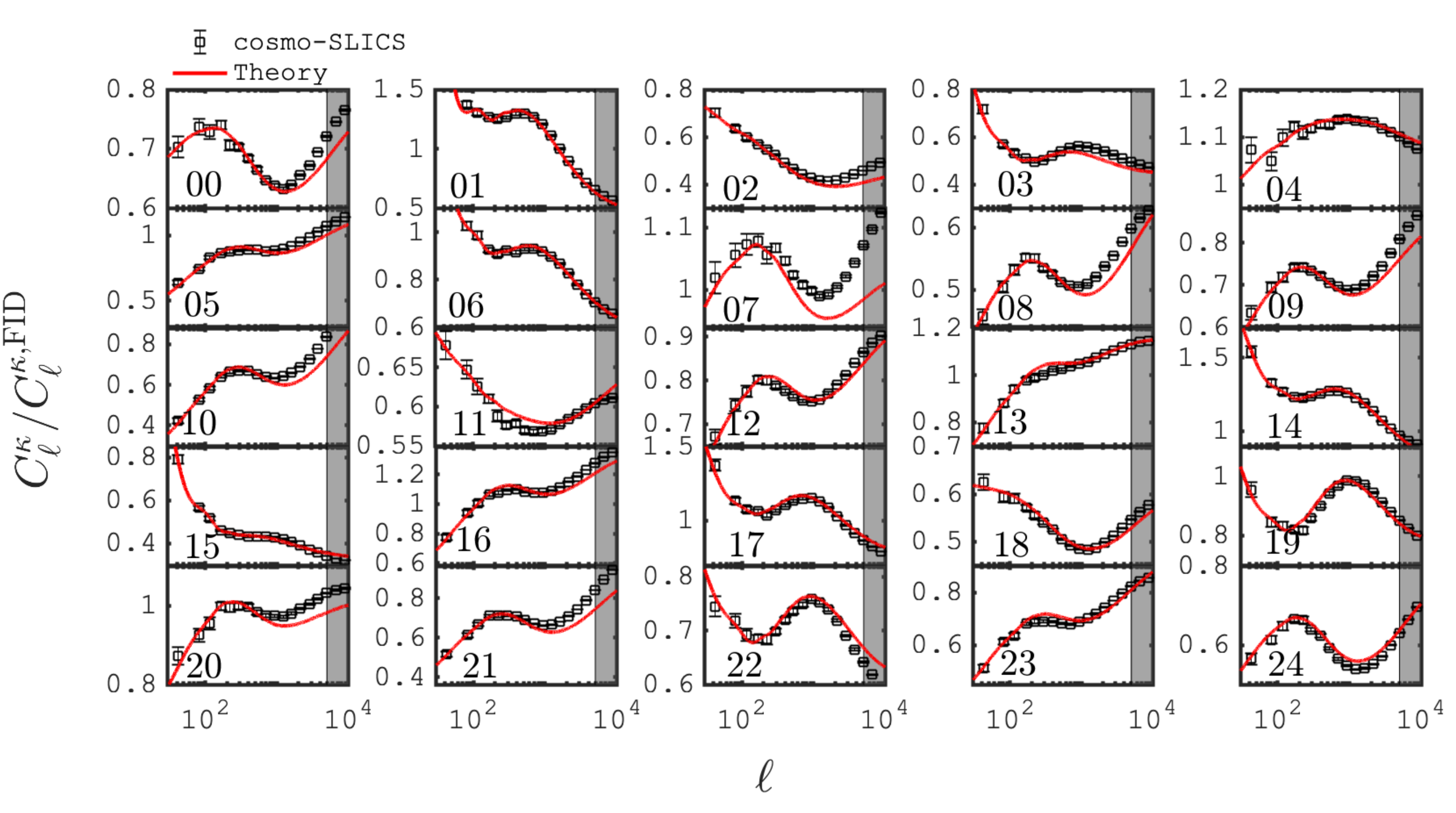}
\caption{Ratio between the lensing convergence power spectra from all 25 $w$CDM cosmological models and that from  model-FID. The symbols are from the simulations, the red lines from the  theoretical predictions. These measurements show the average over the 800 {\it pseudo}-independent line-of-sights, and the error bars represent the error on the mean.}
\label{fig:ratio_C_ell_all}
\end{center}
\end{figure*}

The overall accuracy of the $N$-body simulations is generally well captured by the matter power spectrum $P(k)$, which provides a per-scale assessment of the resolution, and which is straightforward to compare with publicly available fit functions or emulators.  In Sec. \ref{subsec:accuracy} we explained why ratios of $P(k)$ provide  noise-free estimates, and we provided an example in Figs. \ref{fig:deriv_pk} and \ref{fig:deriv_Cell}, where we compared model-12 to model-FID in the form of $P_{\rm 12}(k)/P_{\rm FID}(k)$ and $C_{\ell}^{\kappa,12}/C_{\ell}^{\kappa,{\rm FID}}$, respectively. In this Appendix, we further examine the agreement between our theoretical predictions and the {\it cosmo}-SLICS. 

We present in Fig. \ref{fig:ratio_ratio} the ratio between the simulation estimate of $P_{\rm model}(k)/P_{\rm FID}(k)$ and the corresponding {\sc HaloFit} calculations, where the `model' subscript cycles through all 25 $w$CDM cosmologies. The redshift dumps vary between cosmological models, hence we show here a comparison at $z=120$ (blue), $z\sim0.6$ (red) and $z\sim0.0$ (black). We notice that although some models display an excellent agreement over the full range of scales and redshifts (e.g. models-04 or -22), most exhibit deviations of order 5-10\% in the non-linear regime, some even stronger (models-01, -03, -19 and -21 in particular). Model-01 takes on particularly extreme values of $\sigma_8$ ($=1.34$) and $\Omega_{\rm m}$ ($0.10$), models-03 and -19 have high values for their dark energy equation of states, with $w_0 \sim -0.5$), while that same parameter becomes very low in model-21 ($w_0=-1.99$).  Also, models-01, -15, -06, -14, -03 and -17 take very values of  $\Omega_{\rm m}$, and we see discrepancies even at $z_i = 120$. This seems to points to a miss-match in the BAO amplitude imposed in the simulations by the CAMB transfer function, and that computed by the CAMB code. Very likely this has to do with the fact that the code treats cold dark matter and baryons the same way, while CAMB does not, causing this shift.  Many {\it cosmo}-SLICS models fall outside the calibration range of {\sc HaloFit}, where the predictions are less robust; generally the match  between the ratios degrades in the non-linear regime.

We also note that in some cases, the black and the red lines split at high-$k$, meaning that the two seeds evolve slightly differently (see, for example, model-01). This is not expected and points to residual systematics in the simulations, most likely caused by numerical errors and affecting the $P(k)$ at the 1-2 \% level. This is much smaller that the overall difference with respect to {\sc HaloFit} (at the 10-20 \% level), hence is sub-dominant. 

We show the accuracy of our weak lensing light-cones for all models in Fig. \ref{fig:ratio_C_ell_all}, where we compare the ratio between our $w$CDM power spectra and the $\Lambda$CDM case, model-FID. The measurements from the {\it cosmo}-SLICS are in excellent agreement with the predictions over a wide range of scales. Some discrepancies are observed in the non-linear regime, where both the theory and simulations are known to be less  accurate.

\begin{figure}
\begin{center}
\includegraphics[width=3.5in]{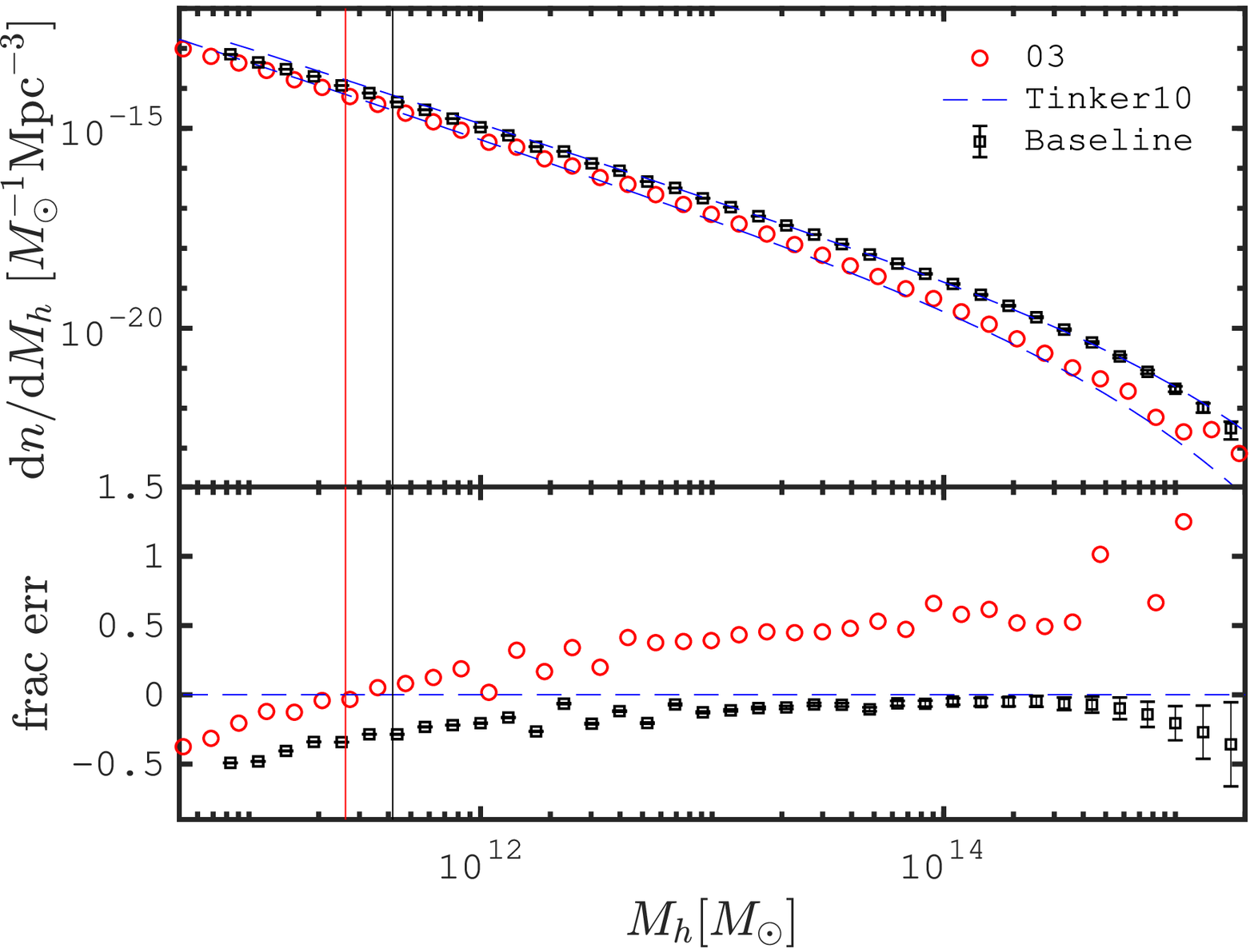}
\caption{ Comparison between the halo mass function measured from the simulations (symbols) and the fit function from \citet[][shown with the blue dashed lines]{Tinker2010a}. The red circles present the measurements from model-03 at redshift $z=0.043$, while the black squares are from the SLICS simulations (hence the error bars).  The lower panel shows the fractional error between simulations and models, where the latter is taken as the reference. The vertical lines mark the mass of dark matter haloes containing 100 particles, which varies between cosmologies due to changes in the particle mass.}
\label{fig:dndm}
\end{center}
\end{figure}

Finally, we compare in Fig. \ref{fig:dndm} the halo mass function measured in the simulations, with that computed from the \citet{Tinker2010a} fit function. We show our results for the $\Lambda$CDM case in black, extracted from the SLICS simulations, and for the $w$CDM model-03, in red, both taken at redshift $z=0.04$. Model-03 is particularly interesting here as it corresponds to the uppermost blue line in the bottom panel of Fig. \ref{fig:CovCompZ}, which exhibits strong differences in variance between simulations and theory. We see that the lack of variance observed in the analytical model can be directly linked to an undershoot of the halo mass function, which is systematically lower than in the simulations. Given that the \citet{Tinker2010a} fit function was only calibrated with $\Lambda$CDM simulations\footnote{The \citet{Tinker2010a} fit to the halo mass function is calibrated over the range $\Omega_{\rm m} \in [0.2, 0.3]$, $\sigma_8 \in [0.75, 0.9]$, $h \in [0.7, 0.73]$, $\Omega_{\rm b} \in [0.040, 0.045]$ and $n_{\rm s} \in [0.94, 1.0]$. }, it is not too surprising to see such large deviations when the dark energy equation of state deviates significantly from $w_0=-1.0$. The {\it cosmo}-SLICS open up a possibility to recalibrate the halo model fit functions in that context, which we leave to future work.


\section{Covariance Estimation with a matched-pair of $N$-body runs}
\label{sec:CovEstimator}

\begin{figure}
\begin{center}
\includegraphics[width=3.5in]{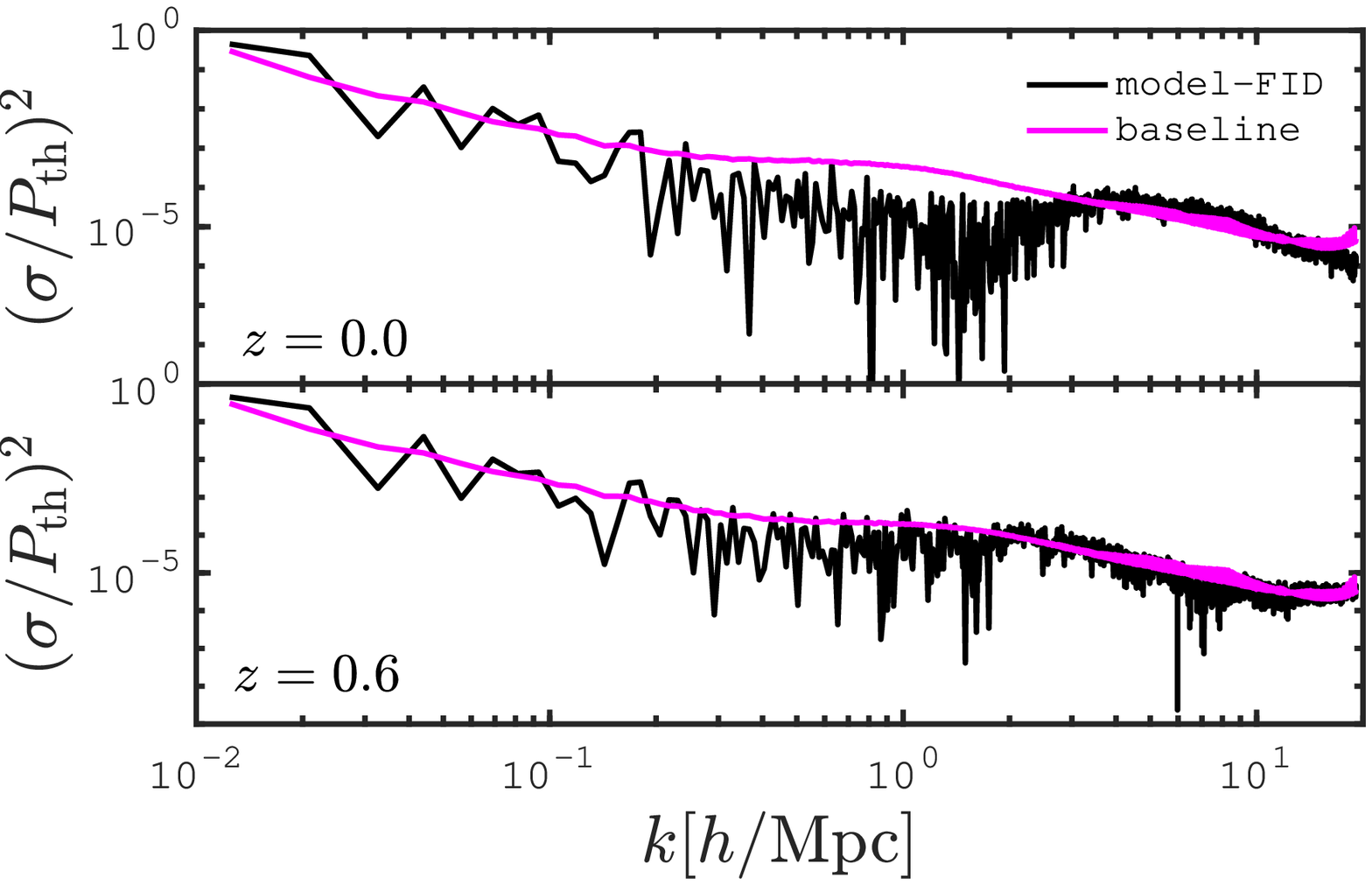}
\caption{Comparison between the signal-to-noise, $\left(\sigma/P(k)\right)^2$, extracted from the SLICS simulations and that estimated from the matched-pair. Upper and lower panels show different redshifts.}
\label{fig:EstimatorTest}
\end{center}
\end{figure}

\begin{figure}
\begin{center}
\includegraphics[width=3.5in]{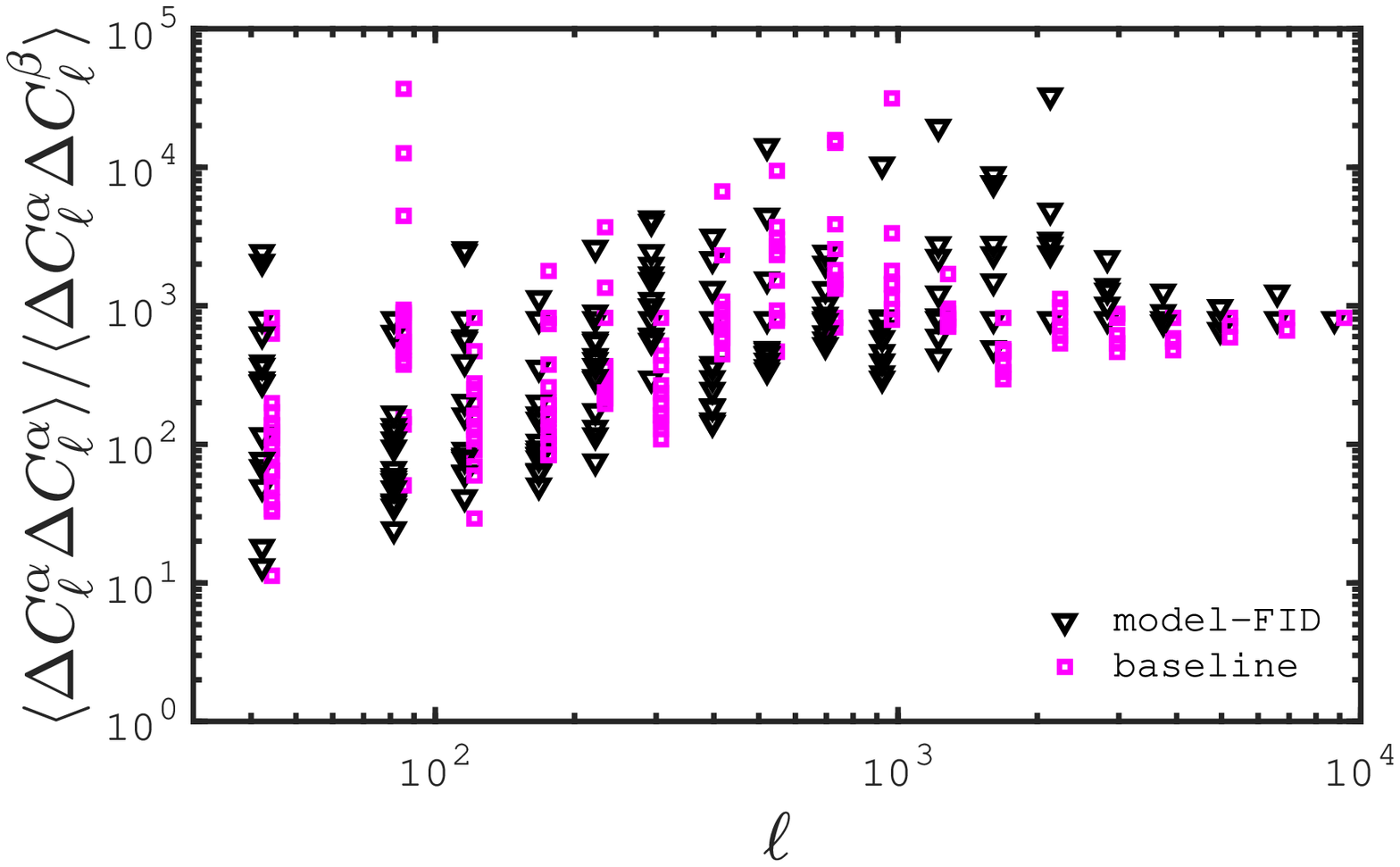}
\caption{Ratio between the elements of the standard covariance matrix, $\bigg\langle \Delta C_{\ell}^{\kappa, \alpha} \Delta C_{\ell'}^{\kappa, \alpha} \bigg\rangle$, and those from the `cross-sample' covariance, $\bigg\langle \Delta C_{\ell}^{\kappa, \alpha} \Delta C_{\ell'}^{\kappa, \beta} \bigg\rangle$, where $\alpha, \beta$ label individual light-cones, and $\alpha \ne \beta$. These matrices contain $18^2$ elements, hence for every $\ell$-mode we plot the 18 $\ell'$ components of the baseline (in magenta squares, offset for clarity) and model-FID estimator (in black triangles).}
\label{fig:residualCov}
\end{center}
\end{figure}

The model-FID covariance estimation described in Section \ref{sec:Covariance} is a hybrid method between the ensemble
approach from independent measurements (two here) and an internal resampling technique. Ray-tracing effectively selects a part of the total simulated volume to extract a light-cone, hence extracting multiple light-cones is equivalent to drawing multiple sub-sets of the simulated data while allowing for repetitions, parent to the bootstrap approach. In this Section we expand on the method and further investigate why it works so well in this context. 

To restate the set-up, the matched-pair are constructed from two $N$-body simulations evolved at the same cosmology, in which the random seeds are chosen such that the initial fluctuations in the matter power spectrum are Gaussian, they cancel to better than 5\%, and oscillate about the mean with crossing at (almost) every $k$-mode.  More than one solution exists that can satisfy these conditions, and we used an empirical approach to draw our matched-pair from an ensemble of initial conditions. We show in Fig. \ref{fig:EstimatorTest} the variance extracted from this pair, compared to the baseline variance, and observe that large and small scales are in excellent agreement, however the model-FID variance is low over the range $k \in [0.2 - 3.0]$ at $z=0$. The level of agreement at this stage is surprisingly high, and some other choice of pairs (i.e. not matched) produce a variance that deviate significantly more, both at large and small scales \citep[see figure 4 in][]{HDP13}. The small discrepancies are subsequently suppressed during the line-of-sight projection that leads to weak lensing observables.

Each member of the pair was ray-traced 400 times, for a total of 800 {\it pseudo}-independent light-cones per pair. The matched-pair covariance estimator can be written from equation (\ref{eq:Cov_sim}), which we repeat here for completeness:
\begin{eqnarray}
{\rm Cov}_{\rm sim}^{\kappa} = \frac{1}{N-1}\sum_{i=1}^{N} \left[\widehat{{C}_{\ell}^{\kappa,i}} - \langle{C}_{\ell}^{\kappa}\rangle \right] \left[\widehat{{C}_{\ell'}^{\kappa,i}} - \langle{C}_{\ell'}^{\kappa}\rangle \right].
\label{eq:Cov_sim_appendix}
\end{eqnarray}
In contrast with the baseline estimate, there is an implicit caveat here, which is that the different realizations are not perfectly independent. This approximation converges to an unbiased estimator in the limits where the mean $\langle{C}_{\ell}^{\kappa}\rangle$ matches the ensemble mean, and where the residual correlations between the multiple light-cones are small. The first condition naturally emerges from the matched-pair by construction, while the second is satisfied when:
\begin{eqnarray}
\bigg\langle \Delta C_{\ell}^{\kappa, \alpha} \Delta C_{\ell'}^{\kappa, \beta} \bigg\rangle \ll \bigg\langle \Delta C_{\ell}^{\kappa, \alpha} \Delta C_{\ell'}^{\kappa, \alpha} \bigg\rangle, \hbox{\hspace{10mm} for $\alpha \ne \beta$},
\label{eq:residual}
\end{eqnarray}
where $\Delta C_{\ell}^{\kappa, \alpha}$ is the mean-subtracted lensing power spectrum measured in light-cone $\alpha$, and the angular brackets refer to the ensemble average over our realizations. 

The term on the right-hand side of equation (\ref{eq:residual}) corresponds to $(N+1)/N$ times the usual lensing covariance matrix, while the term on the left-hand side measures the cross-light-cone covariance matrix. We measured these two terms both from the model-FID and from the baseline, for all $\ell$ and $\ell'$ pairs, averaging over all possible combination of $\alpha$ and $\beta$. We found that in the weakest case, the right-hand side is about ten times larger; for most matrix elements the ratio $\bigg\langle \Delta C_{\ell}^{\kappa, \alpha} \Delta C_{\ell'}^{\kappa, \beta} \bigg\rangle / \bigg\langle \Delta C_{\ell}^{\kappa, \alpha} \Delta C_{\ell'}^{\kappa, \alpha} \bigg\rangle$ is larger than 100, as reported in Fig. \ref{fig:residualCov}. Interestingly, we observe that the model-FID and the baseline scatter plots  are very similar, leading us to the conclusion that the residual correlations across light-cones are negligible.  
\\

%
%

\section{Analytical Covariance Calculations}
\label{sec:CovAna}

In the following we describe the details of the analytical covariance calculation. The code is the same as used in the cosmology analyses of the Kilo-Degree Survey \citep{KiDS450,KiDS450_QE,2017arXiv170605004V,KV450}, with similar implementations also used as default in DES and HSC (\citealp{DES1_Troxel}; \citealp{2018arXiv180909148H}; see also \citealp{cosmolike} for analogous implementation details). We follow \citet{Takada2013a,SSC,2001ApJ...554...56C} closely in our notation.

The matter trispectrum in equation (\ref{eq:trispectrum}) is given by the sum of the terms
\begin{align}
T^{1{\rm h}}(\boldsymbol{k}_1,\boldsymbol{k}_2,\boldsymbol{k}_3,\boldsymbol{k}_4) &= I_4^0(k_1,k_2,k_3,k_4) \;; \\ \nonumber
T^{2{\rm h}}_{22}(\boldsymbol{k}_1,\boldsymbol{k}_2,\boldsymbol{k}_3,\boldsymbol{k}_4) &= P_{\rm lin}(k_{12}) I_2^1(k_1,k_2) I_2^1(k_3,k_4) + 2\, \mbox{perm.}\;; \\ \nonumber
T^{2{\rm h}}_{13}(\boldsymbol{k}_1,\boldsymbol{k}_2,\boldsymbol{k}_3,\boldsymbol{k}_4) &= P_{\rm lin}(k_{1}) I_1^1(k_1) I_3^1(k_2,k_3,k_4) + 3\, \mbox{perm.}\;; \\ \nonumber
T^{3{\rm h}}(\boldsymbol{k}_1,\boldsymbol{k}_2,\boldsymbol{k}_3,\boldsymbol{k}_4) &= B_{\rm PT}(\boldsymbol{k}_1,\boldsymbol{k}_2,\boldsymbol{k}_{34}) I_1^1(k_1) I_1^1(k_2) I_2^1(k_3,k_4)  \\ \nonumber   & \hspace{47mm}  + 5\, \mbox{perm.}\;; \\ \nonumber
T^{4{\rm h}}(\boldsymbol{k}_1,\boldsymbol{k}_2,\boldsymbol{k}_3,\boldsymbol{k}_4) &= T_{\rm PT}(\boldsymbol{k}_1,\boldsymbol{k}_2,\boldsymbol{k}_{3},\boldsymbol{k}_{4}) I_1^1(k_1) I_1^1(k_2) I_1^1(k_3) I_1^1(k_4)\;,
\end{align}
where $P_{\rm lin}$ is the linear matter power spectrum and where $B/T_{\rm PT}$ are the tree-level matter bispectrum and trispectrum, respectively (see e.g. equation 30 in \citealp{Takada2013a} for explicit expressions). Here, halo model integrals were defined as
\begin{eqnarray}
I_\mu^\beta(k_1,k_2, \dots, k_\mu) = \int_0^\infty {\rm d}M\, \frac{{\rm d}n}{{\rm d}M}\, b_\beta \left( \frac{M}{\bar{\rho}_{\rm m}} \right)^\mu \prod_{i=1}^{\mu} \tilde{u}_M(k_i) \;,
\end{eqnarray}
with $\bar{\rho}_{\rm m}$ the mean matter density in the Universe and $\tilde{u}_M$ the Fourier transform of an NFW halo matter density profile (see equation 11 in \citealp{2001ApJ...546...20S}). For the latter we assumed the mass-concentration relation by \citet{2008MNRAS.390L..64D}. Moreover, we set $b_\beta=0$ for $\beta \geq 2$, $b_0=1$, and $b_1=b_{\rm h}(M)$, the halo bias. The expression for the halo bias is consistently matched to the halo mass function, ${\rm d}n/{\rm d}M$. By default, we adopted the fit functions by \citet{Tinker2010a}, but tested the models by \citet{2001MNRAS.323....1S} and \citet{1974ApJ...187..425P} as well. In the results shown in this work we have skipped the two 2-halo contributions to the trispectrum as they have negligible impact on the power spectrum covariance and are time-consuming to compute.

To calculate equation (\ref{eq:Cov_SSC}), we determined the response of the matter power spectrum to a background mode in the halo model as
\begin{eqnarray}
\frac{\partial P(k)}{\partial \delta_{\rm b}} = \left( \frac{68}{21}\! -\! \frac{1}{3} \frac{{\rm d} \ln \left[k^3 I_1^1(k)^2 P_{\rm lin}(k)\right]}{{\rm d} \ln k} \right) I_1^1(k)^2 P_{\rm lin}(k) \hspace{8mm} \nonumber \\
+ I_2^1(k,k).
\end{eqnarray}
The variance of background modes within the survey footprint is given by
\begin{eqnarray}
\label{eq:sigma_survey}
\sigma_{\rm b}^2(\chi,{\cal M}) = \frac{1}{A_{\rm survey}} \int \frac{{\rm d}^2 \ell}{(2 \pi)^2} |\widetilde{\cal M}(\boldsymbol{\ell})|^2 P_{\rm lin}(\ell/\chi,\chi)\;, 
\end{eqnarray}
where $\widetilde{\cal M}$ is the Fourier transform of the survey mask. Since the simulated survey area is small, the flat-sky approximation in equation (\ref{eq:sigma_survey}) is adequate. As we assumed a simple square geometry, the Fourier transform can be determined analytically as
\begin{eqnarray}
\widetilde{\cal M}(\boldsymbol{\ell}) = A_{\rm survey}\, {\rm sinc} \left( \frac{\ell_x}{2} \sqrt{A_{\rm survey}} \right)\, {\rm sinc} \left( \frac{\ell_y}{2} \sqrt{A_{\rm survey}} \right) \;,
\end{eqnarray}
where ${\rm sinc}(x)=\sin x /x$, and where $\ell_{x,y}$ are the Cartesian components of the vector $\boldsymbol{\ell}$. Note that all halo model terms and polyspectra carry a redshift dependence that we have only made explicit as an argument where necessary.

In the Gaussian term (equation \ref{eq:GaussCov}) we based the calculation on the full non-linear matter power spectrum, using the fit function of \citet{Takahashi2012}. We evaluated the lensing efficiencies at the exact redshift of the simulated convergence map, which varies slightly with cosmology. The covariance elements were evaluated at a single effective angular frequency at the logarithmic centre of each bin.

\end{appendix}

\label{lastpage}

\end{document}